\pdfoutput=1
\documentclass[12pt,reqno]{amsart}
\usepackage{amsmath,amsfonts,amssymb,amscd,amsthm,amsbsy, color}
\usepackage[pdftex]{graphicx}
\usepackage{comment}
\usepackage[format=plain,
justification=raggedright,singlelinecheck=false]{caption}

\usepackage{wrapfig}
\usepackage{fancybox}
\usepackage[maxfloats=100]{morefloats}

\usepackage{multimedia}
\usepackage{graphicx,epsfig}

\usepackage{natbib}

\numberwithin{equation}{section}

\newtheorem{theorem}{Theorem}
\newtheorem{meta-thm}[theorem]{Meta-Theorem}
\newtheorem{lemma}[theorem]{Lemma}

\newtheorem{remark}[theorem]{Remark}
\newtheorem{definition}[theorem]{Definition}

\newcommand\beq[1]{ \begin{equation}\label{#1} }
\newcommand{\eeq}{ \end{equation} }

\newcommand\beqa[1]{ \begin{eqnarray} \label{#1}}

\newcommand{\eeqa}{ \end{eqnarray} }
\newcommand{\beqano}{ \begin{eqnarray*} }
\newcommand{\eeqano}{ \end{eqnarray*} }
\newcommand\equ[1]{{\rm (\ref{#1})}}

\def\A{{\mathcal A}}

\def\M{{\mathcal M}}

\def\F{{\mathcal F}}

\def\G{{\mathcal G}}

\def\T{{\mathcal T}}

\def\integer{{\mathbb Z}}

\begin{document}

\title[Dynamics of resonances and equilibria of Low Earth Objects]
{Dynamics of resonances and equilibria of Low Earth Objects}


\author[A. Celletti]{Alessandra Celletti}

\address{
Department of Mathematics, University of Rome Tor Vergata, Via della Ricerca Scientifica 1,
00133 Rome (Italy)}

\email{celletti@mat.uniroma2.it}

\author[C. Gale\c s]{C\u at\u alin Gale\c s}

\address{
Department of Mathematics, Al. I. Cuza University, Bd. Carol I 11,
700506 Iasi (Romania)}
\email{cgales@uaic.ro}

\thanks{Corresponding author: \sl E-mail address: \rm cgales@uaic.ro (C\u at\u alin Gale\c s)}


\baselineskip=18pt              




\begin{abstract}
The nearby space surrounding the Earth is densely populated by
artificial satellites and instruments, whose orbits are
distributed within the Low-Earth-Orbit region (LEO), ranging
between 90 and 2\,000 $km$ of altitude. As a consequence of
collisions and fragmentations, many space debris of different
sizes are left in the LEO region. Given the threat raised by the
possible damages which a collision of debris can provoke with
operational or manned satellites, the study of their dynamics is
nowadays mandatory. This work is focused on the existence of
equilibria and the dynamics of resonances in LEO. We base our results on a simplified
model which includes the geopotential and the atmospheric drag.
Using such model, we make a qualitative study of the
resonances and the equilibrium positions, including their location
and stability. The dissipative effect due to the atmosphere
provokes a tidal decay, but we give examples of different
behaviors, precisely a straightforward passage through the
resonance or rather a temporary capture. We also investigate the
effect of the solar cycle which is responsible of fluctuations of
the atmospheric density and we analyze the influence of Sun and
Moon on LEO objects.
\end{abstract}


\maketitle


\section{Introduction}\label{sec:intro}
The region of space above the Earth is plenty of satellites with
different purposes: Earth's observation including remote sensing
and meteorological satellites, the International Space Station
(ISS), the Space Shuttle, the Hubble Space Telescope. All of them
are moving in the so-called Low-Earth-Orbit (hereafter LEO)
region, which ranges between 90 and 2\,000 $km$ of altitude above
the Earth's surface. Satellites in LEO are characterized by a high
orbital speed and might possess different inclinations, even
reaching very high values as in the case of polar orbits, among
which Sun-synchronous satellites can be found. Satellites can be
permanently located in LEO or they can just cross that region as in
the case of \sl highly elliptical orbits, \rm characterized by a
large eccentricity that leads to big excursions, possibly across
the LEO region.

Being easy to reach, LEOs are convenient for building space
platforms and installing instruments. The disadvantages of placing
a satellite in LEO are due to the closeness to the Earth and to
the air drag. Indeed, the Earth's oblateness has a key role and
must be accurately modeled by including a suitable number of
coefficients of the series expansion of the geopotential (compare
with \cite{FV,Liu}). On the other hand, the presence of the
atmosphere provokes an air drag, which acts as a dissipative force
(see, e.g., \cite{BV2004,Chao,Del1991,Gaias2015}). Its strength
depends on the altitude, since the air density decreases with the
distance from the Earth's surface and it may change due to the
Solar activity (for density models we refer to
\cite{Jacchia,Hedin0, Hedin,ISO}). The drag provokes a tidal decay
of the satellite on time scales which depend on the altitude of
the satellite, hence on the density of the atmosphere. Beside the
gravitational attraction of the Earth, the air drag and the
Earth's oblateness, a comprehensive model includes also the
influence of the Moon, the attraction of the Sun and the Solar
radiation pressure (see \cite{Kaula1962,CGfrontier,CGPbif,EH}). We
refer to \cite{ADRRVDQM, CGmajor, CGminor, CGexternal, CEGGP2016,
DRADVR15, Gedeon, GDGR2016, LDV, Rosengren2013, RS2, VDLC} (and
references therein) for a description of the dynamics at distances
from
the Earth higher than LEO.\\

The large number of satellites in LEO unavoidably provokes a huge
amount of space junk, as a consequence of collisions
between satellites or due to the fact that the satellites' remnants are left
there at the end of their
operational life. The spatial density of the debris has a peak
around 800 $km$, as a consequence of the collisions between the satellites
Iridium and Cosmos in 2009 and the breakup of Fengyun-1C in 2007.
Collisions with space debris might provoke dramatic events, due to
the high speed during the impact. The U.S. Space Surveillance Network
tracks in LEO about 400\,000 debris between 1 $cm$ and 10 $cm$, and 14\,000
debris larger than 10 $cm$. Objects of 1 $cm$ size might damage
a spacecraft and even break the ISS shields; debris of 10 $cm$ size
might provoke a fragmentation of a satellite.\\

More than half of the total amount of space debris
is in LEO, thus increasing the interest for the dynamical behavior
of objects in this region, which is the main goal of the present work.
The knowledge of the dynamics of the space debris can considerably
contribute to the development of mitigation measures, most notably
through the design of suitable post-mission disposal orbits
(see, e.g., \cite{DSBS}).
Among the possible mitigation strategies, one can provoke a re-entry of the debris in the lower atmosphere
or rather a transfer to an orbit with a different lifetime.
It is therefore of paramount importance to know whether an object is
located in a regular, resonant or chaotic region, as well as to know
how much time will spend in such regions. This paper aims to contribute
to give an answer to these question.

This work extends the research performed in \cite{CGmajor,CGminor}, where analytical and
numerical methods, mostly adopting Hamiltonian formalism, have been used to study
the dynamics of objects within resonances located at large distances from the
Earth (the so-called geostationary and GPS regions at distances, respectively, equal
to 42\,164 $km$ and 26\,560 $km$). We also mention \cite{EH, FV, LDV, VDLC} for
accurate modeling and analytical studies of space debris dynamics. With respect
to \cite{CGmajor,CGminor}, the current work presents the novelty that, dealing with LEO,
the model becomes more complicated, due to the effect of the geopotential, being
the Earth very close, and moreover the dynamics is dissipative because of
the air drag.

The dynamics is described through
a set of equations of motion which include the geopotential, the
atmospheric drag and the contribution of Sun and Moon. In particular,
we study four specific resonances located in LEO at different altitudes;
such resonances are due to a commensurability between the orbital period of
the debris and the period of rotation of the Earth. The geopotential is
expanded in spherical harmonics, although only a limited number of
coefficients is taken into account, precisely those which contribute to shape
the dynamics, being the dominant terms in a specific region of orbital parameters.
The atmospheric drag is modeled through a set of equations which are first
averaged with respect to the mean anomaly and then translated in terms of
the Delaunay actions. A qualitative study of the resonances is based on
the construction of a \sl toy model, \rm which provides a sound analytical
support to the numerical investigation of the problem. We are thus able
to draw conclusions about the role of the dissipation, the location and
stability character of the equilibria, the occurrence of temporary
capture into a resonance or rather a straight passage through it.
Once the results for the toy model are obtained, we pass to investigate a
problem which includes the change of the local density of the atmosphere
due to the effect of the solar cycle and the gravitational influence
given by Sun and Moon. The study leads to interesting results, which
can be used in concrete cases to make a thorough analysis of the
dynamics of space debris and even to design possible disposal orbits, or rather to provide practical solutions for control
and maintenance of LEO satellites. Due to dissipative effects, frequent maneuvers are required to keep the
orbital altitude. Our study reveals strong evidence that there exists equilibrium points in LEO that might be used in practice by parking operational satellites in their close vicinity,
thus reducing the cost of maintenance.\\

This paper is organized as follows.
In Section~\ref{sec:equations_of_motion} we provide the equations of motion
in Delaunay action-angle coordinates derived from a Hamiltonian including the
Keplerian part and the effect of the oblateness of the Earth. The geopotential
is expanded in Section~\ref{sec:geopotential_Ham} using a classical development
in terms of the spherical harmonic coefficients. A model for the atmospheric
drag is provided in Section~\ref{sec:diss_effect_drag}. Resonances, equilibria
and their stability are analyzed in Section~\ref{sec:qualitative_resonance},
while the effect of the the solar cycle and lunisolar perturbations are
studied in Section~\ref{sec:results}.

\section{Equations of motion}\label{sec:equations_of_motion}

We consider a small body, say $S$, located in the LEO region around the Earth.
We study its perturbed motion, taking into account the oblateness of the Earth, the rotation of our planet
and the atmospheric drag. To introduce the equations of motion, we use the action--angle Delaunay variables, denoted as $(L,G,H,M,\omega,\Omega)$, which are related to the orbital elements $(a,e,i,M,\omega,\Omega)$ by the expressions
\begin{equation}\label{LGH_aei}
L=\sqrt{\mu_E a}\,,\qquad  G=L \sqrt{1-e^2}\,, \qquad H=G \cos i\,,
\end{equation}
where $a$ is the semimajor axis,
$e$ the eccentricity, $i$ the inclination, $M$ the mean anomaly, $\omega$ the argument of perigee,
$\Omega$ the longitude of the ascending node and $\mu_E=\G m_E$ with $\G$ the gravitational constant and $m_E$ the mass of the Earth.

We denote by $\mathcal{H}$ the geopotential Hamiltonian (see \cite{CGmajor}), which can be written as
\beq{H}
\mathcal{H}(L,G,H,M,\omega,\Omega,\theta)=-{\mu^2_E\over {2L^2}}+\mathcal{H}_{earth}(L,G,H,M,\omega,\Omega,\theta)\ ,
\eeq
where $\theta$ is the sidereal time, $-{\mu^2_E\over {2L^2}}$ is the Keplerian part
and $\mathcal{H}_{earth}$ represents the perturbing function
(for which an explicit approximate expression is given in Section~\ref{sec:geopotential_Ham}).

We denote by $F_{_L}$, $F_{_G}$, $F_{_H}$ the components of
the dissipative effects due to the atmospheric drag, whose explicit expressions are given in Section~\ref{sec:diss_effect_drag}.
Then, the dynamical equations of motion are given by
\begin{equation} \label{canonical_eq}
\begin{split}
\dot{M}=\frac{\partial \mathcal{H}}{\partial L}\,,\qquad \quad & \qquad   \dot{\omega}=\frac{\partial \mathcal{H}}{\partial G}\,,
\ \quad \qquad \qquad \dot{\Omega}=\frac{\partial \mathcal{H}}{\partial H}\, ,\\
 \dot{L}=-\frac{\partial \mathcal{H}}{\partial M}+F_{_L}\,, & \qquad \dot{G}=
 -\frac{\partial \mathcal{H}}{\partial \omega}+F_{_G}\,, \qquad \dot{H}=
 -\frac{\partial \mathcal{H}}{\partial \Omega}+F_{_H}\ .
\end{split}
\end{equation}

\section{The geopotential Hamiltonian} \label{sec:geopotential_Ham}

Following \cite{Kaula}, we expand $\mathcal{H}_{earth}$ as
\beq{Rearth}
\mathcal{H}_{earth}=- {{\mu_E}\over a}\ \sum_{n=2}^\infty \sum_{m=0}^n \Bigl({R_E\over a}\Bigr)^n\ \sum_{p=0}^n \overline{F}_{nmp}(i)\
\sum_{q=-\infty}^\infty G_{npq}(e)\ \overline{S}_{nmpq}(M,\omega,\Omega,\theta)\ ,
\eeq
where $R_E$ is the Earth's radius,  $\overline{F}_{nmp}$ the normalized inclination function defined as
$$
\overline{F}_{nmp}=\sqrt{\frac{(2-\delta_{0n}) (2n+1)(n-m)!}{(n+m)!}}\, F_{nmp}\ ,
$$
where $\delta_{0n}$ is the Kronecker function, the inclination and eccentricity functions
$F_{nmp}$, $G_{npq}$ are computed by well--known recursive formulae
(see, e.g., \cite{Kaula, Chao, CGmajor}), while
   $\overline{S}_{nmpq}$ is expressed as
\beq{Snmpq}
\overline{S}_{nmpq}=\left[%
\begin{array}{c}
  \overline{C}_{nm} \\
  -\overline{S}_{nm} \\
 \end{array}%
\right]_{n-m \ odd}^{n-m \ even} \cos \Psi_{nmpq}+ \left[%
\begin{array}{c}
  \overline{S}_{nm} \\
  \overline{C}_{nm} \\
 \end{array}%
\right]_{n-m \ odd}^{n-m \ even} \sin \Psi_{nmpq}\ ,
\eeq
where $\overline{C}_{nm}$ and $\overline{S}_{nm}$ are, respectively, the cosine and sine normalized coefficients of the spherical harmonics potential terms (see Table~\ref{table:CS} for concrete values) and
\beq{psi}
\Psi_{nmpq}=(n-2p) \omega+(n-2p+q)M+m(\Omega-\theta)\ .
\eeq

The normalized coefficients $\overline{C}_{nm}$ and $\overline{S}_{nm}$ are related to the geopotential
coefficients $C_{nm}$ and $S_{nm}$ through the expressions (see \cite{Kaula, MG}):
$$
\left(%
\begin{array}{c}
  \overline{S}_{nm} \\
  \overline{C}_{nm} \\
 \end{array}%
\right)= \sqrt{\frac{(n+m)!}{(2-\delta_{0n}) (2n+1)(n-m)!}}
\left(%
\begin{array}{c}
  {S}_{nm} \\
  {C}_{nm} \\
 \end{array}%
\right).
$$
As we shall see later, we consider resonant motions which involve the rate of variations
of the mean anomaly and the sidereal angle through a linear combination with integer coefficients
(see Definition~\ref{def:resonance} below). We shall be  interested in specific resonances,
which will correspond to linear combinations involving the index $m$ with $m\geq 11$ (see Table~\ref{table:res_location}
below).

Since we deal with harmonic terms with large order (precisely
$m\geq 11$), we use the normalized coefficients, which have the advantage of being
more uniform in magnitude than  the unnormalized coefficients. In fact, the size of the normalized
coefficients is expressed approximately by the empirical Kaula's rule (see \cite{Kaula}):
$\overline{C}_{nm}, \, \overline{S}_{nm} \simeq 10^{-5}/n^2$, and therefore they decay less rapidly with $n$.
This allows us to avoid some computational complications which might appear when working with very small numbers,
such as $C_{nm}$, $S_{nm}$ for large $n$,
or very big numbers, which are involved in the computation of $F_{nmp}$.

As common in geodesy, we introduce also the quantities $\overline{J}_{nm}$ defined by
$$
\overline{J}_{nm} = \sqrt{\overline{C}_{nm}^2+\overline{S}_{nm}^2}   \quad \textrm{if} \ m\neq 0\ , \qquad
\overline{J}_{n0} \equiv \overline{J}_n= -\overline{C}_{n0}
$$
and the quantities $\lambda_{nm}$ defined through the relations
\begin{equation}\label{lambda_nm}
\overline{C}_{nm}=-\overline{J}_{nm} \cos(m \lambda_{nm}) \ , \qquad \overline{S}_{nm}=-\overline{J}_{nm} \sin(m \lambda_{nm}) \ .
\end{equation}

The coefficients $ \overline{J}_{nm}$ in units of $10^{-6}$ as well as the values of $\lambda_{nm}$, involved in the study of the resonances, are given
in Table 1; they are computed according to the Earth's gravitational model EGM2008 (\cite{EGM2008}).

\begin{table}[h]
\begin{tabular}{|c|c|c|c||c|c|c|c||c|c|c|c|}
  \hline
  \\
  $n$ & $m$  & $\overline{J}_{nm}$ & $\lambda_{nm}$ & $n$ & $m$  & $\overline{J}_{nm}$ & $\lambda_{nm}$ &  $n$ & $m$  & $\overline{J}_{nm}$ & $\lambda_{nm}$ \\
  \hline
2 &0&484.1651&$0^{\circ}$   & 15& 11&  0.0186 & $-7_{\cdot}^{\circ}82$  &    19 &    11  &      0.0193 & $19_{\cdot}^{\circ}31$ \\
3 &0& -0.9572 & $0^{\circ}$ & 15 & 12   &  0.036 &     $-2_{\cdot}^{\circ}14$  &   19 & 12  &   0.0098 & $-6_{\cdot}^{\circ}29$\\
4 & 0 & -0.54 & $0^{\circ}$  & 15 & 13  &  0.0287 &   $ 0_{\cdot}^{\circ}70$  &    19 & 13    &  0.0295 &  $5_{\cdot}^{\circ}78$ \\
5 & 0&  -0.0687 &  $0^{\circ}$ & 15 & 14&  0.0249 &   $7_{\cdot}^{\circ}29$  &  19 & 14 & 0.0137 &      $4_{\cdot}^{\circ}98$  \\
6 &  0  &  0.15 & $0^{\circ}$ & 16 & 11 &  0.0194 & $15_{\cdot}^{\circ}50$  &  20 & 11 & 0.024 &    $11_{\cdot}^{\circ}55$ \\
7 & 0&  -0.0905 & $0^{\circ}$  & 16 & 12    &  0.0207 &  $16_{\cdot}^{\circ}58$  & 20 & 12 &   0.0193 &  $-5_{\cdot}^{\circ}86$  \\
11 & 11 &  0.0836 &  $11_{\cdot}^{\circ}23$ &  16 & 13& 0.0138 &$14_{\cdot}^{\circ}18$  & 20 & 13 &  0.0282 &  $14_{\cdot}^{\circ}91$  \\
12 & 11 &  0.013 &  $13_{\cdot}^{\circ}70$  & 16 & 14&  0.0432 & $4_{\cdot}^{\circ}53$    & 20 &  14 & 0.0184 &  $ 9_{\cdot}^{\circ}19 $\\
12 & 12 & 0.0114 &  $6_{\cdot}^{\circ}47$  & 17 & 11& 0.0195 & $-3_{\cdot}^{\circ}15$   & 21 &   12 & 0.0151 &  $-6_{\cdot}^{\circ}44$  \\
13 & 11 & 0.0448 &  $0_{\cdot}^{\circ}56$  & 17 & 12 & 0.0353 & $17_{\cdot}^{\circ}96$  & 21 &   13 & 0.0239 &  $-2_{\cdot}^{\circ}75$ \\
13 & 12 &  0.0933 & $-5_{\cdot}^{\circ}87$  & 17 & 13 &  0.026 &     $17_{\cdot}^{\circ}74$   & 21 & 14 & 0.0216 &   $ 14_{\cdot}^{\circ}28 $\\
13 & 13 &  0.0916 & $-3_{\cdot}^{\circ}70$   & 17 & 14 & 0.0184 &  $-2_{\cdot}^{\circ}79 $   & 22 & 13 & 0.026 &   $-3_{\cdot}^{\circ}74$ \\
14 & 11 &  0.0421 &  $10_{\cdot}^{\circ}17$ &  18 &  11 & 0.0072 & $-1_{\cdot}^{\circ}56$ & 22 & 14 & 0.0137 &  $ 15_{\cdot}^{\circ}53$ \\
14 & 12 & 0.0323 & $8_{\cdot}^{\circ}77$  & 18 & 12 & 0.034 & $2_{\cdot}^{\circ}43$  &  23 & 14  & 0.0071 & $12_{\cdot}^{\circ}01$ \\
14 & 13 & 0.0555 & $18_{\cdot}^{\circ}04$ & 18 & 13 & 0.0355 & $6_{\cdot}^{\circ}14$ &  & & &  \\
14 & 14 & 0.0521 & $0_{\cdot}^{\circ}38$ & 18 & 14&  0.0153 &  $4_{\cdot}^{\circ}08$  &  & & &\\
  \hline
 \end{tabular}
 \vskip.1in
 \caption{The values of $\overline{J}_{nm}$ (in units of $10^{-6}$) and the quantities $\lambda_{nm}$ computed from
 \cite{EGM2008}.}
\label{table:CS}
\end{table}

\subsection{Approximation of the Hamiltonian}\label{sec:secres}
The expansion of the disturbing function $\mathcal{H}_{earth}$ in \equ{Rearth} contains an infinite number of
trigonometric terms, but the long term variation of the orbital elements is mainly governed by the
secular and resonant terms. Moreover, for the gravitational resonances located in the GEO and MEO regions,
we pointed out in \cite{CGmajor,CGexternal,CGminor} that just some of these terms are really relevant for the dynamics.

In the present work, we perform the study of the effects of the \sl gravitational resonances \rm (also called \sl tesseral \rm resonances,
see \cite{Gedeon,EH}), within the LEO region. The precise definition of resonance is given as follows.

\vskip.1in

\begin{definition}\label{def:resonance}
A tesseral (or gravitational) resonance of order $j:k$ with $j$, $k\in\integer\backslash\{0\}$
occurs when the orbital period of the debris and the rotational period of the Earth are commensurable
of order $j:k$. In terms of the orbital elements, a $j:k$ gravitational resonance occurs if
$$
k\ \dot{M}-j\ \dot{\theta} = 0\ , \qquad j,k \in \mathbb{N}\ .
$$
\end{definition}

Following \cite{CGmajor,CGexternal,CGminor}, we approximate
$\mathcal{H}_{earth}$ by
$$
\mathcal{H}_{earth}=\mathcal{H}^{sec}_{earth}+\mathcal{H}_{earth}^{res}+\mathcal{H}_{earth}^{nonres}\cong
\sum_{n=2}^N \sum_{m=0}^n \sum_{p=0}^n \sum_{q=-\infty}^{\infty} \mathcal{T}_{nmpq} \ ,
$$
where $\mathcal{H}^{sec}_{earth}$, $\mathcal{H}_{earth}^{res}$, $\mathcal{H}_{earth}^{nonres}$ denote, respectively, the secular, resonant and non--resonant
contributions to the Earth's potential, the approximation index $N\in\integer_+$ will be given later,
while the coefficients $\mathcal{T}_{nmpq}$ are defined by:
\begin{equation}\label{T_nmpq_term}
\mathcal{T}_{nmpq}=-\frac{\mu_E R_E^n}{a^{n+1}}\ \overline{F}_{nmp}(i)G_{npq}(e) \overline{S}_{nmpq}(M, \omega, \Omega , \theta)\ .
\end{equation}

In the following we describe the secular part of the expansion \equ{Rearth} by computing the average over the
fast angles, say $\mathcal{H}_{earth}^{sec}$, and the resonant part associated to a given $j:k$ tesseral resonance,
say $\mathcal{H}_{earth}^{resj:k}$.

Since the value of the oblateness coefficient  $\overline{J}_2=\overline{J}_{20}$ is much larger than the value of any other zonal coefficient (see Table~\ref{table:CS}),
we consider the same secular part for all resonances;
the explicit expression of the secular part will be given in Section~\ref{sec:secular}.

Concerning the resonant part, say $\mathcal{H}_{earth}^{res\,j:k}$,  it is essential to retain a minimum number
of significant terms in practical computations. The criteria for selecting these terms are described in
Section~\ref{sec:relevant}.

\subsubsection{The secular part of $\mathcal{H}_{earth}$}\label{sec:secular}
With reference to the expression for $\overline{S}_{nmpq}$ given in \equ{Snmpq}-\equ{psi}, the secular terms correspond to $m=0$ and $n-2p+q=0$. From Table~\ref{table:CS}, it is clear
that $\overline{J}_2\gg \overline{J}_n$ for all $n \in \mathbb{N}$, $n>2$.
Therefore, in the secular part the most important harmonic is $\overline{J}_2$. Moreover,
from Table~\ref{table:CS} it follows that $|\overline{J}_3|$ and $|\overline{J}_4|$ are larger than $|\overline{J}_n|$, $n>4$.  Since we are interested in orbits having small eccentricities, for our purposes it is enough to consider just a few harmonic terms in the expansion of the secular part.
In practical computations, for all resonances considered in the forthcoming sections, we approximate the secular part with the following expression,
computed e.g. in \cite{CGmajor}:
\beqa{Rsec}
\mathcal{H}_{earth}^{sec}&=&\frac{\sqrt{5} \mu_E R^2_E  \overline{J}_{2}}{a^3} \Bigl(\frac{3}{4} \sin^2 i -\frac{1}{2}\Bigr) (1-e^2)^{-3/2} \nonumber\\
&+&\frac{2 \sqrt{7}\mu_E R^3_E  \overline{J}_{3}}{a^4} \Bigl(\frac{15}{16} \sin^3 i -\frac{3}{4} \sin i\Bigr) e (1-e^2)^{-5/2} \sin \omega \nonumber \\
&+&\frac{3 \mu_E R^4_E \overline{J}_{4}}{a^5} \Bigl[\Bigl(-\frac{35}{32} \sin^4 i +\frac{15}{16} \sin^2 i\Bigr) \frac{3e^2}{2}(1-e^2)^{-7/2} \cos(2\omega) \nonumber \\
&+&
\Bigl(\frac{105}{64} \sin^4 i -\frac{15}{8} \sin^2 i+\frac{3}{8}\Bigr) (1+\frac{3e^2}{2})(1-e^2)^{-7/2} \Bigr]\ .
\eeqa
It is important to stress that the numerical results, obtained by taking into account the above approximation
of the secular part, may be analytically explained by considering only the influence of $\overline{J}_2$;
this will lead to consider a \sl toy model, \rm which well describes the dynamics,
as it will be explained in Section~\ref{sec:qualitative_resonance}.
The results based on the toy model will allow to draw conclusions about the importance
of $\overline{J}_2$ with respect to the other harmonics.

Clearly, in view of \eqref{LGH_aei}, $\mathcal{H}_{earth}^{sec}$ can be written as a function of $L$, $G$, $H$ and $\omega$.

\subsubsection{The resonant part of $\mathcal{H}_{earth}$}\label{sec:resonant}
From \equ{Snmpq}-\equ{psi} we see that the terms associated to a resonance of order $j:k$ correspond to $j(n-2p+q)=k\, m$.
We consider the resonant part corresponding to the following resonances located in the close vicinity of the Earth:
11:1, 12:1, 13:1 and 14:1. As we will show in Table~\ref{table:res_location} below, the resonances
11:1, 12:1, 13:1, 14:1 range from an altitude equal to 2\,146.61 $km$ down to an altitude equal to
880.55 $km$.

Hence, we consider $k=1$ and, within all possible combinations, the solution for which $j=m$ and $n-2p+q=1$
is relevant for our purposes.

Since  the majority of infinitesimal bodies of the LEO region moves on almost circular orbits,
we focus our analysis on small eccentricities with $e\in [0,0.02]$.
For such orbits, just some harmonic resonant terms are significant for the dynamics;
their selection will be made by using an analytical argument.
In fact, we will see that the resonant part can be approximated with a large degree of accuracy
by the sum of some terms, whose formal expression is:
\begin{equation} \label{Resonant_part}
\mathcal{H}_{earth}^{res\,m:1}=  \left\{
\begin{array}{lc}
  \sum_{\alpha=0}^N \, A^m_\alpha(L,G,H) \cos(\sigma_{m 1} -m\, \lambda_{m+2\alpha, \, m})\,,& \textrm{if } m=11 \textrm{ or } m=13\,, \\
  \sum_{\alpha=0}^N \, A^m_\alpha(L,G,H) \sin(\sigma_{m 1} -m \,\lambda_{m+2\alpha+1, \, m})\,, & \textrm{if } m=12 \textrm{ or } m=14\,, \\
 \end{array}%
\right .
\end{equation}
where the resonant angle is defined by
\begin{equation}\label{sigma_angle}
\sigma_{m 1}=M-m \theta+ \omega +m \Omega\,,
\end{equation}
$N$ is a natural number sufficiently  large so that the approximation of the resonant part includes
all harmonic terms with high magnitude (in this work we take $N=4$), $A^m_\alpha(L,G,H)$
might be computed by using \eqref{Rearth} and \eqref{LGH_aei}, once $\overline{F}_{nmp}$ and $G_{npq}$ are known, while the values of the constants $\lambda_{nm}$ are given in Table~\ref{table:CS}.

In a more compact notation,  $\mathcal{H}_{earth}^{res\,m:1}$ is written as:
\begin{equation}\label{Resonant_part_2}
\mathcal{H}_{earth}^{res\,m:1} = \mathcal{A}_0^{(m)}(L,G,H) \cos (\sigma_{m 1} -\varphi_0^{(m)}(L,G,H))\,,
\end{equation}
where  $\mathcal{A}_0^{(m)}(L,G,H)$ and $\varphi_0^{(m)}(L,G,H)$ are defined through the relations
\begin{equation}\label{A_varphi_11_13}
\begin{split}
& \mathcal{A}_0^{(m)}(L,G,H) \cos \varphi_0^{(m)}(L,G,H)=\sum_{\alpha=0}^N A_\alpha^m(L,G,H) \cos (m \lambda_{m+2\alpha , m})\,,\\
& \mathcal{A}_0^{(m)}(L,G,H) \sin \varphi_0^{(m)}(L,G,H)=\sum_{\alpha=0}^N A_\alpha^m(L,G,H) \sin (m \lambda_{m+2\alpha , m}) \qquad \textrm{if } m=11 \textrm{ or } m=13
\end{split}
\end{equation}
and
\begin{equation}\label{A_varphi_12_14}
\begin{split}
& \mathcal{A}_0^{(m)}(L,G,H) \cos \varphi_0^{(m)}(L,G,H)=-\sum_{\alpha=0}^N A_\alpha^m(L,G,H) \sin (m \lambda_{m+2\alpha+1 , m})\,,\\
& \mathcal{A}_0^{(m)}(L,G,H) \sin \varphi_0^{(m)}(L,G,H)=\sum_{\alpha=0}^N A_\alpha^m(L,G,H) \cos (m \lambda_{m+2\alpha+1 , m}) \qquad
\textrm{if } m=12 \textrm{ or } m=14\ .
\end{split}
\end{equation}
To provide the analytical explanation of how the relevant harmonic terms
can be selected, we need two essential comments on the index $q$ labeling the term
$\mathcal{T}_{nmpq}$ (see \eqref{T_nmpq_term}). First, we notice that the coefficients $G_{npq}(e)$
decay as powers of the eccentricity, precisely $G_{npq}(e)= \mathcal{O}(e^{|q|})$
(see \cite{Kaula, CGmajor}). Henceforth, the term $\mathcal{T}_{nmpq}$ is of order $|q|$ in the eccentricity.
On the other hand, in view of \eqref{Snmpq}, \eqref{psi}, \eqref{lambda_nm} and \eqref{sigma_angle},
it follows that the argument of the resonant term $\mathcal{T}_{nmpq}$ has the form $\sigma_{m1}-q \omega+const$.
Therefore, we conclude that the resonant harmonic terms can be grouped into terms of the same order in the eccentricity
and having the same argument (modulo a constant).

Let us denote by $\mathcal{M}^m_q$ the set of the resonant terms associated to the resonance $m:1$
and having the same index $q$, namely
\beq{Mmq}
\M_q^m\equiv \{\T_{nmpq}:\ n-2p+q=1\ ,\ n\in\mathbb{N}\ ,\ \ p\in\mathbb{N}\ , \ \ n \geq m\ , \ \ p\leq n  \}\ .
\eeq
The sets $\M_q^m$ with $q=-1,0,1$ and for the resonances 11:1, 12:1, 13:1, 14:1 are given in Table~\ref{tab:resonant_terms}.
The introduction of the set $\M_q^m$ is motivated by the fact that, from a dynamical point of view,
the terms belonging to $\mathcal{M}^m_q$ combine to give rise to a single resonant island at the same altitude.
Indeed, as it was pointed out
in \cite{CGmajor} and \cite{CGminor}, each resonance splits into a multiplet of resonances;
the exact location of the resonance for each component of the multiplet is obtained as the solution
of the relation $\dot{\sigma}_{m1}-q \dot{\omega}=0$.
However, since the elements of the set $\mathcal{M}^m_q$ have the same argument $\sigma_{m1}-q \omega$
(modulo a constant), a single resonant island is obtained when $n$ and $p$ vary,
even if $\mathcal{M}^m_q$ includes terms which are all different from each other.
Using \equ{Rearth}, \equ{Snmpq}, \equ{psi}, \equ{T_nmpq_term}, \equ{sigma_angle}, we have
the following result.

\begin{lemma}\label{lem:Mmq}
The sum of the terms of the set $\mathcal{M}^m_q$ in \equ{Mmq} can be written formally as
$$
\sum_{\mathcal{T} \in \mathcal{M}_q^m} \mathcal{T}=\mathcal{A}^{(m)}_q(L,G,H) \cos(\sigma_{m 1} -q \omega -\varphi_q^{(m)} (L,G,H))\ ,
$$
where $\mathcal{A}^{(m)}_q(L,G,H)$ and $\varphi_q^{(m)} (L,G,H)$ can be explicitly computed for each set
$\mathcal{M}^m_q$, once its elements are known.
\end{lemma}

Without loss of generality, we assume that
$\mathcal{A}^{(m)}_q(L,G,H)$ is non-negative for every $L$, $G$, $H$, possibly shifting the
argument of the trigonometric function.

\vskip.2in

\begin{table}[h]
\begin{tabular}{|c|c|c|}
  \hline
  $m:1$ & $\mathcal{M}^m_q$ & terms \\
  \hline
   & $\mathcal{M}^{1\!1}_0$ & $\mathcal{T}_{1\!1\,1\!1\,5\,0},\, \mathcal{T}_{1\!3\,1\!1\,6\,0},\, \mathcal{T}_{1\!5\,1\!1\,7\,0},\, \mathcal{T}_{1\!7\,1\!1\,8\,0},\, \mathcal{T}_{1\!9\,1\!1\,9\,0}$ \\
  11:1 & $\mathcal{M}^{1\!1}_{-1}$ & $\mathcal{T}_{1\!2\,1\!1\,5\,-1},\, \mathcal{T}_{1\!4\,1\!1\,6\,-1},\, \mathcal{T}_{1\!6\,1\!1\,7\,-1},\, \mathcal{T}_{1\!8\,1\!1\,8\,-1},\, \mathcal{T}_{2\!0\,1\!1\,9\,-1}$ \\
   & $\mathcal{M}^{1\!1}_1$ & $\mathcal{T}_{1\!2\,1\!1\,6\,1},\, \mathcal{T}_{1\!4\,1\!1\,7\,1},\, \mathcal{T}_{1\!6\,1\!1\,8\,1},\, \mathcal{T}_{1\!8\,1\!1\,9\,1},\, \mathcal{T}_{2\!0\,1\!1\,1\!0\,1}$ \\
    \hline
   & $\mathcal{M}^{1\!2}_0$ & $\mathcal{T}_{1\!3\,1\!2\,6\,0},\, \mathcal{T}_{1\!5\,1\!2\,7\,0},\, \mathcal{T}_{1\!7\,1\!2\,8\,0},\, \mathcal{T}_{1\!9\,1\!2\,9\,0},\, \mathcal{T}_{2\!1\,1\!2\,1\!0\,0}$ \\
  12:1 & $\mathcal{M}^{1\!2}_{-1}$ & $\mathcal{T}_{1\!2\,1\!2\,5\,-1},\, \mathcal{T}_{1\!4\,1\!2\,6\,-1},\, \mathcal{T}_{1\!6\,1\!2\,7\,-1},\, \mathcal{T}_{1\!8\,1\!2\,8\,-1},\, \mathcal{T}_{2\!0\,1\!2\,9\,-1}$ \\
   & $\mathcal{M}^{1\!2}_1$ & $\mathcal{T}_{1\!2\,1\!2\,6\,1},\, \mathcal{T}_{1\!4\,1\!2\,7\,1},\, \mathcal{T}_{1\!6\,1\!2\,8\,1},\, \mathcal{T}_{1\!8\,1\!2\,9\,1},\, \mathcal{T}_{2\!0\,1\!2\,1\!0\,1}$ \\
    \hline
 & $\mathcal{M}^{1\!3}_0$ & $\mathcal{T}_{1\!3\,1\!3\,6\,0},\, \mathcal{T}_{1\!5\,1\!3\,7\,0},\, \mathcal{T}_{1\!7\,1\!3\,8\,0},\, \mathcal{T}_{1\!9\,1\!3\,9\,0},\, \mathcal{T}_{2\!1\,1\!3\,1\!0\,0}$ \\
  13:1 & $\mathcal{M}^{1\!3}_{-1}$ & $ \mathcal{T}_{1\!4\,1\!3\,6\,-1},\, \mathcal{T}_{1\!6\,1\!3\,7\,-1},\, \mathcal{T}_{1\!8\,1\!3\,8\,-1},\, \mathcal{T}_{2\!0\,1\!3\,9\,-1},\, \mathcal{T}_{2\!2\,1\!3\,1\!0\,-1}$ \\
   & $\mathcal{M}^{1\!3}_1$ & $ \mathcal{T}_{1\!4\,1\!3\,7\,1},\, \mathcal{T}_{1\!6\,1\!3\,8\,1},\, \mathcal{T}_{1\!8\,1\!3\,9\,1},\, \mathcal{T}_{2\!0\,1\!3\,1\!0\,1},\, \mathcal{T}_{2\!2\,1\!3\,1\!1\,1},\,$ \\
  \hline
  & $\mathcal{M}^{1\!4}_0$ & $ \mathcal{T}_{1\!5\,1\!4\,7\,0},\, \mathcal{T}_{1\!7\,1\!4\,8\,0},\, \mathcal{T}_{1\!9\,1\!4\,9\,0},\, \mathcal{T}_{2\!1\,1\!4\,1\!0\,0},\, \mathcal{T}_{2\!3\,1\!4\,1\!1\,0}$ \\
  14:1 & $\mathcal{M}^{1\!4}_{-1}$ & $ \mathcal{T}_{1\!4\,1\!4\,6\,-1},\, \mathcal{T}_{1\!6\,1\!4\,7\,-1},\, \mathcal{T}_{1\!8\,1\!4\,8\,-1},\, \mathcal{T}_{2\!0\,1\!4\,9\,-1},\, \mathcal{T}_{2\!2\,1\!4\,1\!0\,-1}$ \\
   & $\mathcal{M}^{1\!4}_1$ & $ \mathcal{T}_{1\!4\,1\!4\,7\,1},\, \mathcal{T}_{1\!6\,1\!4\,8\,1},\, \mathcal{T}_{1\!8\,1\!4\,9\,1},\, \mathcal{T}_{2\!0\,1\!4\,1\!0\,1},\, \mathcal{T}_{2\!2\,1\!4\,1\!1\,1},\,$ \\
  \hline
\end{tabular}
 \vskip.1in
 \caption{The sets $\mathcal{M}^{m}_{0}$, $\mathcal{M}^{m}_{-1}$, $\mathcal{M}^{m}_{1}$ for the
 resonances 11:1, 12:1, 13:1, 14:1.}\label{tab:resonant_terms}
\end{table}

\vskip.1in

\begin{figure}[h]
\centering \vglue0.1cm \hglue0.2cm
\includegraphics[width=6truecm,height=5truecm]{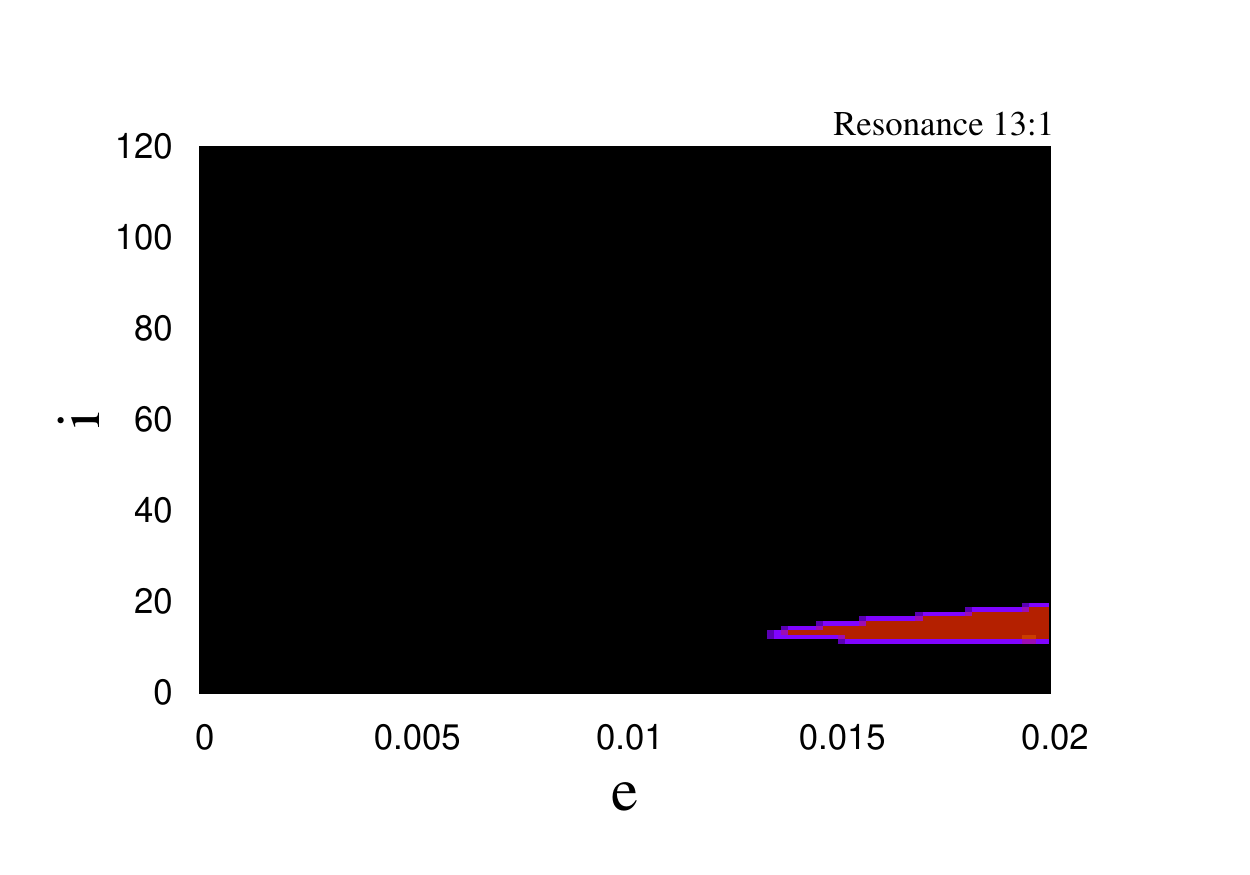}
\includegraphics[width=6truecm,height=5truecm]{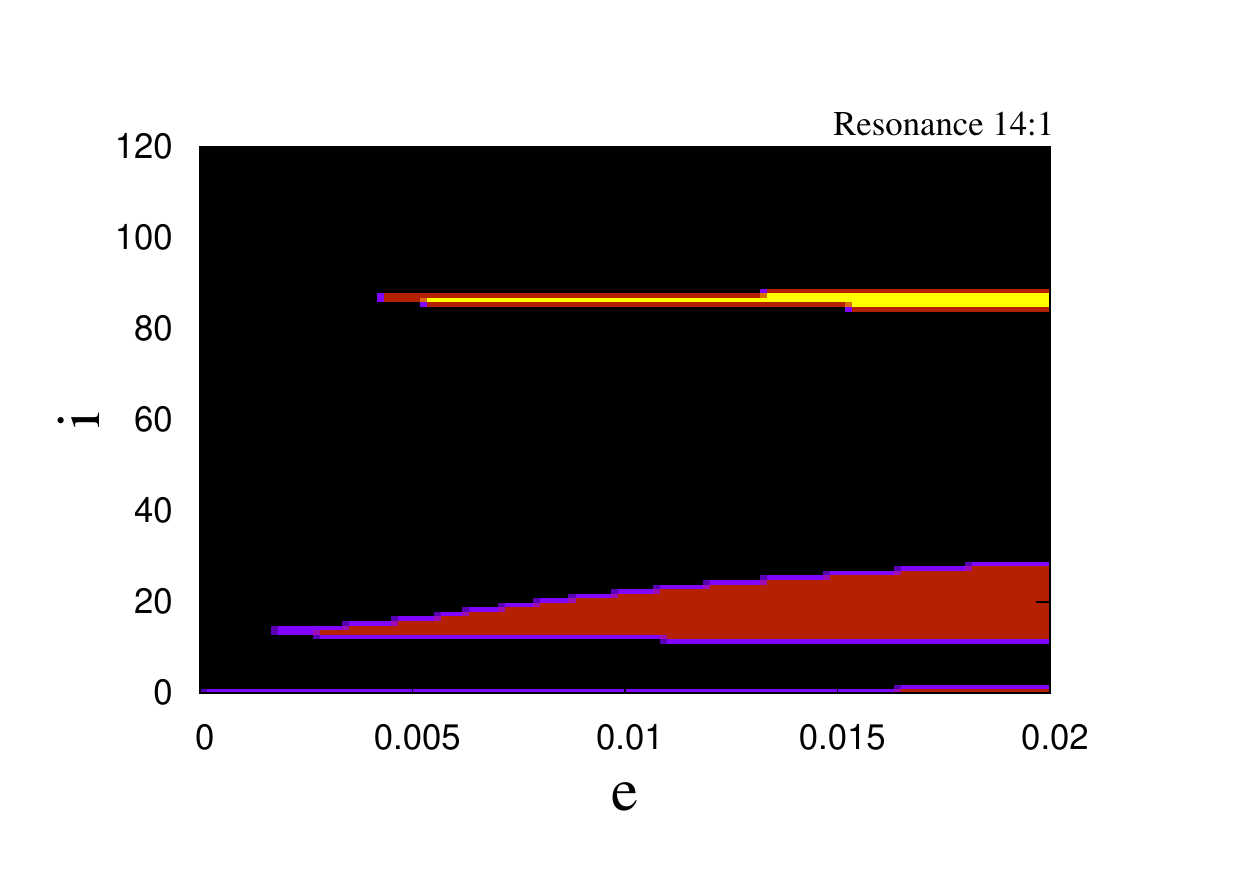}
\vglue0.4cm
\caption{Dominant sets for the 13:1 (left) and 14:1 (right)  resonances as a function of eccentricity and inclination: $\mathcal{M}^m_0$ -- black, $\mathcal{M}^m_{-1}$ -- brown, $\mathcal{M}^m_1$ -- yellow, where $m=13,\, 14$ and the sets $\mathcal{M}^m_0$, $\mathcal{M}^m_{-1}$, $\mathcal{M}^m_1$ are defined in Section~\ref{sec:resonant}.} \label{fig:big_terms}
\end{figure}

\subsection{The most relevant terms of the Hamiltonian}\label{sec:relevant}
Our next task  is to retain those sets $\mathcal{M}^m_q$ which are important for the dynamics,
as well as to keep only the most relevant elements of each selected set.
Since our analysis involves small eccentricities, one expects that $\mathcal{M}^m_0$ will play the most important
role, while the influence of the other sets, precisely $\mathcal{M}^m_{-1}$, $\mathcal{M}^m_{1}$, will
be weaker. Concerning the elements of the set $\mathcal{M}^m_q$, it is important to stress that
the coefficients of degree $n$ decay as $(R_E/a)^n$, so the role of
the harmonic terms with higher degree becomes increasingly less influent.
However, since we are considering resonances which are very close to the Earth, the quantity
$(R_E/a)^n$ decays slowly for increasingly higher values of $n$.
In conclusion, to get a reliable model, the set $\mathcal{M}^m_q$ should contain as many harmonic terms as possible.
However, due to computational limitations, in this paper the maximum number of elements of $\mathcal{M}^m_q$ is 5,
which is a good compromise between accuracy and complexity. It is meaningful to consider a larger
number of coefficients when dealing with specific concrete cases.

To give an explicit example, let us take the set $\mathcal{M}^{11}_0$. Comparing the coefficient $(R_E/a)^{1\!1}$
of the term $\mathcal{T}_{1\!1\, 1\!1\, 5\,0}$ (see \eqref{T_nmpq_term} and Table~\ref{tab:resonant_terms})
with the coefficient $(R_E/a)^{2\!1}$ of $\mathcal{T}_{2\!1\, 1\!1\, 1\!0\,0}$
(namely, the first term of $\mathcal{M}^{11}_0$ neglected in our computations), we find that for $a=8524.75$ $km$
(see Table~\ref{table:res_location} below) the term $\mathcal{T}_{2\!1\, 1\!1\, 1\!0\,0}$
is 18 times smaller than $\mathcal{T}_{1\!1\, 1\!1\, 5\,0}$, thus showing that
the neglected harmonic terms are smaller in magnitude than those considered in our model.
Of course, the conclusion is valid for all other sets, although with different ratios.
We report in Table~\ref{tab:resonant_terms} the terms of the sets $\mathcal{M}^m_0$, $\mathcal{M}^m_{-1}$ and
$\mathcal{M}^m_1$ that we are going to consider for each resonance.

Once the elements of $\mathcal{M}^m_q$ are selected, it remains to discriminate which are the most important ones.
Making use of Lemma~\ref{lem:Mmq}, we introduce the following definition, which gives a hierarchy between the
sets $\M_q^m$.

\begin{definition} \label{def_dominant}
Let $\mathcal{H}_{earth}^{res\,m:1}$ be the resonant part of $\mathcal{H}_{earth}$, corresponding to the resonance $m:1$.
For given values of  the orbital elements $(a,e,i)$, equivalently for given values of $(L,G,H)$,
we say that a set $\mathcal{M}^m_q$, for some $q \in \mathbb{Z}$,
is {\it dominant} with respect to the other sets $\mathcal{M}^m_{\widetilde{q}}$,
where $\widetilde{q} \in \mathbb{Z}$ with $\widetilde{q} \neq q$, if $\mathcal{A}^{(m)}_q(L,G,H) \geq \mathcal{A}^{(m)}_{\widetilde{q}}(L,G,H)$ for all $\widetilde{q} \in  \mathbb{Z}$.
\end{definition}

A plot of the dominant sets according to Definition~\ref{def_dominant} for the resonances $13:1$ and $14:1$
is provided in Figure~\ref{fig:big_terms}, within the orbital elements' intervals
$e\in [0,0.02]$ and $i \in [0^o, 120^o]$. The black, brown and yellow colors are, respectively,
used to show the regions where $\mathcal{M}^m_0$, $\mathcal{M}^m_{-1}$ and $\mathcal{M}^m_{1}$ dominate.
Similar plots are also obtained for the 11:1 and 12:1 resonances, but
in these cases the regions associated  to $\M_{-1}^m$ are very small and
those related to $\M_1^m$ are negligible.
From the analysis of Figure~\ref{fig:big_terms}, we conclude that $\mathcal{M}^m_0$ is dominant in almost all regions of the $(e,i)$ - plane, except for some small inclinations and for $i=86.18^o$ in the case of the 14:1 resonance. Taking into account the fact that the amplitudes
of the two resonant islands associated to $\mathcal{M}^m_{-1}$ and $\mathcal{M}^m_{1}$ are small
(at most few hundred meters as it will be shown in Section~\ref{sec:qualitative_resonance}),
we may approximate the resonant part $\mathcal{H}_{earth}^{res\,m:1}$ by
the sum of the terms of $\mathcal{M}^m_0$. Therefore, from Table~\ref{tab:resonant_terms}
and collecting \eqref{Snmpq}, \eqref{psi}, \eqref{lambda_nm}, \eqref{T_nmpq_term}, \eqref{sigma_angle}, it follows that $\mathcal{H}_{earth}^{res\,m:1}$ can be written in the form \eqref{Resonant_part}
(or equivalently in the form \eqref{Resonant_part_2}) for a suitable integer $N$, which counts the number
of terms generated by $\M_0^m$.
Section~\ref{sec:qualitative_resonance} will confirm that the analytical model, constructed on the basis of this approximation,
leads to reliable results.  In fact, the numerical investigation will be performed by taking into account the effects of all three sets $\mathcal{M}^m_0$, $\mathcal{M}^m_{-1}$ and $\mathcal{M}^m_{1}$,
but we will obtain results that
can be easily explained in terms of an analytical model which includes just the influence of  $\mathcal{M}^m_0$.

Since the normalized inclination functions $\overline{F}_{nmp}$
involve very long expressions (often more than half
page for each function), we avoid giving the explicit forms of
the terms $\mathcal{T}_{nmpq}$ and of the functions
$A^{m}_{\alpha}(L,G,H)$,  $\mathcal{A}_0^{(m)}(L,G,H)$ and
$\varphi_0^{(m)}(L,G,H)$. The reader can compute these quantities by
using the recursive formulae for the functions $F_{nmp}$,
$G_{npq}$ (see \cite{Kaula,CGmajor}) and by using the relations
presented in Section~\ref{sec:secres}.

\section{Dissipative effects: the atmospheric drag} \label{sec:diss_effect_drag}

During its motion within the Earth's atmosphere, an infinitesimal object (satellite or space debris) encounters air molecules, whose change of momentum gives rise to a dissipative  force
oriented opposite to the motion of the body and known as atmospheric drag.
The atmospheric drag force depends on the local density of the atmosphere, the velocity of the object
relative to the atmosphere and the cross--sectional area in the direction of motion.

The purpose of this Section is to derive the functions $F_{_L}$, $F_{_G}$, $F_{_H}$, characterizing the atmospheric drag perturbations in the dynamical equations \eqref{canonical_eq}.
To this end, we use the following averaged equations of variation of the orbital elements (see~\cite{Liu, Chao}):

\beqa{dissae}
\dot a&=&-{1\over {2\pi}}\int_0^{2\pi}B\, \rho\, v{a\over {1-e^2}}\ \Big[1+e^2+2e\cos f-\omega_E\cos i\sqrt{{a^3(1-e^2)^3}\over {\mu_E}}\Big]\ dM\nonumber\\
&\equiv&\F^{(a)}(a,e,i)\ ,\nonumber\\
\dot e&=&-{1\over {2\pi}}\int_0^{2\pi}B\, \rho\, v\ \Big[e+\cos f-{{r^2\omega_E\cos i}\over{2\sqrt{\mu_E a(1-e)^2}}}\Big(2(e+\cos f)-e\sin^2f\Big)\Big]\, dM\nonumber\\
&\equiv&\F^{(e)}(a,e,i)\ ,
\eeqa
where $f$ is the true anomaly, $\omega_E$ (coinciding with $\dot{\theta}$) is
the Earth's rotation rate, $\rho$ the atmospheric density, $B$
the ballistic coefficient, while the body's speed relative to the atmosphere is given by
\begin{equation}\label{speed}
v=\sqrt{{\mu_E\over a(1-e^2)}(1+e^2+2e\cos f)}\ \Big(1-{{(1-e^2)^{3\over 2}}\over {1+e^2+2e\cos f}}\, {\omega_E\over n^*}\cos i\Big)\ ,
\end{equation}
where $n^*$ is the mean motion of the satellite.
Notice that $r$, $f$ (hence $v$) are functions of $M$.
We stress that the atmospheric drag affects just $\dot{a}$ and $\dot{e}$ and not the other variables
(namely, the inclination and the angle variables).

We recall that the ballistic coefficient is expressed in terms of the cross--sectional area $A$ with respect to the relative
wind and in terms of the mass $m$ of the object
through the formula $B=C_D\, A/m$, where $C_D$ is the drag coefficient.
For a debris, the coefficient $B$ can vary by a factor 10 depending on the satellite's orientation
(see Table 8-3 in \cite{LW} for a list of estimated ballistic coefficients associated to various LEO satellites;
note that this table provides $1/B$). Although the ballistic coefficient of a satellite slightly modifies in time,
in all simulations we suppose that $B$ is constant.
This assumption is motivated by the fact that we are interested in studying the equilibrium points,
and therefore in such dynamical configuration
the small variation of $B$ can be neglected in a first approximation.
Moreover, in order to show the existence of the equilibrium points even
for strong dissipative effects, in our simulations we shall often
use large values for the ballistic coefficient, up to $2200 \, cm^2/kg$, although
the value for a satellite is much smaller, typically $25 \leq B\leq 500 \, cm^2/kg$ (see \cite{ISO}).

\vskip.2in
\begin{table}
\begin{tabular}{|c|c|c|c|c|}
  \hline
  Altitude & Atm. scale & Minimum density  & Mean density & Maximum density \\
  $h_0$  ($km$) & height $H_0$ ($km$) &  ($kg/m^3$) &  ($kg/m^3$) &  ($kg/m^3$) \\
  \hline
  700 & 99.3 & $5.74\cdot 10^{-15}$ & $2.72\cdot 10^{-14}$ & $1.47\cdot 10^{-13}$\\
  800 & 151 &  $2.96\cdot 10^{-15}$ &  $9.63\cdot 10^{-15}$ &  $4.39\cdot 10^{-14}$\\
  1000 & 296 & $1.17\cdot 10^{-15}$ & $2.78\cdot 10^{-15}$ & $8.84\cdot 10^{-15}$\\
  1250 & 408 & $4.67\cdot 10^{-16}$ & $1.11\cdot 10^{-15}$ & $2.59\cdot 10^{-15}$ \\
  1500 & 516 & $2.30\cdot 10^{-16}$ & $5.21\cdot 10^{-16}$ & $1.22\cdot 10^{-15}$ \\
  2000 & 829 & $-$ & $-$ & $-$ \\
  \hline
 \end{tabular}
 \vskip.1in
 \caption{The scaling height $H_0$ as well as the minimum, mean and maximum densities at
 the reference altitude $h_0$, from MSIS atmospheric model (\cite{Hedin}, see also \cite{LW}).}\label{table:rho}
\end{table}

To  complete the discussion of equations \equ{dissae}, let us mention that
the atmospheric density can be computed from density models such as
that developed by Jacchia
(\cite{Jacchia}), the Mass Spectrometer Incoherent Scatter - MSIS model (\cite{Hedin0, Hedin}) and other models
(see \cite{ISO}). Following the dynamical density MSIS model, the local density is a function of various parameters
such as the altitude of the body,
the solar flux, the Earth's magnetic index, etc. (see~\cite{Hedin0, Hedin}).
Of particular interest is the variation of density as effect of the solar activity,
which fluctuates with an 11--year cycle.

In this work we use the numbers provided by the MSIS model.
Therefore, we assume that the local density varies with the altitude above the surface, say $h=r-R_E$, with $r$ the distance from the Earth's center, and we use the the following barometric formula:
\begin{equation}\label{rho_h}
\rho(h)=\rho_0\ \exp \biggl(-{{h-h_0}\over {H_0}}\biggr)\ ,
\end{equation}
where $\rho_0$ is the (minimum, mean or maximum) density, estimated for (minimum, mean or maximum) solar activity
at the reference altitude $h_0$, while $H_0$ is the scaling height at $h_0$.
Reference empirical values are given in Table~\ref{table:rho} (see also \cite{LW} for a more detailed list
of values and further explanations).

Although our investigation involves small eccentricities, say up to $e=0.02$,
the difference in altitude between apogee and perigee is not negligible and it amounts to
about 300 $km$ (see Table~\ref{table:res_location}). In fact, comparing the altitudes reported in Tables~\ref{table:res_location} and \ref{table:rho}, it is clear that $\rho=0$ for the 11:1 resonance, while for the other resonances one should use the formula \eqref{rho_h} with the corresponding values for $\rho_0$, $h_0$ and $H_0$ taken from Table~\ref{table:rho}.

\vskip.2in

\begin{table}[h]
\begin{tabular}{|c|c|c|c|c|}
  \hline
  $m:1$ & $a$  & Altitude & Perigee altitude  & Apogee altitude  \\
   & ($km$) & ($km$) &  for $e=0.02$ ($km$)  &  for $e=0.02$ ($km$)  \\
  \hline
 11:1 & 8524.75 & 2146.61 & 1976.25  & 2317.25 \\
 12:1 & 8044.32 & 1666.18 & 1505.43 & 1827.21 \\
 13:1 & 7626.31 & 1248.17 & 1095.78 & 1400.84 \\
 14:1 & 7258.69 & 880.55 & 735.52 & 1025.86 \\
 \hline
 \end{tabular}
 \vskip.1in
 \caption{The semimajor axis and the altitude corresponding to some resonances of order $m:1$, as well
 as the  perigee and apogee altitudes of a resonant elliptic orbit with $e=0.02$.
 The altitudes are computed by considering the reference value $R_E=6378.14$ $km$ for the Earth's
 radius. }\label{table:res_location}
\end{table}

\vskip.2in



Once the framework has been settled, we can approximate in the
computations the true anomaly $f$ (entering \equ{dissae}, \equ{speed})
and the altitude $h$ (entering in \equ{rho_h}) by the following well known
series (\cite{Roy,Alebook}):
\begin{equation}\label{anomaly_rovera}
\begin{split}
& f = M+2e \sin M+\frac{5 e^2}{4}\sin(2M)+O(e^3)\ ,\\
& h=a(1-e \cos f)-R_E= a\Bigl\{1-e \cos M+\frac{e^2}{2}\Bigl[1- \cos(2M)\Bigr]\Bigr\}-R_E +O(e^3)\ ,
\end{split}
\end{equation}
where $O(e^3)$ denotes terms of order 3 in the eccentricity.
Casting together the relations \eqref{dissae}, \eqref{speed}, \eqref{rho_h}  and \eqref{anomaly_rovera}, by the algebraic manipulator \verb"Mathematica"$^\copyright$ we compute the integrals appearing in the right
hand side of $\eqref{dissae}$. In this way, we deduce that the right hand sides
in the first of $\eqref{dissae}$, thereby denoted as
$\mathcal{F}^{(a)}$ and $\mathcal{F}^{(e)}$, are functions of $a$, $e$, $i$, while $\rho_0$ and $B$ are parameters. As in Section~\ref{sec:resonant} we do not provide the explicit form of $\mathcal{F}^{(a)}(a,e,i)$ and $\mathcal{F}^{(e)}(a,e,i)$,
since they involve long expressions. The reader can self-compute these functions by a simple implementation of the above
formulae, possibly using an algebraic manipulator.
Once $\mathcal{F}^{(a)}$ and $\mathcal{F}^{(e)}$ are computed as a function of the orbital elements, it is
trivial to express them in terms of the Delaunay actions:
$\mathcal{F}^{(a)}=\mathcal{F}^{(a)}(L,G,H)$ and $\mathcal{F}^{(e)}=\mathcal{F}^{(e)}(L,G,H)$.

Since the atmospheric drag does not affect the inclination, from \eqref{LGH_aei} we obtain
\beqano
\dot L&=&{{1}\over {2}} \sqrt{\frac{\mu_E}{a}}\ \dot a\,,\nonumber\\
\dot G&=&{{1}\over {2}} \sqrt{\frac{\mu_E (1-e^2)}{a}}\ \dot a-e \sqrt{{{ \mu_E a}\over {1-e^2}}} \ \dot e\,,\nonumber\\
\dot H&=&\Bigl({{1}\over {2}} \sqrt{\frac{\mu_E (1-e^2)}{a}}\ \dot a-e \sqrt{{{ \mu_E a}\over {1-e^2}}} \ \dot e \Bigr)
\cos i \ .\nonumber\\
\eeqano
Using the relations $a=L^2/\mu_E$, $e=\sqrt{1-G^2/L^2}$ and $\cos i=H/G$, we deduce that the functions $F_{_L}$, $F_{_G}$, $F_{_H}$, characterizing the atmospheric drag perturbations
in \equ{canonical_eq}, are given by

\begin{equation}\label{dissipative_functions}
\begin{split}
&F_{_L}={{\mu_E}\over {2 L}}\ {\mathcal{F}^{(a)}}(L,G,H)\,,\\
&F_{_G}={{\mu_E G}\over {2L^2}}\ \mathcal{F}^{(a)}(L,G,H)-{{ L^2}\over {G}} \sqrt{1-\frac{G^2}{L^2}}\
\mathcal{F}^{(e)}(L,G,H)\,,\\
&F_{_H}=\frac{H}{G}\Bigl({{\mu_E G}\over {2L^2}}\ \mathcal{F}^{(a)}(L,G,H)-
{{ L^2}\over {G}} \sqrt{1-\frac{G^2}{L^2}}\ \mathcal{F}^{(e)}(L,G,H)\Bigr)\ .
\end{split}
\end{equation}
In conclusion, to study the main dynamical features of
tesseral resonances, we have introduced
(Sections~\ref{sec:equations_of_motion}, \ref{sec:geopotential_Ham}
and \ref{sec:diss_effect_drag}) a mathematical model characterized
by the equations \eqref{canonical_eq}, where the secular part of
the Hamiltonian \eqref{H} is given by \eqref{Rsec}, the resonant
part of $\mathcal{H}$  is obtained as the sum of the resonant
harmonic terms of Table~\ref{tab:resonant_terms}, while the
dissipative part is described by the functions $F_{_L}$, $F_{_G}$,
$F_{_H}$ defined by \eqref{dissipative_functions}. Hereafter, this
model will be called the \emph{dissipative model of LEO
resonances}, or simply DMLR.

\section{A qualitative study of resonances}\label{sec:qualitative_resonance}
This section presents a qualitative study of the resonances.  Precisely, it
includes an analysis of the conservative and dissipative effects,
an estimate of the amplitude of the resonances, a study related to the
existence, location and stability of the equilibrium points. Some
analytical results based on a toy model that will be introduced in
Section~\ref{sec:toy} are confirmed by numerical simulations
obtained by using the DMLR.

We stress that, although the degree $n$ of the resonant terms is large ($n\geq 11$),
which implies that the magnitude of these terms is small, the effects of the conservative part can be quantified;
in particular, for some inclinations the resonant regions have
a width larger than one or two kilometers. Since at high altitudes the drag effect is sufficiently low, even if the solar activity reaches its maximum, for such inclinations one can show that equilibrium points exist.

\subsection{The toy model}\label{sec:toy}

To give an analytical support to the numerical results that will
be performed on the DMLR, we construct in parallel a
simplified model, to which we refer as the \sl toy model, \rm
allowing to explain the main features of the dynamics. In this
model the secular part contains just the $\bar{J}_2$ term (first
term of \eqref{Rsec}), the resonant part is defined by
\eqref{Resonant_part_2} and the dissipative functions are given
by~\eqref{dissipative_functions_circular} below. Following
\cite{Chao}, for nearly circular orbits, the function
$\mathcal{F}^{(a)}$ can be simplified as
\begin{equation}\label{adot_circular}
\mathcal{F}^{(a)}=-B \rho n^* a^2 \Bigl(1-\frac{\omega_E}{n^*} \cos i\Bigr)^2\ ,
\end{equation}
where $\rho$ is assumed to be constant at a fixed altitude of the orbit and
$n^*=\sqrt{\mu_E/a^3}$. As mentioned before, the variation of the
eccentricity can be considered a small quantity; therefore,
in the simplified model we take $\mathcal{F}^{(e)}=0$.
Using \eqref{LGH_aei}, \eqref{dissipative_functions} and \eqref{adot_circular} we get
\begin{equation}\label{dissipative_functions_circular}
\begin{split}
F_{_L}=-\frac{1}{2} B \rho \mu_E\Bigl(1-\frac{\omega_E L^3 H}{\mu_E^2 G} \Bigr)^2\,,\qquad
F_{_G}=\frac{G}{L} F_{_L}\ ,\qquad F_{_H}=\frac{H}{L} F_{_L}\ .
\end{split}
\end{equation}

\begin{figure}[h]
\centering
\vglue0.1cm
\hglue0.1cm
\includegraphics[width=6truecm,height=5truecm]{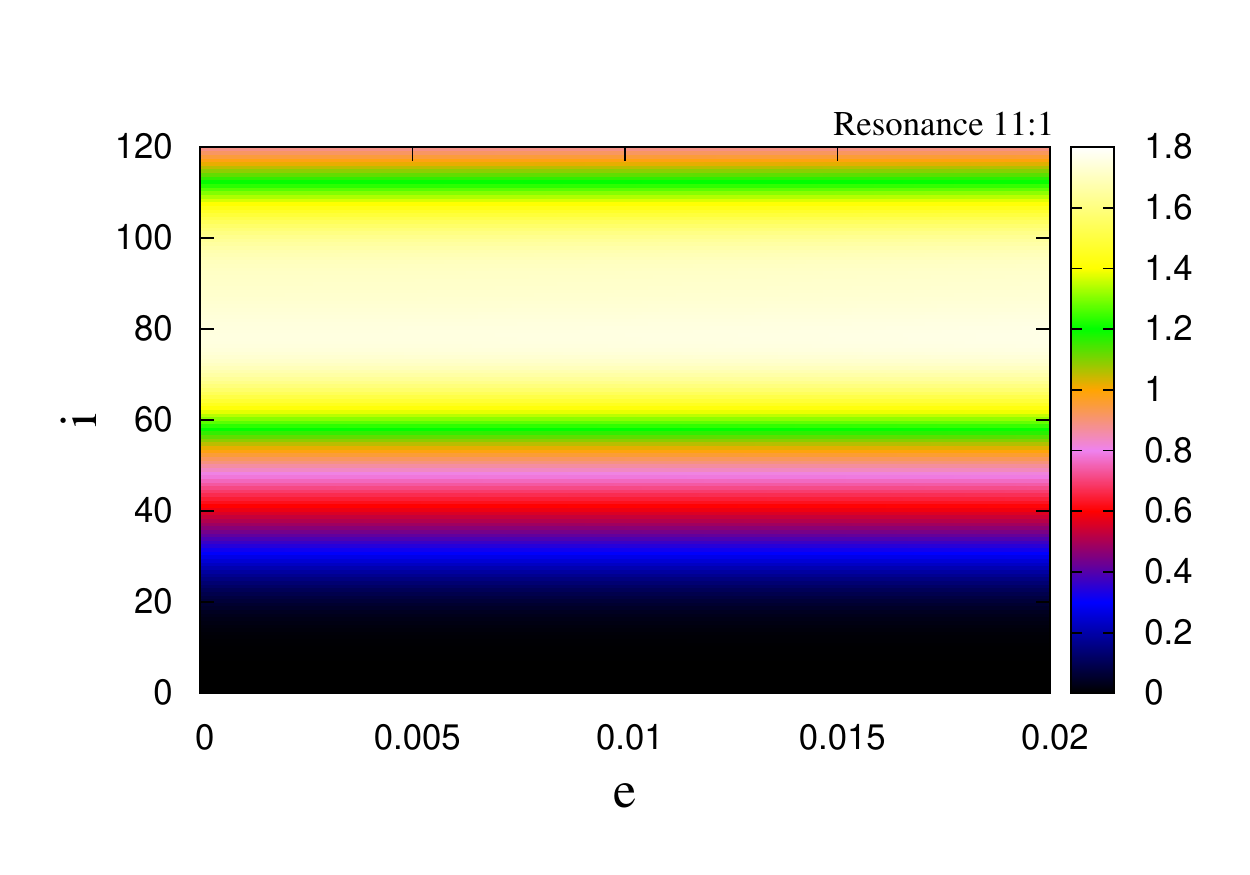}
\includegraphics[width=6truecm,height=5truecm]{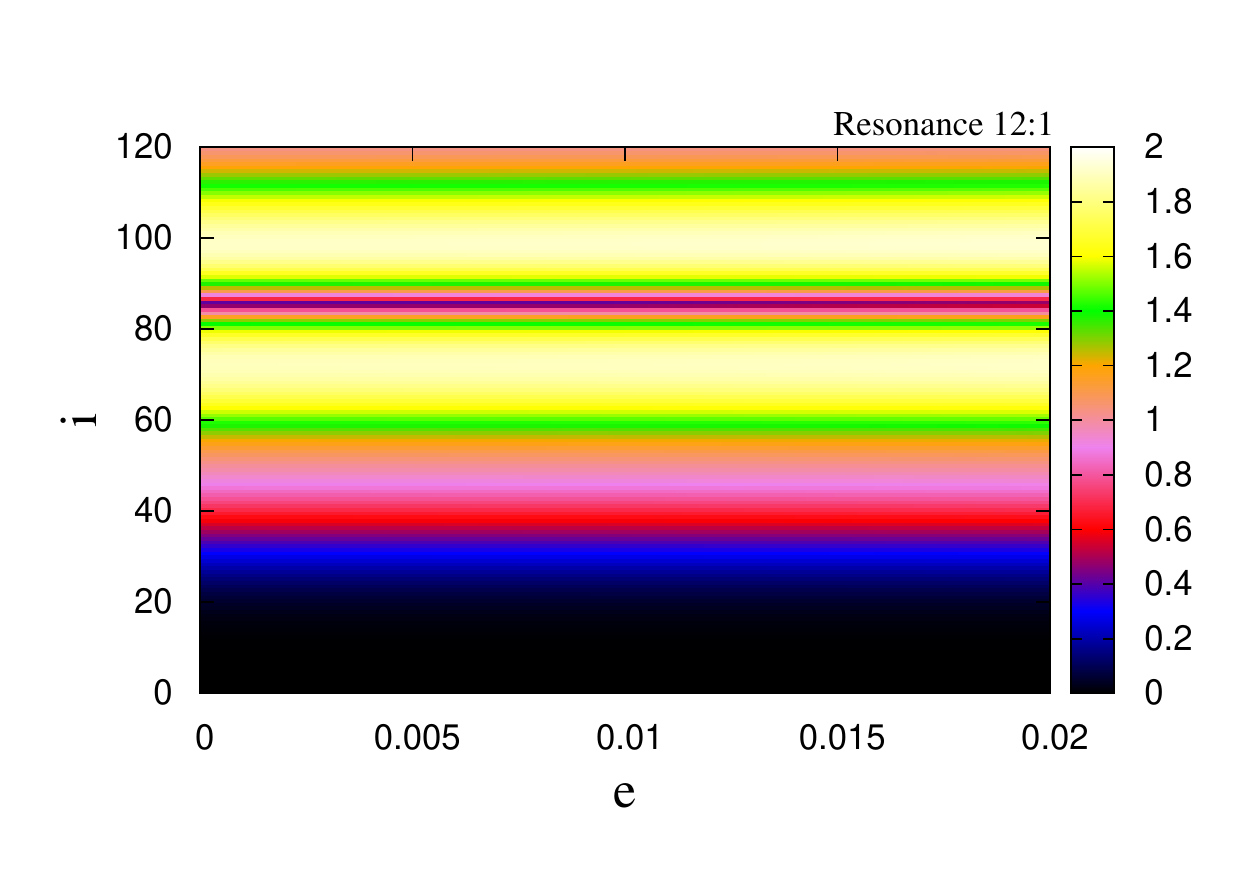}\\
\vglue-0.6cm
\includegraphics[width=6truecm,height=5truecm]{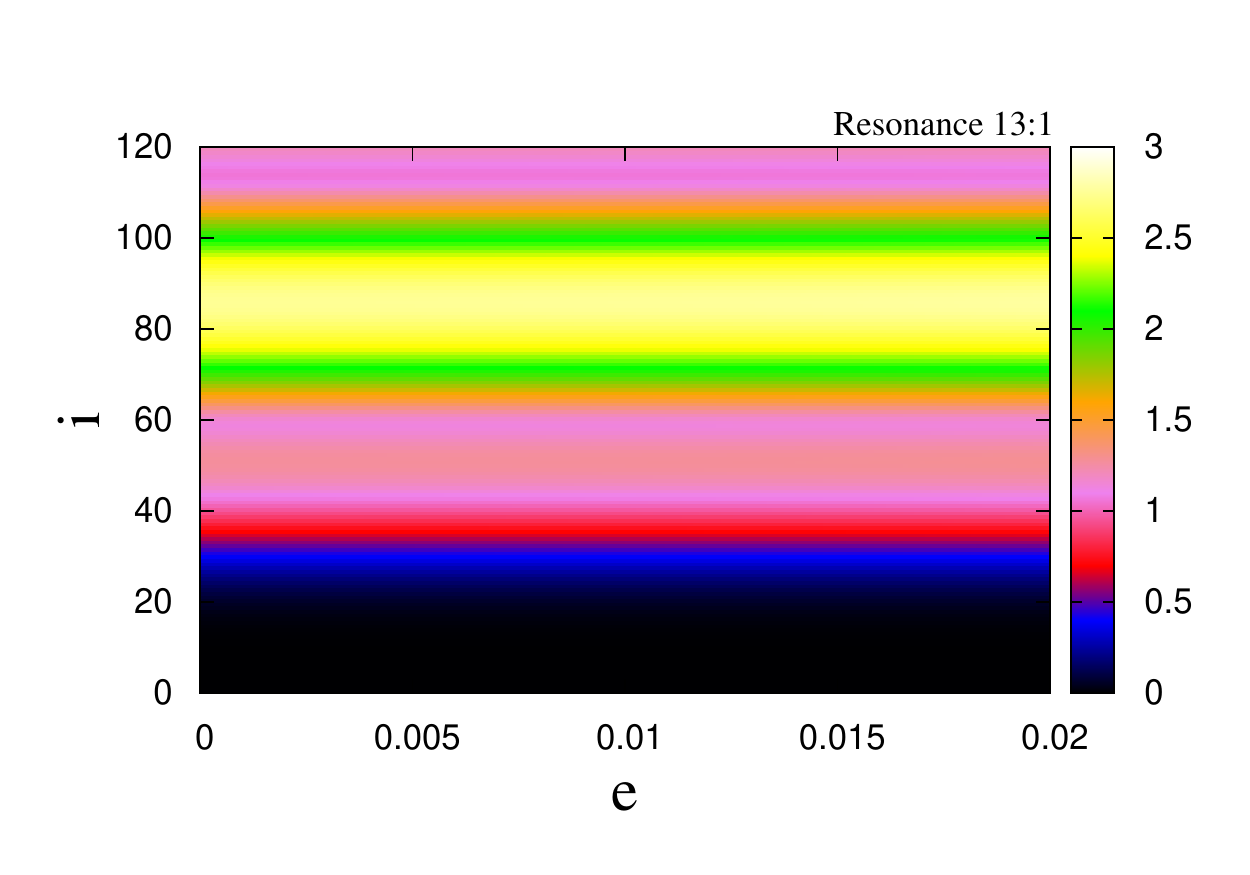}
\includegraphics[width=6truecm,height=5truecm]{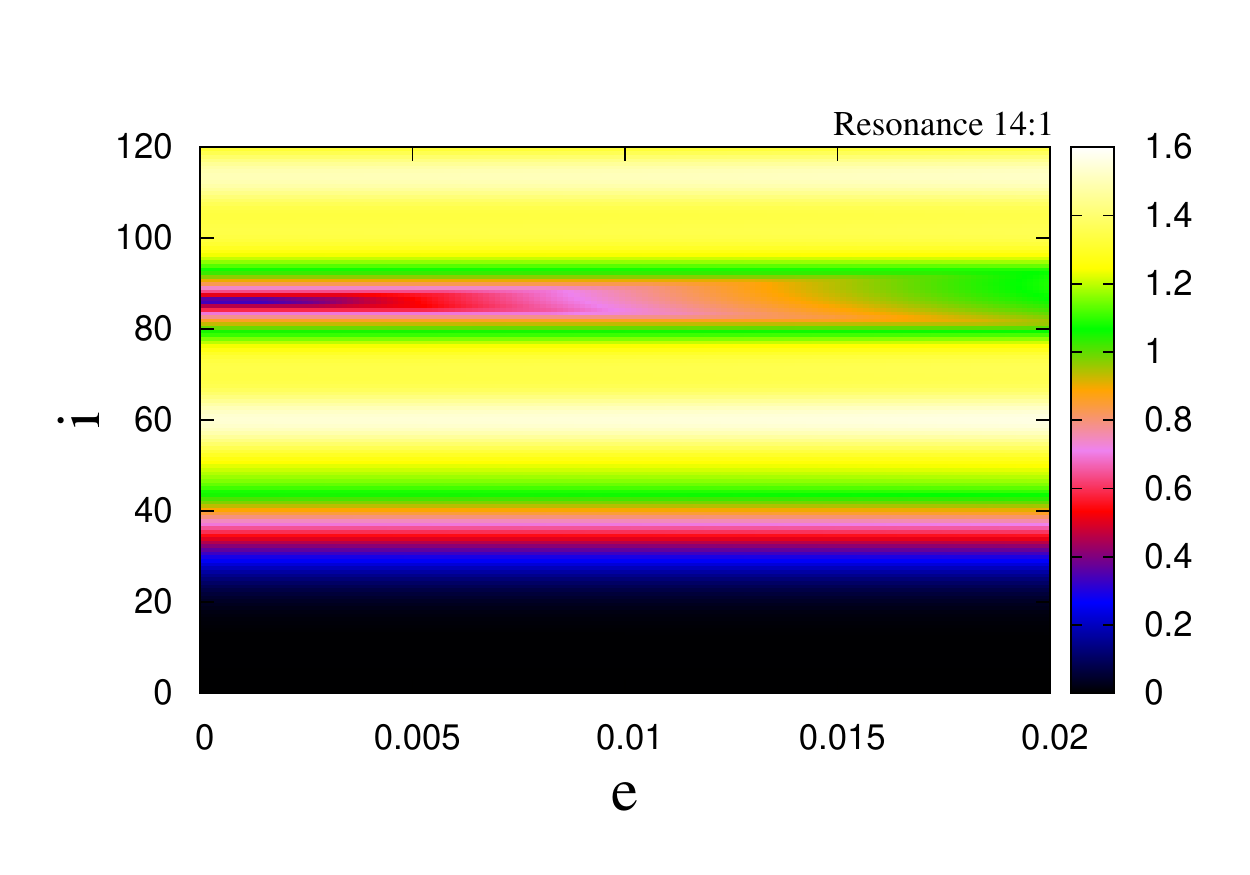}
\vglue0.4cm
\caption{The amplitude of the resonances for different values of the eccentricity (within 0 and 0.02 on the
horizontal axis)
and the inclination (within $0^o$ and $120^o$ on the vertical axis); the color bar provides the measure of the amplitude in kilometers.
In order from top to bottom, left to right: 11:1, 12:1, 13:1, 14:1.} \label{fig:amplitude}
\end{figure}

In view of \eqref{LGH_aei}, \eqref{H}, \eqref{Rsec}, \eqref{Resonant_part_2},  the conservative part of
the toy model is given by
\begin{equation}\label{H_toy}
\mathcal{H}^{m:1}_{toy}(L,G,H,\sigma_{m1})=-\frac{\mu_E^2}{2 L^2}+\frac{\alpha}{L^3 G^3} \Bigl(1-3 \frac{H^2}{G^2}\Bigr)
+\mathcal{A}_0^{(m)} (L,G,H) \cos (\sigma_{m1}-\varphi_0^{(m)} (L,G,H))\,,
\end{equation}
where
$$
\alpha=\frac{\sqrt{5}R_E^2 \overline{J}_2 \mu_E^4}{4}
$$
and $\sigma_{m1}$, $\mathcal{A}_0^{(m)}$, $\varphi_0^{(m)}$ are given by
\eqref{sigma_angle}, \eqref{A_varphi_11_13}, \eqref{A_varphi_12_14}.

Let us now perform a canonical change of coordinates, similar to that presented in~\cite{CGmajor}, which transforms the variables $(L,G,H, M, \omega, \Omega)$ into $(\widetilde{L}, \widetilde{G}, \widetilde{H}, \sigma_{m1}, \omega, \Omega)$, where $\sigma_{m1}$
is given by \eqref{sigma_angle}, $\omega$ and $\Omega$ are kept unaltered and
\begin{equation}\label{canonical_transformation}
\widetilde{L}=L\,, \qquad \widetilde{G}=G-L\,, \qquad \widetilde{H}=H-m L\,.
\end{equation}
In terms of the new variables, the Hamiltonian \equ{H_toy} takes the form
\begin{equation}\label{toy_ham_canonical}
\widetilde{\mathcal{H}}^{m:1}_{toy}(\widetilde L,\widetilde G,\widetilde H,\sigma_{m1})=
\widetilde{h}^{(m)}(\widetilde L,\widetilde G,\widetilde H)
+\varepsilon \widetilde{\mathcal{A}}^{(m)}(\widetilde L,\widetilde G,\widetilde H)
\cos (\sigma_{m1}-\widetilde{\varphi}^{(m)}(\widetilde L,\widetilde G,\widetilde H))\ ,
\end{equation}
where
\begin{equation}\label{hAphi_tilde}
\begin{split}
&\widetilde{h}^{(m)}(\widetilde{L}, \widetilde{G}, \widetilde{H})=-\frac{\mu_E^2}{2 \widetilde{L}^2}-m \widetilde{L}+\frac{\alpha}{\widetilde{L}^3 (\widetilde{G}+\widetilde{L})^3} \Bigl(1-3 \frac{(\widetilde{H}+m \widetilde{L})^2}{(\widetilde{G}+ \widetilde{L})^2}\Bigr)\,,\\
&\varepsilon \widetilde{\mathcal{A}}^{(m)}(\widetilde{L}, \widetilde{G}, \widetilde{H})=\mathcal{A}_0^{(m)}(\widetilde{L}, \widetilde{G}+\widetilde{L}, \widetilde{H}+m \widetilde{L})\,,\\
&\widetilde{\varphi}^{(m)}(\widetilde{L}, \widetilde{G}, \widetilde{H})=\varphi_0^{(m)}(\widetilde{L}, \widetilde{G}+\widetilde{L}, \widetilde{H}+m \widetilde{L})\,
\end{split}
\end{equation}
and $\varepsilon$ is a small coefficient introduced for convenience,
so that $\widetilde{h}^{(m)}$ and  $\widetilde{\mathcal{A}}^{(m)}$ have comparable
sizes, when measured at the same point.

Strictly speaking, the quantity $\widetilde{\varphi}^{(m)}(\widetilde{L}, \widetilde{G}, \widetilde{H})$
depends on the variable $\widetilde{L}$ and not on $\widetilde{L}_{res}$, which is the
value of $\widetilde{L}$ at the resonance. However, the numerical tests show that the error is very small,
of the order of few arcseconds, if $\widetilde{\varphi}^{(m)}(\widetilde{L}, \widetilde{G}, \widetilde{H})$
is replaced by $\widetilde{\varphi}^{(m)} (\widetilde{L}_{res},\widetilde{G},\widetilde{H})$.
Since we are interested in obtaining a reduced model, allowing to explain the results provided by the DMLR,
we take $\widetilde{\varphi}^{(m)}$ as constant in $\widetilde{L}$ and write  $\widetilde{\varphi}^{(m)}=\widetilde{\varphi}^{(m)}(\widetilde{G}, \widetilde{H})$ in order to underline this aspect.

Before analyzing the dissipative part, let us study first the conservative effects. Therefore, we disregard for the moment the influence of the drag force and we focus our attention on  the Hamiltonian \eqref{toy_ham_canonical}.
Since $\omega$ and $\Omega$ are cyclic variables,
it results that $\widetilde{G}$ and $\widetilde{H}$ are constants, so that
the dynamics is described by a pendulum type Hamiltonian. In particular, following the method described
in \cite{CGmajor}, the width of the resonances can be easily computed for the pendulum-like model.
We refer to \cite{CGminor} for the formulae necessary to compute the amplitudes of the islands
associated to \equ{H_toy}.
Figure~\ref{fig:amplitude} provides the amplitudes of the 11:1, 12:1, 13:1 and 14:1 resonances as the eccentricity varies
between $0$ and $0.02$, while the inclination ranges between $0^o$ and $120^o$. The color bar indicates the size of the amplitude in kilometers.

Figure~\ref{fig:amplitude} shows that for inclinations less than $30^o$ the amplitude is small, at most $350\, m$,  while for larger inclinations, the amplitude could reach about two (or three for the 13:1 resonance) kilometers.
Having in mind these results, we can anticipate what happens when the dissipative effects
are taken into account for the 12:1, 13:1, 14:1 resonances: we expect the equilibrium points to persist for those inclinations which lead (in the conservative case) to large amplitudes,
even if the ballistic coefficient is high. On the contrary, for small inclinations -- since the amplitude is small --  one has the opposite situation:
the magnitude of the drag force is large in comparison with the resonant part and, therefore,
we anticipate that the equilibrium points do not exist. These statements are proved analytically in
Sections~\ref{sec:existence}, \ref{sec:type}.

For the moment, let us go back to the equations of motion
and discuss about the dissipative part. Collecting \eqref{canonical_eq}, \eqref{dissipative_functions_circular}, \eqref{canonical_transformation} and \eqref{toy_ham_canonical}, we obtain:
\begin{equation} \label{toy_canonical_eq}
\begin{split}
\dot{\sigma}_{m1}=\frac{\partial \widetilde{\mathcal{H}}^{m:1}_{toy}}{\partial \widetilde{L}}\,,\qquad \quad & \qquad   \dot{\omega}=\frac{\partial \widetilde{\mathcal{H}}^{m:1}_{toy}}{\partial \widetilde{G}}\,, \ \quad \qquad \qquad \dot{\Omega}=\frac{\partial \widetilde{\mathcal{H}}^{m:1}_{toy}}{\partial \widetilde{H}}\,,\\
\dot{\widetilde{L}}=-\frac{\partial \widetilde{\mathcal{H}}^{m:1}_{toy}}{\partial \sigma_{m1}}-
\eta D^{(m)}_{_L}(\widetilde L,\widetilde G,\widetilde H)\,, & \qquad
\dot{\widetilde{G}}=-\eta D^{(m)}_{_G}(\widetilde L,\widetilde G,\widetilde H)\,, \qquad
\dot{\widetilde{H}}=-\eta D^{(m)}_{_H}(\widetilde L,\widetilde G,\widetilde H)\,,
\end{split}
\end{equation}
where the dissipative effects are described by the time depending parameter $\eta=\rho B$ and the functions
$D^{(m)}_{_L}$, $D^{(m)}_{_G}$, $D^{(m)}_{_H}$ are defined as
\beqa{diss_D}
D^{(m)}_{_L}(\widetilde L,\widetilde G,\widetilde H)&=&\frac{\mu_E}{2} \biggl(1-\frac{\omega_E \widetilde{L}^3
(\widetilde{H}+m \widetilde{L})}{\mu_E^2 (\widetilde{G}+\widetilde{L})} \biggr)^2\ ,\nonumber\\
D^{(m)}_{_G}(\widetilde L,\widetilde G,\widetilde H)&=&\frac{\widetilde{G}}{\widetilde{L}}
D^{(m)}_{_L}(\widetilde L,\widetilde G,\widetilde H)\ ,\nonumber\\
D^{(m)}_{_H}(\widetilde L,\widetilde G,\widetilde H)&=&\frac{\widetilde{H}}{\widetilde{L}}
D^{(m)}_{_L}(\widetilde L,\widetilde G,\widetilde H)\ .
\eeqa
Since $\eta$ is a small quantity, from $\eqref{toy_canonical_eq}$, it follows that $\widetilde{G}$ and $\widetilde{H}$
modify slightly in time, due to the effect of the dissipation. Being interested in equilibria located
in the $(\sigma_{m1}, \widetilde{L})$ plane, and also in obtaining a very reduced model
apt to explore the dynamics of infinitesimal bodies close to resonances,
we define a \sl dissipative toy model \rm governed by the following
differential equations:
\beqa{toy_canonical_eq_final}
\dot{\sigma}_{m1}&=&\widetilde{h}^{(m)}_{,L}(\widetilde L,\widetilde G,\widetilde H)+ \varepsilon
\widetilde{\mathcal{A}}^{(m)}_{,L}(\widetilde L,\widetilde G,\widetilde H)\
\cos(\sigma_{m1}-\widetilde{\varphi}^{(m)}(\widetilde G,\widetilde H))\ ,\nonumber\\
 \dot{\widetilde{L}}&=& \varepsilon \widetilde{\mathcal{A}}^{(m)}(\widetilde L,\widetilde G,\widetilde H)
 \sin (\sigma_{m1}-\widetilde{\varphi}^{(m)}(\widetilde G,\widetilde H))-
 \eta D^{(m)}_{_L}(\widetilde L,\widetilde G,\widetilde H)\ ,
\eeqa
where $\widetilde{G}$ and $\widetilde{H}$ are considered constants;
let us stress this aspect by replacing them in the following by $\widetilde{G}_0$ and $\widetilde{H}_0$.
Also, we will use the customary differentiation convention stating that subscripts preceded
by a comma denote partial differentiation with respect to the corresponding variable.

Since the main goal of this section is to present a qualitative description of
the interplay between the resonances and the dissipative effects (including the existence,
type and location of the equilibrium points as a function of various parameters),
we shall consider the parameter $\eta$ as a constant, leaving to Section~\ref{sec:results} the study
of the case of a variable $\eta$, which corresponds to study the effects of
the solar cycle.

\begin{figure}[h]
\centering \vglue0.1cm \hglue0.2cm
\includegraphics[width=5truecm,height=4truecm]{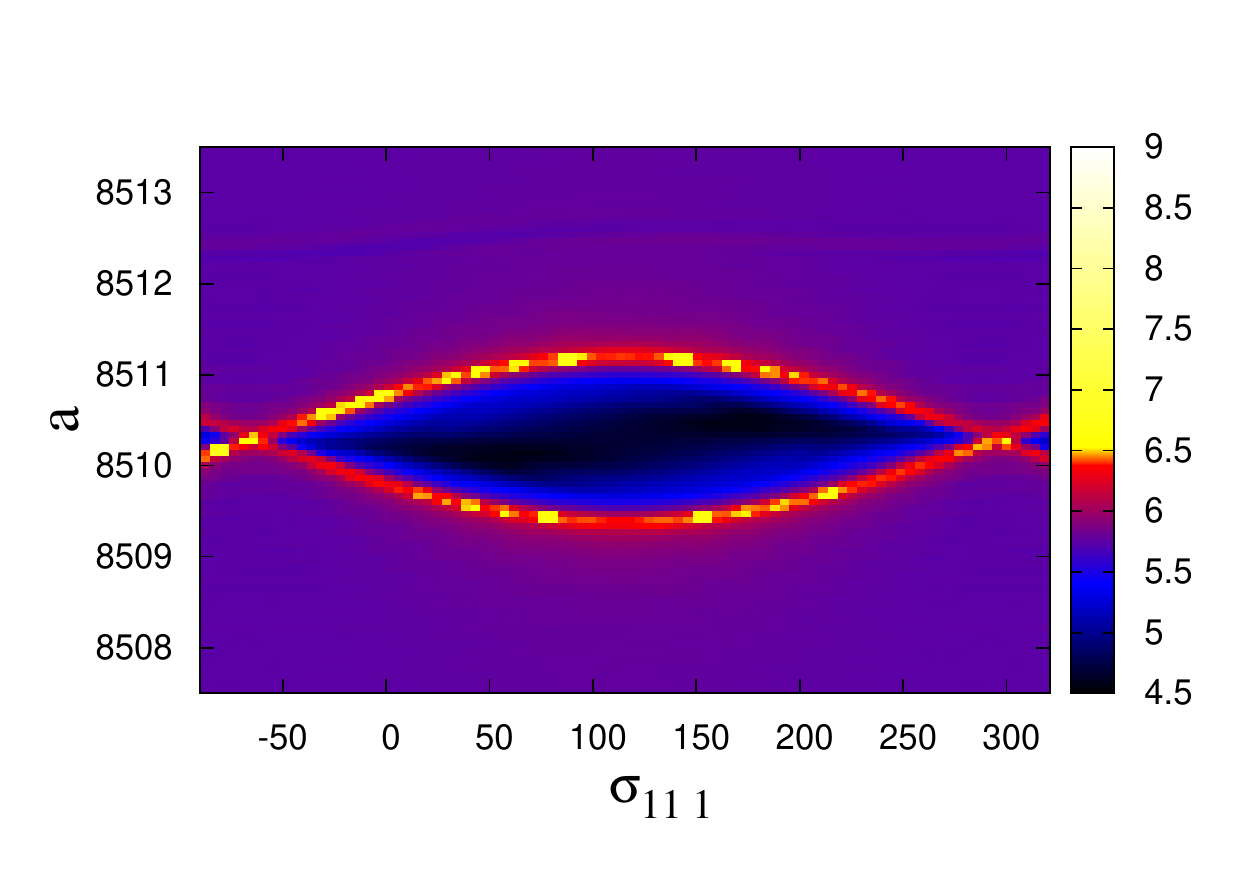}
\includegraphics[width=5truecm,height=4truecm]{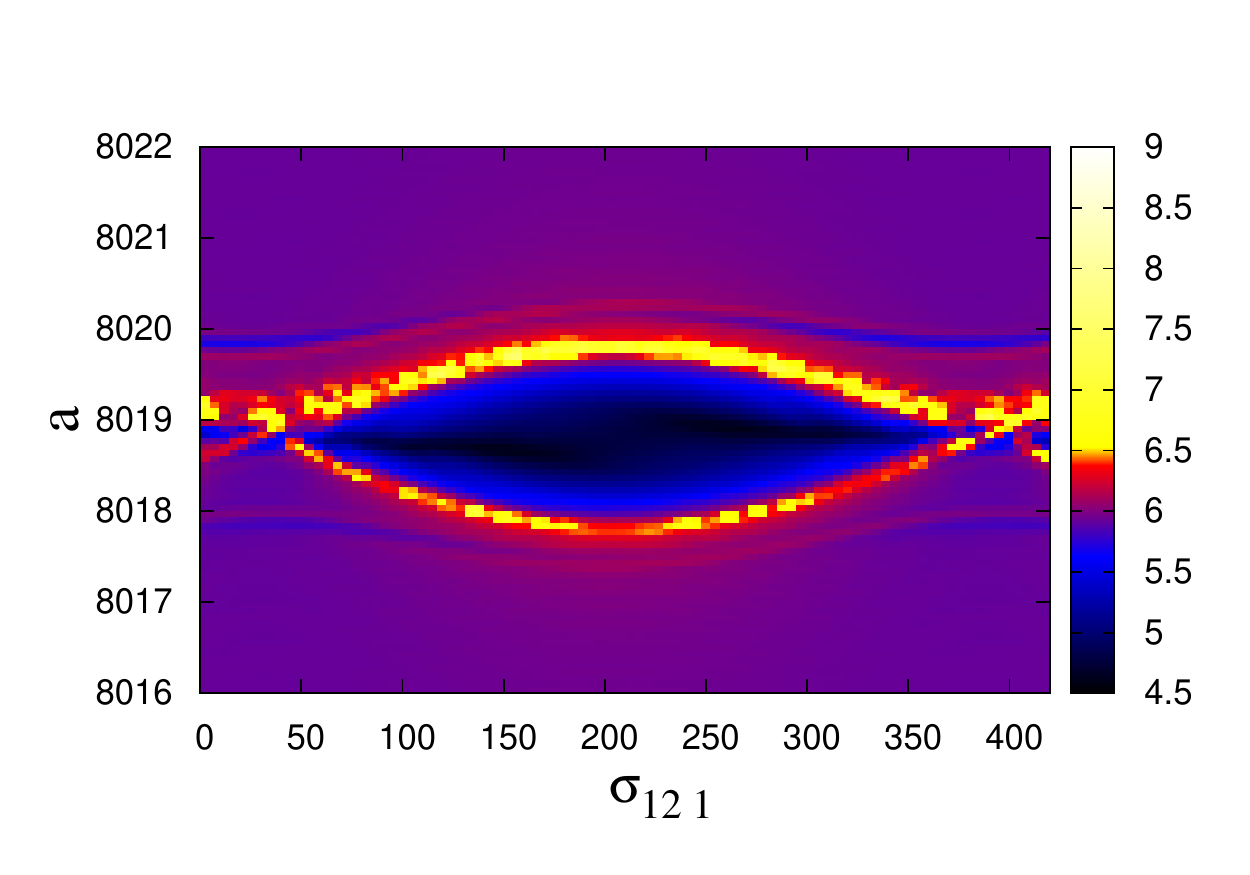}
\includegraphics[width=5truecm,height=4truecm]{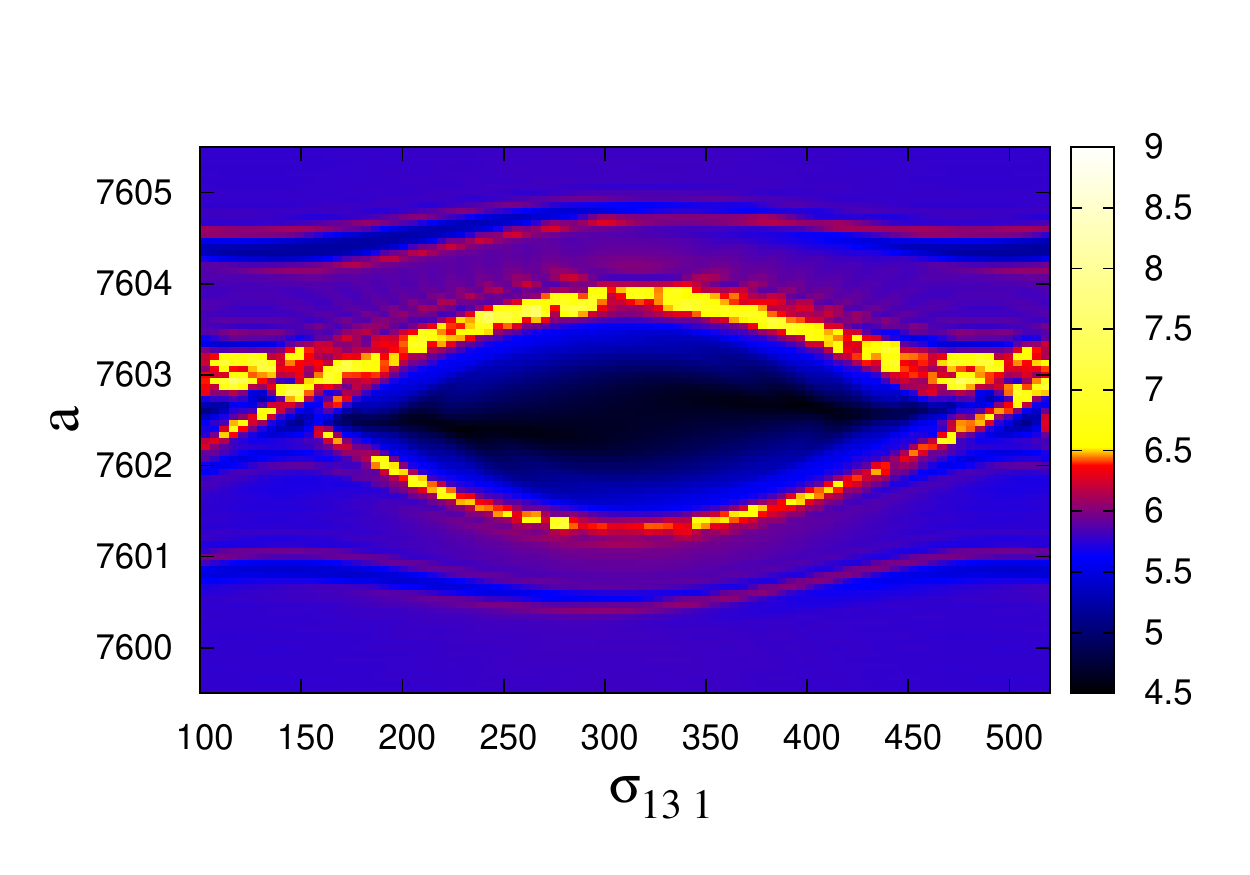}\\
\vglue-0.6cm
\includegraphics[width=5truecm,height=4truecm]{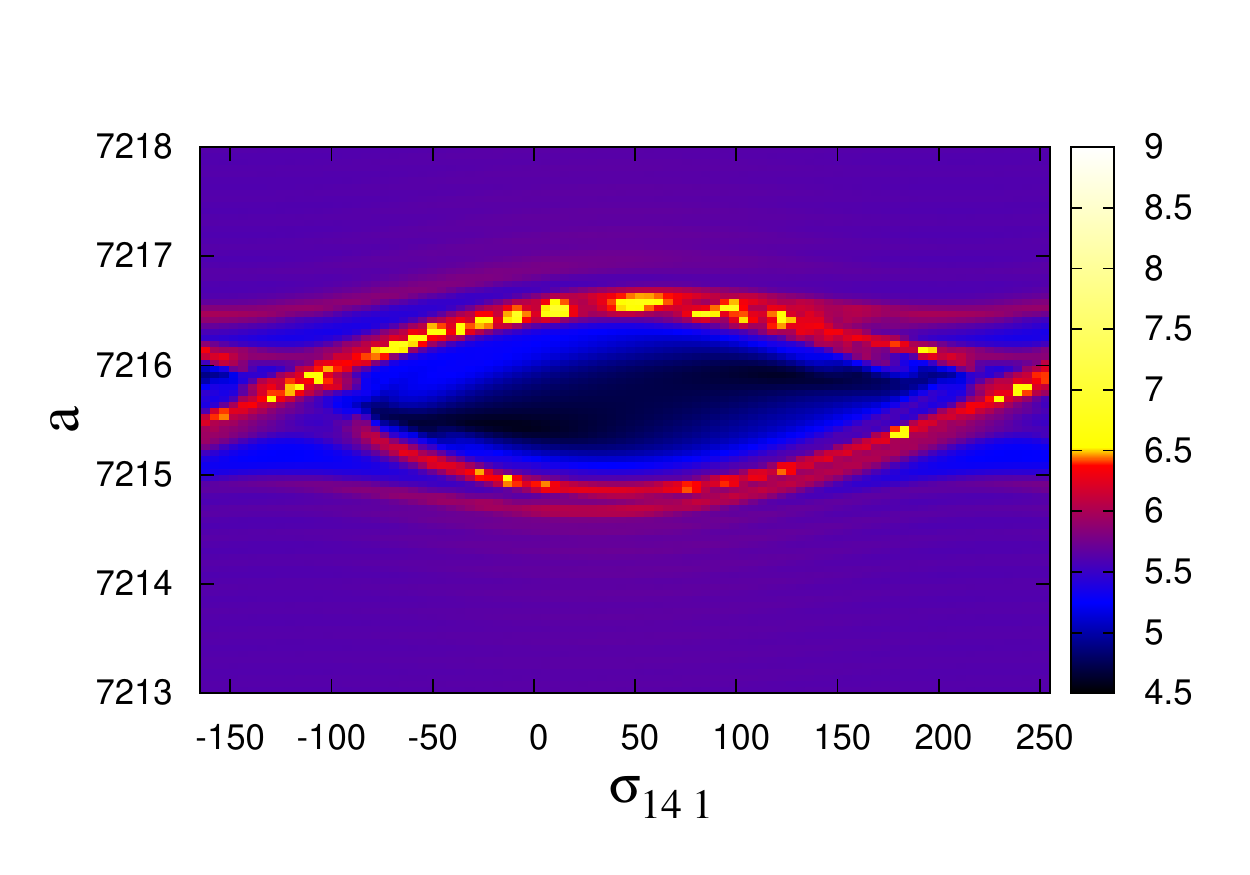}
\includegraphics[width=5truecm,height=4truecm]{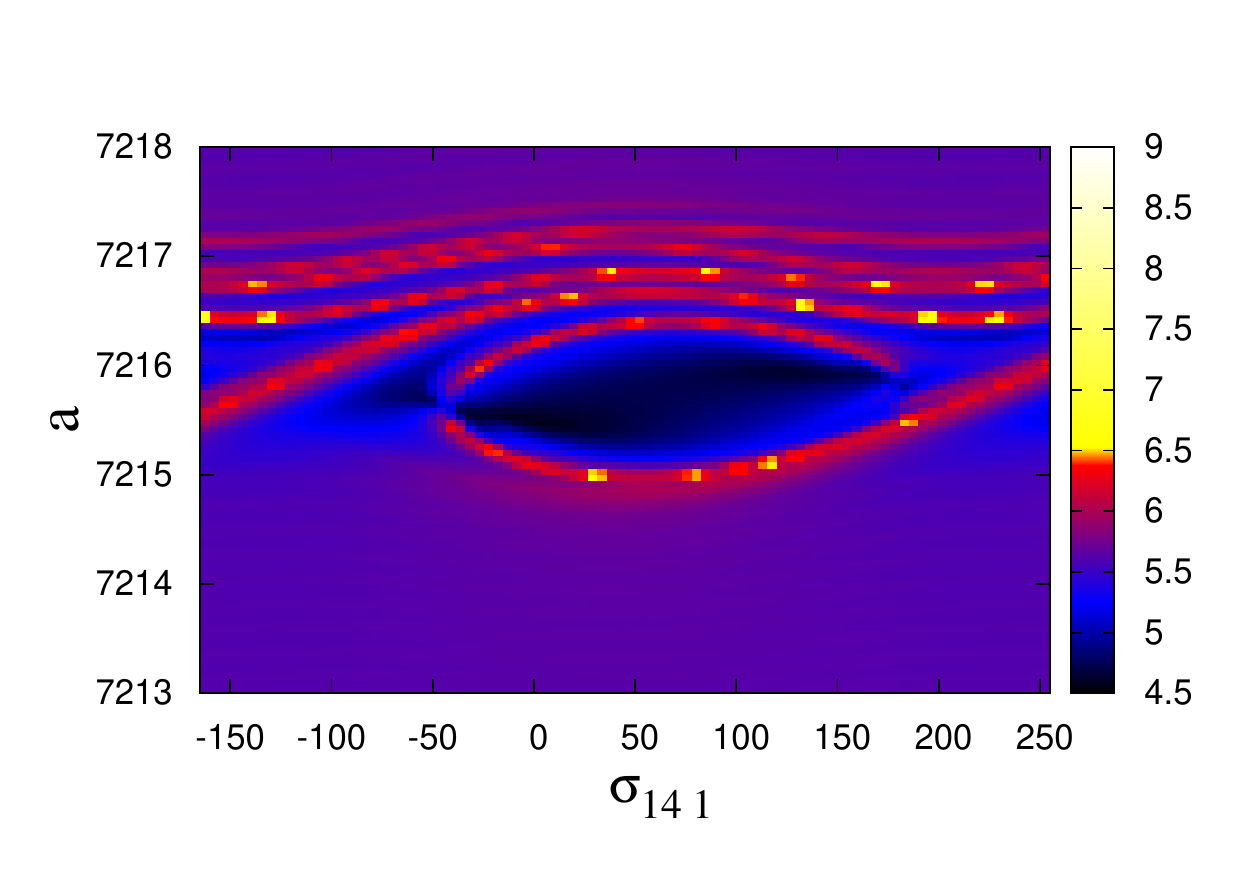}
\vglue0.4cm
\caption{FLI (using the DMLR) for the 11:1, 12:1, 13:1, 14:1 resonances for $e=0.005$, $\omega=0^o$, $\Omega=0^o$.
Top left: 11:1 resonance for $i=80^o$; top middle: 12:1 resonance for $i=70^o$, mean atmospheric density and
$B=220$ $[cm^2/kg]$; top right: 13:1 resonance for $i=75^o$, mean atmospheric density and $B=220$ $[cm^2/kg]$;
bottom: 14:1 resonance for $i=60^o$, mean atmospheric density and $B=30$ $[cm^2/kg]$ (left panel),
respectively  $B=220$ $[cm^2/kg]$  (right panel). The time span is $1500$ sidereal days
(about $4$ years).} \label{fig:cartography}
\end{figure}

In order to validate the toy model and to show numerically the
existence of the equilibrium points, we present in Figure~\ref{fig:cartography}
some results obtained by using the DMLR
described in the previous sections, including also the air
resistance effect for the 12:1, 13:1 and 14:1 resonances.
Plotting the Fast Lyapunov Indicator\footnote{The Fast Lyapunov Indicator is a measure
of the regular and chaotic dynamics; it was introduced in \cite{froes} and it amounts,
in  short, to the Lyapunov exponent computed on  finite times.}, hereafter denoted as FLI
(see, e.g, \cite{froes,GLF2002,GL2013,CGmajor}) for some given values of the
parameters (i.e. eccentricity, inclination, ballistic coefficient,
etc.), we can infer a very good agreement between the equilibria of
the toy model and those of the DMLR. Indeed, the equilibrium points are clearly
revealed for small dissipations (or for the non--dissipative case of
the 11:1 resonance), the resonant islands have the amplitude as
predicted by the conservative toy model (compare with
Figures~\ref{fig:amplitude} and \ref{fig:cartography}) and, as we
will see in the next sections, the dissipative toy model is able
to predict the existence and location of the equilibrium points.

\begin{figure}[h]
\centering \vglue0.1cm \hglue0.2cm
\includegraphics[width=6truecm,height=5truecm]{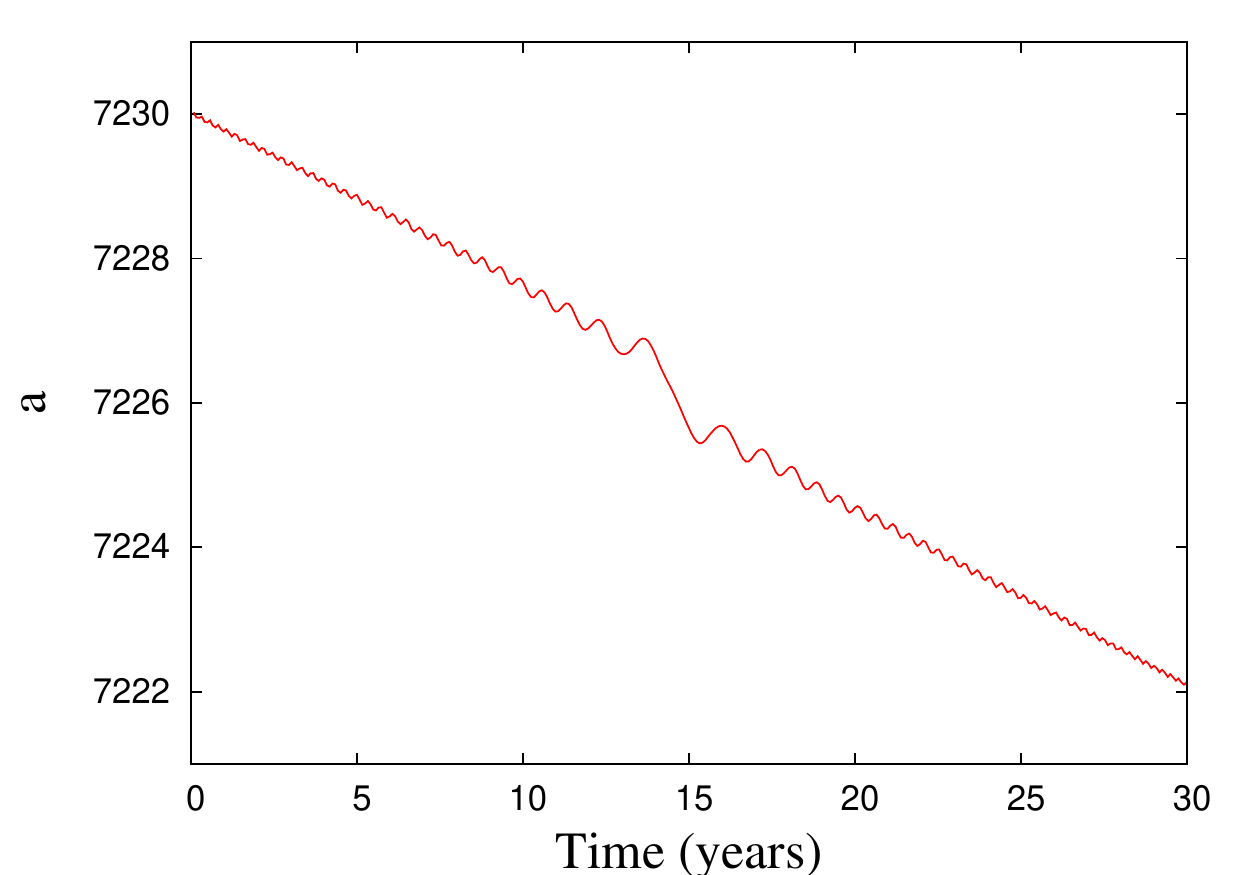}
\includegraphics[width=6truecm,height=5truecm]{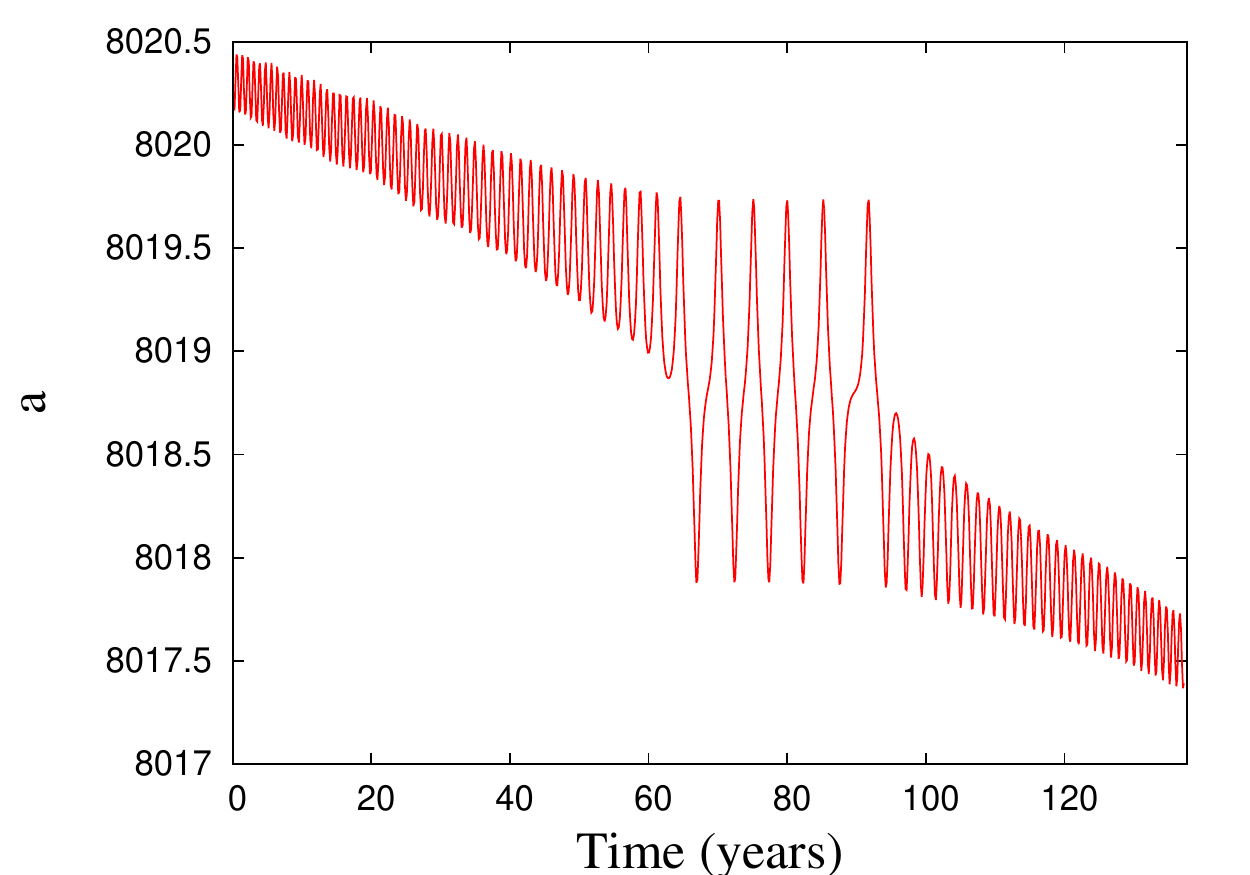}
\vglue0.8cm
\caption{Passage through the 14:1 resonance (left) and temporary capture into the 12:1 resonance (right). The plots are obtained for $e=0.005$, $i=70^o$, $\omega=0^o$, $\Omega=0^o$, $\sigma_{m1}=100^o$ and $B=220$ $[cm^2/kg]$. } \label{fig:pass_capture}
\end{figure}

Since the upper left panel of Figure~\ref{fig:cartography} is obtained for a conservative model,
more precisely a pendulum-type Hamiltonian, the stable and unstable points,
as well as the separatrix are clearly marked. Since the other plots of Figure~\ref{fig:cartography}
take also into account the dissipative effect, the separatrix of each plot
is not longer a single line, as for a conservative system;
the gradual decrease of the orbits' altitude due to dissipation leads the paths located above
the resonant region to reach, after some time, the separatrix. This is the reason why in
all other plots of Figure~\ref{fig:cartography} we notice a larger chaotic region above the resonant island,
than below it. The plots are obtained by integrating the equations of motion for an interval of 1500 sidereal days.
Due to the orbital decay process, a longer time span integration is considered, hence a much larger
chaotic region is obtained above the resonant zone. Once an orbit reaches
the resonant region, two scenarios are possible: either it passes through resonance, or
it is captured into resonance.
Numerical simulations show that the capture is a rare and temporary phenomenon, depending on
different factors including that $\eta$ varies in time as effect of the solar cycle (see Section~\ref{sec:results}).
In any case, even if  the object is captured temporarily by a resonance,
it does not usually reach the center of the island, where the  spiral point is located.
Figure~\ref{fig:pass_capture}, obtained by using the DMLR, shows an example of the two different phenomena:
a passage through the 14:1 resonance and a temporary capture in the 12:1 resonance.

\subsection{Existence of equilibrium points} \label{sec:existence}
Using the toy model introduced in Section~\ref{sec:toy}, we can prove the following result.

\begin{theorem} \label{Theorem:existence}
For fixed values of $e\in [0,0.02]$ and $i\in [0^o,120^o]$ (or equivalently, given $\widetilde{G}_0$
and $\widetilde{H}_0$ in the corresponding intervals), let  $(\sigma_{m1}^{(0)}$, $\widetilde{L}_0)$ be an equilibrium point for
the model described by the Hamiltonian \equ{toy_ham_canonical}. Let $\widetilde{\mathcal{A}}^{(m)}$
be as in \equ{hAphi_tilde} and $D^{(m)}_{_L}$ as in \equ{diss_D}; assume that
$\eta$, $\varepsilon$ satisfy the inequalities:
\beqa{existence_condition}
&&\left| \frac{\eta D^{(m)}_{_L} (\widetilde{L}_0, \widetilde{G}_0, \widetilde{H}_0)}{\varepsilon
\widetilde{\mathcal{A}}^{(m)}(\widetilde{L}_0, \widetilde{G}_0, \widetilde{H}_0)} \right| +  \frac{  2 \varepsilon \Bigl( \widetilde{\A}_{,L}^{(m)}(\widetilde{L}_0, \widetilde{G}_0, \widetilde{H}_0)\Bigr)^2
 +2\eta \left| \widetilde{\A}_{,L}^{(m)}(\widetilde{L}_0, \widetilde{G}_0, \widetilde{H}_0)  D_{L,L}^{(m)}(\widetilde{L}_0, \widetilde{G}_0, \widetilde{H}_0)\Bigr)\right|}{ | \widetilde{h}_{,LL}^{(m)} (\widetilde{L}_0, \widetilde{G}_0, \widetilde{H}_0) | \ \widetilde{\A}^{(m)}(\widetilde{L}_0, \widetilde{G}_0, \widetilde{H}_0)} \leq 1-\delta\ ,\nonumber\\
\nonumber\\
&&\qquad\qquad\varepsilon^2<\gamma\ \delta
\eeqa
for some constants $0<\delta<1$ and $\gamma>0$. Then, the dissipative toy model described by the equations \equ{toy_canonical_eq_final} admits equilibrium points. At first order in $\eta$, the point
($\sigma_{m1}^{(1)}$, $\widetilde{L}_1$) defined by
\begin{equation}\label{eq_L_sigma}
\sigma_{m1}^{(1)}= \sigma_{m1}^{(0)}+ \frac{D^{(m)}_{_L} (\widetilde{L}_0, \widetilde{G}_0, \widetilde{H}_0)}
{\varepsilon  \widetilde{ \mathcal{A}}^{(m)}(\widetilde{L}_0, \widetilde{G}_0, \widetilde{H}_0)
\cos (\sigma_{m1}^0-\widetilde{\varphi}^{(m)}) }\eta \ , \qquad \widetilde{L}_1=\widetilde{L}_0
\end{equation}
is an equilibrium point for the dissipative model.
\end{theorem}

{\bf Proof.} Since  $(\sigma_{m1}^{(0)}$, $\widetilde{L}_0)$ is an equilibrium point for the conservative
model \equ{toy_ham_canonical}, one has
\beqa{eq_points_cons_cond}
\widetilde{h}_{,L}^{(m)}(\widetilde{L}_0, \widetilde{G}_0, \widetilde{H}_0)+
\varepsilon \widetilde{\mathcal{A}}_{,L}^{(m)}(\widetilde{L}_0, \widetilde{G}_0, \widetilde{H}_0)
\cos (\sigma_{m1}^{(0)}-\widetilde{\varphi}^{(m)}( \widetilde{G}_0, \widetilde{H}_0))&=&0\ ,\nonumber\\
\qquad \sin(\sigma_{m1}^{(0)}-\widetilde{\varphi}^{(m)}( \widetilde{G}_0, \widetilde{H}_0))&=&0\ .
\eeqa
The relations \eqref{eq_points_cons_cond} represent an uncoupled system of two equations. The second of \eqref{eq_points_cons_cond} provides two values for $\sigma_{m1}^{(0)}$ in the interval $[0^o, 360^o)$. Once $\sigma_{m1}^{(0)}$ is known, $\widetilde{L}_0$ is found
by solving the first of \eqref{eq_points_cons_cond} for fixed values of $\widetilde{G}_0$, $\widetilde{H}_0$.

It is important to stress that, since $\varepsilon$ is small,  $\widetilde{L}_0$ has the form
$ \widetilde{L}_{0}=\widetilde{L}^{sec} + \varepsilon L_0^* +O(\varepsilon^2)$, where $ L_0^*$ is independent of $\varepsilon$ and
 $\widetilde{L}^{sec}$ satisfies the equation
$$\widetilde{h}_{,L}^{(m)}(\widetilde{L}^{sec}, \widetilde{G}_0, \widetilde{H}_0)=0\ .$$
Inserting $ \widetilde{L}_{0}=\widetilde{L}^{sec} + \varepsilon L_0^* +O(\varepsilon^2)$ in the first of \eqref{eq_points_cons_cond} and expanding to the first order in $\varepsilon$,
we find that  $\widetilde{L}_0$ has the form
\begin{equation}\label{Lsec}
 \widetilde{L}_{0}=\widetilde{L}^{sec} \pm \varepsilon \ \frac{\widetilde{\mathcal{A}}_{,L}^{(m)}
 (\widetilde{L}^{sec}, \widetilde{G}_0, \widetilde{H}_0)}{\widetilde{h}_{,LL}^{(m)}(\widetilde{L}^{sec}, \widetilde{G}_0, \widetilde{H}_0)} +O(\varepsilon^2)\ ,
\end{equation}
where  the signs $\pm$ correspond to the two solutions of the second of \eqref{eq_points_cons_cond}.
In obtaining \eqref{Lsec},  we took into account the fact that $\widetilde{h}^{(m)}$ in \equ{hAphi_tilde} assures that $\widetilde{h}_{,LL}^{(m)}$ cannot be zero for the resonances
and parameter values considered in this work (notably, $J_2$ is sufficiently small).

On the other hand, for the dissipative toy model we have the following coupled equations for the determination of an equilibrium point, say $(\sigma_{m1}^{(d)}$, $\widetilde{L}_d)$:
\beqa{eq_points_dissipative_equations}
\widetilde{h}_{,L}^{(m)}(\widetilde{L}_d, \widetilde{G}_0, \widetilde{H}_0)+
\varepsilon \widetilde{\mathcal{A}}_{,L}^{(m)}(\widetilde{L}_d, \widetilde{G}_0, \widetilde{H}_0)
\cos (\sigma_{m1}^{(d)}-\widetilde{\varphi}^{(m)}( \widetilde{G}_0, \widetilde{H}_0))&=&0\ ,\nonumber\\
\qquad
\varepsilon \widetilde{\mathcal{A}}^{(m)}(\widetilde{L}_d, \widetilde{G}_0, \widetilde{H}_0)
\sin (\sigma_{m1}^{(d)}-\widetilde{\varphi}^{(m)}( \widetilde{G}_0, \widetilde{H}_0))
-\eta  D_L^{(m)} ( \widetilde{L}_d,  \widetilde{G}_0, \widetilde{H}_0)&=&0\ .
\eeqa
 The first of \eqref{eq_points_dissipative_equations} can always be satisfied; that is, for any value of $\sigma_{m1}^{(d)}$ in the interval $[0^o, 360^o)$ we may find a value
 $\widetilde{L}_d$ which verifies this equation. However, the second of \eqref{eq_points_dissipative_equations} is satisfied only if the dissipative effects do not exceed a threshold value. To show this, let us fix an arbitrary value for $\sigma_{m1}^{(d)}$ in the interval $[0^o, 360^o)$ and let $\widetilde{L}_d^\sigma$ be such that $(\sigma_{m1}^{(d)}, \widetilde{L}_d^\sigma)$ satisfies the first of \eqref{eq_points_dissipative_equations}.
Using the same argument as the one used to obtain \eqref{Lsec}, we deduce that  $\widetilde{L}_d^\sigma$ has the form
 \begin{equation}\label{Ldis}
 \widetilde{L}_{d}^\sigma=\widetilde{L}^{sec} - \varepsilon \ \frac{\widetilde{\mathcal{A}}_{,L}^{(m)}
 (\widetilde{L}^{sec}, \widetilde{G}_0, \widetilde{H}_0) \cos (\sigma_{m1}^{(d)}-\widetilde{\varphi}^{(m)}( \widetilde{G}_0, \widetilde{H}_0))}{\widetilde{h}_{,LL}^{(m)}(\widetilde{L}^{sec}, \widetilde{G}_0, \widetilde{H}_0)} +O(\varepsilon^2)\ .
\end{equation}
 Now, we note that if $f$ is a differentiable function of $\widetilde{L}$, then in view of the relation $ \widetilde{L}^{sec}=\widetilde{L}_{0} - \varepsilon L_0^* +O(\varepsilon^2)$, we can write $f(\widetilde{L}^{sec})=f(\widetilde{L}_{0})-\varepsilon L_0^*\ f_{,L}(\widetilde{L}_0)+O(\varepsilon^2)$. Using this argument, from \eqref{Lsec} and \eqref{Ldis} we get
\begin{equation}\label{Ldis0}
 \widetilde{L}_{d}^\sigma=\widetilde{L}_0 - \varepsilon \ \frac{\widetilde{\mathcal{A}}_{,L}^{(m)}
 (\widetilde{L}_0, \widetilde{G}_0, \widetilde{H}_0) \Bigl(\cos (\sigma_{m1}^{(d)}-\widetilde{\varphi}^{(m)}( \widetilde{G}_0, \widetilde{H}_0)) \mp 1\Bigr)}{\widetilde{h}_{,LL}^{(m)}(\widetilde{L}_0, \widetilde{G}_0, \widetilde{H}_0)} +O(\varepsilon^2)\ .
\end{equation}
 Inserting $\widetilde{L}_{d}^\sigma$ given by \eqref{Ldis0} in the second of \eqref{eq_points_dissipative_equations}, then after some computations we get the following equation for the unknown variable $\sigma_{m1}^{(d)}$
 \begin{equation} \label{sincond}
 \sin (\sigma_{m1}^{(d)}-\widetilde{\varphi}^{(m)}) =\frac{\eta D_L^{(m)}}{\varepsilon \widetilde{\A}^{(m)}}+ \frac{\widetilde{\A}_{,L}^{(m)} \ \Bigl(\cos (\sigma_{m1}^{(d)}-\widetilde{\varphi}^{(m)}) \mp 1\Bigr)}{ \widetilde{h}_{,LL}^{(m)} \widetilde{\A}^{(m)}} \Bigl( \varepsilon \widetilde{\A}_{,L}^{(m)} \sin (\sigma_{m1}^{(d)}-\widetilde{\varphi}^{(m)})-\eta D_{L,L}^{(m)} \Bigr) +O(\varepsilon^2)\,,
 \end{equation}
 where all functions are evaluated at $\widetilde{L}_0$, $\widetilde{G}_0$, $\widetilde{H}_0$. In view of \eqref{existence_condition},
bounding the terms of second order in $\varepsilon$ by $C_0\varepsilon^2$ for a suitable constant
$C_0>0$, we have
$$
\Bigl| \frac{\eta D_L^{(m)}}{\varepsilon \widetilde{\A}^{(m)}}+ \frac{\widetilde{\A}_{,L}^{(m)} \ \Bigl(\cos (\sigma_{m1}^{(d)}-\widetilde{\varphi}^{(m)}) \mp 1\Bigr)}{ \widetilde{h}_{,LL}^{(m)} \widetilde{\A}^{(m)}} \Bigl( \varepsilon \widetilde{\A}_{,L}^{(m)} \sin (\sigma_{m1}^{(d)}-\widetilde{\varphi}^{(m)})-\eta D_{L,L}^{(m)} \Bigr)\Bigr| +C_0\varepsilon^2$$
  $$\hspace{3cm} \leq \left| \frac{\eta D^{(m)}_{_L} }{\varepsilon
\widetilde{\mathcal{A}}^{(m)}} \right| +  \frac{  2 \varepsilon \Bigl(\widetilde{\A}_{,L}^{(m)}\Bigr)^2 +2
 \eta |\widetilde{\A}_{,L}^{(m)}  D_{L,L}^{(m)}|}{ |\widetilde{h}_{,LL}^{(m)}| \  \widetilde{\A}^{(m)}}
 +C_0\varepsilon^2 \leq 1-\delta+C_0 \varepsilon^2 \leq 1\ ,
$$
for $\varepsilon$ sufficiently small with respect to $\delta$ as in the second of \eqref{existence_condition} with $\gamma \equiv 1/C_0$.  Therefore, if
\eqref{existence_condition} are satisfied, then the right hand side of \eqref{sincond} is subunitary,
which implies that the dissipative toy model admits equilibrium points.

Assuming that
$\eta$ is sufficiently small, so that \eqref{existence_condition} holds, then at first order in $\eta$, it is natural to look for an equilibrium point of the dissipative toy model
$\eqref{toy_canonical_eq_final}$ of the type $(\sigma_{m1}^{(1)}$, $\widetilde{L}_1)$, where
\begin{equation}\label{Lsig_form}
\sigma_{m1}^{(1)}=\sigma_{m1}^{(0)}+ \eta \sigma_{m1}^*+O(\eta^2)\,, \qquad
\widetilde{L}_1=\widetilde{L}_0+\eta L^*+O(\eta^2)
\end{equation}
with $L^*$ and $\sigma_{m1}^*$ independent of $\eta$. In fact, we shall suppose that $\eta$ is smaller than $\varepsilon$, ensuring thus that  $\sigma_{m1}^{(1)}$ is close to $\sigma_{m1}^{(0)}$.
As a consequence, if $g$ is a differentiable function of $\sigma_{m1}$, then it follows that $g(\sigma_{m1}^{(1)})=g(\sigma_{m1}^{(0)})+\eta \sigma_{m1}^*\ g_{,\sigma}(\sigma_{m1}^{(0)})+O(\eta^2)$.
Inserting \eqref{Lsig_form} in the right hand side of \eqref{toy_canonical_eq_final}
and using \eqref{eq_points_cons_cond}, we obtain after some computations:
\begin{equation}\label{eq_points_diss_cond}
\begin{split}
&\widetilde{h}_{,L}^{(m)} (\widetilde{L}_1, \widetilde{G}_0, \widetilde{H}_0)+\varepsilon
\widetilde{\mathcal{A}}_{,L}^{(m)} (\widetilde{L}_1, \widetilde{G}_0, \widetilde{H}_0) \cos (\sigma_{m1}^{(1)}
-\widetilde{\varphi}^{(m)}(\widetilde{G}_0, \widetilde{H}_0))\\
&= \eta [\widetilde{h}_{,LL}^{(m)}(\widetilde{L}_0, \widetilde{G}_0, \widetilde{H}_0)
+\varepsilon \widetilde{\mathcal{A}}_{,LL}^{(m)}(\widetilde{L}_0, \widetilde{G}_0, \widetilde{H}_0)
\cos(\sigma_{m1}^{(0)}-\widetilde{\varphi}^{(m)}( \widetilde{G}_0, \widetilde{H}_0))] L^*+O(\eta^2)\,,\\
& \varepsilon \widetilde{\A}^{(m)} (\widetilde{L}_1, \widetilde{G}_0, \widetilde{H}_0)
\sin(\sigma_{m1}^{(1)}-\widetilde{\varphi}^{(m)}( \widetilde{G}_0, \widetilde{H}_0)) -
\eta D_{L}^{(m)} (\widetilde{L}_1, \widetilde{G}_0, \widetilde{H}_0)  \\
&=\eta [\varepsilon \widetilde{\mathcal{A}}^{(m)}(\widetilde{L}_0, \widetilde{G}_0, \widetilde{H}_0)
\sigma_{m1}^*\cos (\sigma_{m1}^{(0)}-\widetilde{\varphi}^{(m)}(\widetilde{G}_0, \widetilde{H}_0))-
D_L^{(m)}(\widetilde{L}_0, \widetilde{G}_0, \widetilde{H}_0)] +O(\eta^2)\,.\\
\end{split}
\end{equation}

Taking into account that $\varepsilon$ is a small parameter, it follows that the quantity in brackets at
the right hand side of the first of $\eqref{eq_points_diss_cond}$ is different from zero
for $\varepsilon$ sufficiently small.
Therefore, for $(\sigma_{m1}^{(1)}$, $\widetilde{L}_1)$ to be an equilibrium point (at first
order in $\eta$) for the dissipative toy model,
one should have $L^*=0$ and, consequently,
\begin{equation}\label{Lsigma_star}
 \sigma_{m1}^*= \frac{D_L^{(m)}(\widetilde{L}_0, \widetilde{G}_0, \widetilde{H}_0)}
 {\varepsilon \widetilde{\mathcal{A}}^{(m)}(\widetilde{L}_0, \widetilde{G}_0, \widetilde{H}_0)
 \cos(\sigma_{m1}^{(0)}-\widetilde{\varphi}^{(m)}(\widetilde{L}_0, \widetilde{G}_0, \widetilde{H}_0))}\ .
\end{equation}
From \eqref{Lsig_form} and \eqref{Lsigma_star}, we get \eqref{eq_L_sigma}. $\square$

\vskip.1in

\begin{remark}
Since $\varepsilon$ and $\eta$ are small (for instance, for the 14:1 resonance the parameter
$\varepsilon$ is of the order of $10^{-9}$, while $\eta$ is smaller than $\varepsilon$),
the existence condition can be replaced by the following simplified inequality
$$
\left| \frac{\eta D^{(m)}_{_L} (\widetilde{L}_0, \widetilde{G}_0, \widetilde{H}_0)}{\varepsilon
\widetilde{\mathcal{A}}^{(m)}(\widetilde{L}_0, \widetilde{G}_0, \widetilde{H}_0)} \right|  \leq 1-\delta\ ,
$$
where $\varepsilon$ and $\eta$ satisfy the relation
$$
 \gamma_1 \varepsilon+\gamma_2 \eta+ \gamma_3 \varepsilon^2< \delta,
$$
for some positive constants $\gamma_1$, $\gamma_2$ and $\gamma_3$.
\end{remark}

Besides the existence condition \eqref{existence_condition},  Theorem~\ref{Theorem:existence}  shows that a change in magnitude of the dissipative effects leads to a shift of the equilibrium
points on the $\sigma_{m1}$ axis,  $\widetilde{L}$ (or equivalently the semimajor axis $a$) remaining unchanged. Indeed, in the bottom panels of Figure~\ref{fig:cartography}, obtained for $B=30\, [cm^2/kg]$ (left) and $B=220\, [cm^2/kg]$ (right), the centers of the islands are
located at about $\sigma_{14,1}=48^o$ and $\sigma_{14,1}=60^o$, respectively,  revealing thus the shift of
equilibrium points on the $\sigma_{14,1}$ axis, while confirming that the value of $\widetilde L$ at equilibrium
does not change.

\subsection{Type of equilibrium points} \label{sec:type}

In Section~\ref{sec:existence} we investigated the existence of equilibrium points without
specifying their character. Since the conservative toy model reduces to a pendulum problem,
the equilibrium points are centers and saddles.  Therefore, it remains to clarify the nature
of equilibria for the dissipative toy model. The link between the character of the equilibria
in the conservative and dissipative frameworks is given by the following result.

\begin{theorem}\label{Theorem:type}
For given values of $\widetilde{G}_0$ and $\widetilde{H}_0$, let $(\sigma_{m1}^{(0)}, \widetilde{L}_0)$, $m\in \{11,12,13,14\}$,
be an equilibrium point for the conservative toy model described by the Hamiltonian \equ{toy_ham_canonical},
satisfying \eqref{existence_condition} for some $\delta>0$, $0<\varepsilon<1$. Assume that the existence condition \eqref{existence_condition} is satisfied and
that $\eta<\varepsilon$.
Then, the following statement holds true:
if $(\sigma_{m1}^{(0)}, \widetilde{L}_0)$ is a center (respectively a saddle) for the conservative toy model,
then the equilibrium point at first order in $\eta$,
say $(\sigma_{m1}^{(1)}, \widetilde{L}_1)$, defined by \eqref{eq_L_sigma} is an unstable spiral (respectively
a saddle) for the dissipative toy model described by \equ{toy_canonical_eq_final}.
\end{theorem}

{\bf Proof.}
Using \equ{eq_points_cons_cond}, the Jacobian matrix associated to the conservative case has the form:
$$
{J}_{C}=\left(%
\begin{array}{cc}
  0 & \widetilde{h}^{(m)}_{,LL}
  +\varepsilon  \widetilde{\mathcal{A}}_{,LL}^{(m)}
  \cos(\sigma_{m1}^{(0)}- \widetilde{\varphi}^{(m)})  \\
  \varepsilon  \widetilde{\mathcal{A}}^{(m)} \cos(\sigma_{m1}^{(0)}-
  \widetilde{\varphi}^{(m)}) & 0 \\
 \end{array}%
\right)\ ,
$$
where all functions are computed at $(\widetilde{L}_0, \widetilde{G}_0, \widetilde{H}_0)$.
Since $\varepsilon$ is a small parameter and $\widetilde{h}^{(m)}(\widetilde{L}_0, \widetilde{G}_0, \widetilde{H}_0)$,
$\widetilde{\mathcal{A}}^{(m)}(\widetilde{L}_0, \widetilde{G}_0, \widetilde{H}_0)$ have the same order of magnitude,
the sign of $ \det (J_C)$ is given by the expression
$ - \varepsilon \widetilde{h}^{(m)}_{,LL}(\widetilde{L}_0, \widetilde{G}_0, \widetilde{H}_0)
\widetilde{\mathcal{A}}^{(m)}(\widetilde{L}_0, \widetilde{G}_0, \widetilde{H}_0)
 \cos(\sigma_{m1}^{(0)}- \widetilde{\varphi}^{(m)}(\widetilde{L}_0, \widetilde{G}_0, \widetilde{H}_0))$,
provided $\varepsilon$ is sufficiently small.
Moreover, taking into account that $\widetilde{h}^{(m)}_{,LL}(\widetilde{L}_0, \widetilde{G}_0, \widetilde{H}_0)<0$
(provided $\alpha$ in \equ{hAphi_tilde} is sufficiently small) and (as we mentioned in Section~\ref{sec:resonant})
$\widetilde{\mathcal{A}}^{(m)}(\widetilde{L}_0, \widetilde{G}_0, \widetilde{H}_0)>0$, then  for
$\sigma_{m1}^{(0)}=\widetilde{\varphi}^{(m)}(\widetilde{L}_0, \widetilde{G}_0, \widetilde{H}_0)+2 k \pi $,
$k \in \mathbb{Z}$, one has that $\det (J_C) >0$. As a consequence, $(\sigma_{m1}^{(0)}, \widetilde{L}_0)$
is a center, while for $\sigma_{m1}^{(0)}=\widetilde{\varphi}^{(m)}+\pi +2 k \pi $,
$k \in \mathbb{Z}$, the equilibrium point $(\sigma_{m1}^{(0)}, \widetilde{L}_0)$ is a saddle.

Assuming that the existence condition \eqref{existence_condition} is satisfied, then
for the dissipative case, the Jacobian matrix is
\beq{Jdissipative}
{J}_{D}=\left(%
\begin{array}{cc}
 - \varepsilon  \widetilde{\mathcal{A}}_{,L}^{(m)}  \sin(\sigma_{m1}^{(1)}- \widetilde{\varphi}^{(m)}) &
 \widetilde{h}^{(m)}_{,LL} +\varepsilon  \widetilde{\mathcal{A}}_{,LL}^{(m)} \cos(\sigma_{m1}^{(1)}- \widetilde{\varphi}^{(m)})  \\
  \varepsilon  \widetilde{\mathcal{A}}^{(m)} \cos(\sigma_{m1}^{(1)}- \widetilde{\varphi}^{(m)}) &
  \varepsilon  \widetilde{\mathcal{A}}_{,L}^{(m)} \sin(\sigma_{m1}^{(1)}- \widetilde{\varphi}^{(m)})-\eta D_{L,L}^{(m)} \\
 \end{array}%
\right)\ ,
\eeq
where all functions are evaluated at $\widetilde{L}_1$, $\widetilde{G}_0$, $\widetilde{H}_0$.

Using that $\eta$ is smaller
than $\varepsilon$ (thus ensuring that $\sigma_{m1}^{(1)}$ is close to $\sigma_{m1}^{(0)}$), then if $f$ and $g$ are two differentiable functions of $\widetilde{L}$ and $\sigma_{m1}$, respectively, in view of \eqref{Lsig_form} and the fact that $L^*=0$, we can write
\begin{equation}\label{fgL0L1}
\begin{split}
& f(\widetilde{L}_1)= f(\widetilde{L}_0)+ f_{,L}(\widetilde{L}_0) (\widetilde{L}_1-\widetilde{L}_0)+O(\eta^2)= f(\widetilde{L}_0)+O(\eta^2)\,\\
& g(\sigma_{m1}^{(1)})= g(\sigma_{m1}^{(0)})+g_{,\sigma}(\sigma_{m1}^{(0)}) (\sigma_{m1}^{(1)}-\sigma_{m1}^{(0)})+O(\eta^2)=g(\sigma_{m1}^{(0)})+\eta g_{,\sigma}(\sigma_{m1}^{(0)}) \sigma_{m1}^{*}+O(\eta^2).
\end{split}
\end{equation}

From \eqref{eq_points_cons_cond}, \eqref{Jdissipative} and  \eqref{fgL0L1} it follows that
($tr$ is the trace of the matrix and det its determinant)
\begin{equation}\label{JDtrace}
\begin{split}
&tr (J_D)=  -\eta D_{L,L}^{(m)}+ O (\eta^2)\,,\nonumber\\
&\det(J_D)= -\varepsilon \widetilde{h}^{(m)}_{,LL}  \mathcal{A}^{(m)} \cos(\sigma_{m1}^{(0)}-
\widetilde{\varphi}^{(m)})+ O (\eta \varepsilon)+ O (\eta^2)+ O(\varepsilon^2)\,,\nonumber\\
& \Bigl(tr (J_D)\Bigr)^2/4-\det(J_D)=\varepsilon \widetilde{h}^{(m)}_{,LL} \mathcal{A}^{(m)}
\cos(\sigma_{m1}^{(0)}- \widetilde{\varphi}^{(m)})+ O (\eta \varepsilon)+ O (\eta^2)+ O(\varepsilon^2)\ ,\nonumber
\end{split}
\end{equation}
where all functions are evaluated at $\widetilde{L}_0$, $\widetilde{G}_0$, $\widetilde{H}_0$.
In order to establish the nature of the equilibrium point $(\sigma_{m1}^{(1)}, \widetilde{L}_1)$,
we must know the sign of the above quantities. Therefore, let us discuss in more detail the sign of $D_{L,L}^{(m)}$.
In view of the first of $\eqref{diss_D}$, we get
\begin{equation}\label{D_derivative}
D_{L,L}^{(m)}=-\frac{\omega_E}{\mu_E} \biggl(1-\frac{\omega_E \widetilde{L}^3_0 (\widetilde{H}_0+m \widetilde{L}_0)}
{\mu_E^2 (\widetilde{G}_0+\widetilde{L}_0)} \biggr) \frac{[3 \widetilde{L}^2_0
(\widetilde{H}_0+m \widetilde{L}_0)+m \widetilde{L}_0^3](\widetilde{G}_0+\widetilde{L}_0)-
\widetilde{L}_0^3 (\widetilde{H}_0+m \widetilde{L}_0)}{(\widetilde{G}_0+\widetilde{L}_0)^2}\ .
\end{equation}
To evaluate the sign of the above expression, we take into account that the
eccentricity is a small quantity, say $e=O(\epsilon)$ with $\epsilon$ small. Therefore,
from \eqref{LGH_aei} and \eqref{canonical_transformation},
it follows that $\widetilde{G}_0=O(\epsilon)$, $\widetilde{H}_0=\widetilde{L}_0(\cos i -m)+O(\epsilon)$,
which leads to
$$
[3 \widetilde{L}^2_0 (\widetilde{H}_0+m \widetilde{L}_0)+m \widetilde{L}_0^3](\widetilde{G}_0+\widetilde{L}_0)
-\widetilde{L}_0^3 (\widetilde{H}_0+m \widetilde{L}_0) = \widetilde{L}_0^4 (2\cos i +m)+O(\epsilon) >0
$$
for $m>2$.
Since the term in round brackets at the right hand side of \eqref{D_derivative} is positive for all resonances
within the geostationary distance,
we deduce that $D_{L,L}^{(m)}$ is negative and, as a consequence, $tr (J_D)$ is positive, provided
$\eta$ is sufficiently small.

We are therefore led to the following conclusion. If $(\sigma_{m1}^{(0)}, \widetilde{L}_0)$
is a center for the conservative model and using that $\eta$ is smaller
than $\varepsilon$ (thus ensuring that $\sigma_{m1}^{(1)}$ is close to $\sigma_{m1}^{(0)}$), then
one has $tr (J_D)>0$,  $\det(J_D) > 0$,
$\Bigl(tr (J_D)\Bigr)^2/4-\det(J_D) <0$ and, as a consequence,
$(\sigma_{m1}^{(1)}, \widetilde{L}_1)$ is an unstable spiral for the dissipative toy model.
Otherwise, if $(\sigma_{m1}^{(0)}, \widetilde{L}_0)$ is a saddle for the conservative model,
then $\det(J_D) < 0$, $\Bigl(tr (J_D)\Bigr)^2/4-\det(J_D) >0$,
which means that $(\sigma_{m1}^{(1)}, \widetilde{L}_1)$ is a saddle point for the dissipative toy model. $\square$

\subsection{Location of equilibrium points} \label{sec:location}

\begin{figure}[h]
\centering \vglue0.1cm \hglue0.2cm
\includegraphics[width=6truecm,height=5truecm]{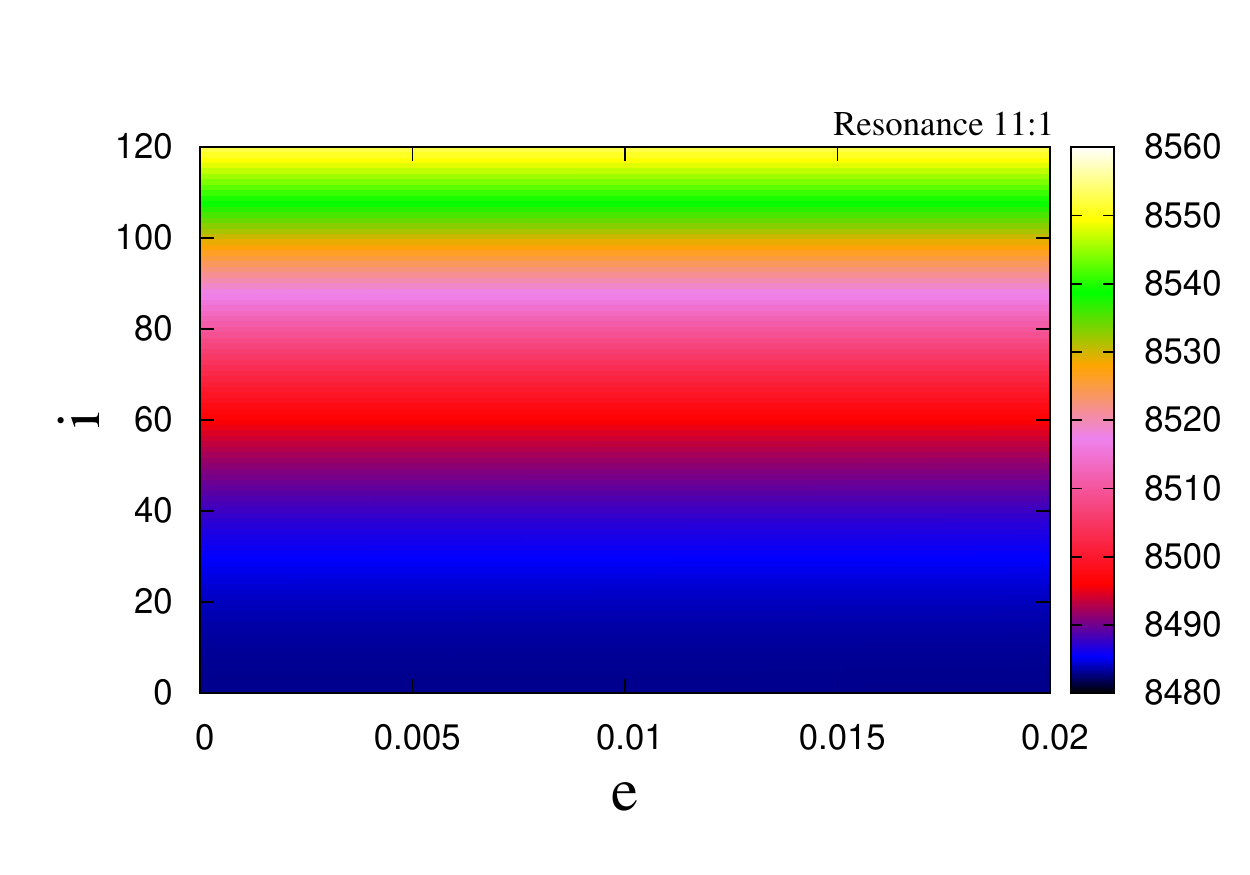}
\includegraphics[width=6truecm,height=5truecm]{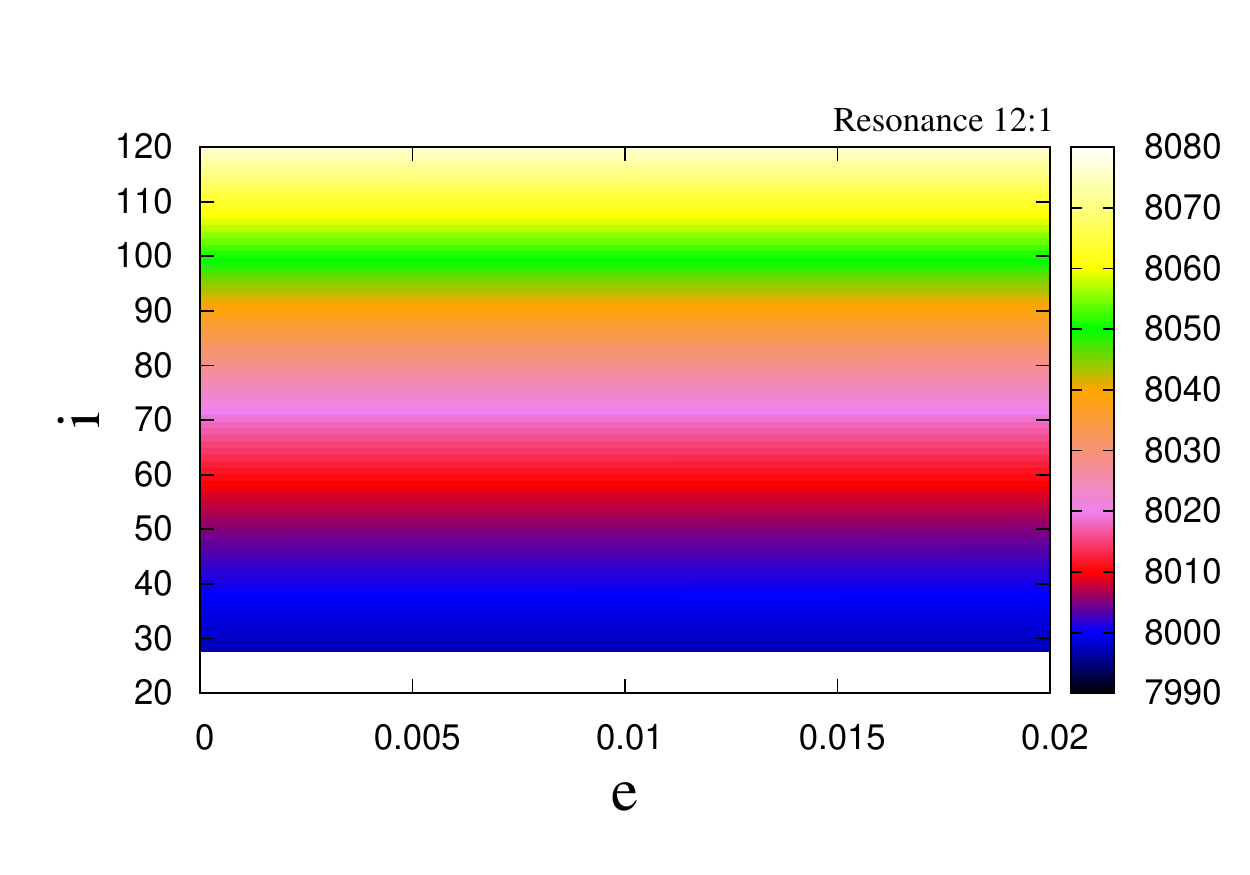}
\includegraphics[width=6truecm,height=5truecm]{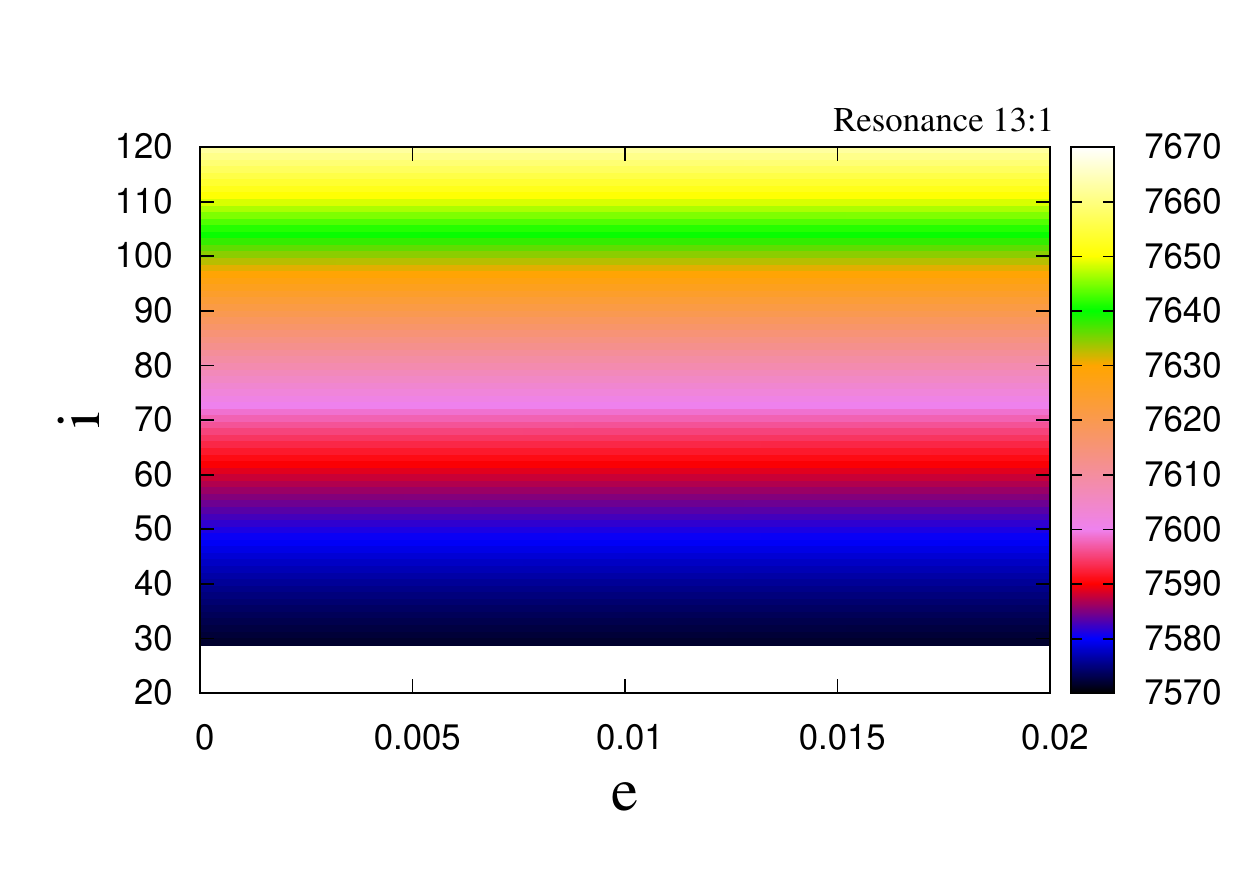}
\includegraphics[width=6truecm,height=5truecm]{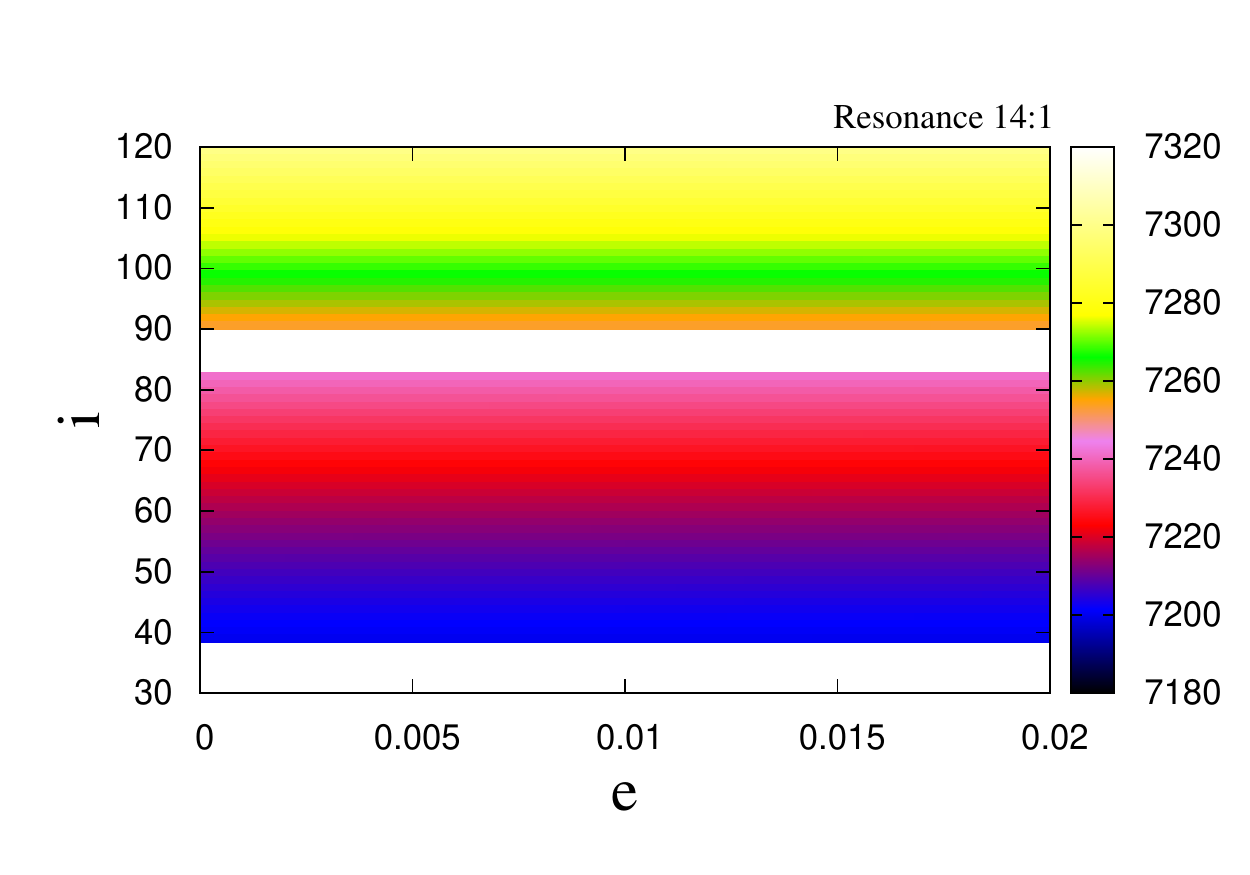}
\vglue0.5cm
\caption{Position of the equilibrium points obtained using \equ{eq_L_sigma}
on the semi-major axis as a function of eccentricity and inclination.
The color bar provides the distance of the equilibrium points from the Earth's center. From top left to bottom right: 11:1, 12:1, 13:1, 14:1
resonances.
Excluding the 11:1 resonance obtained within the conservative case, all other plots are given
for  $B=220\, [cm^2/kg]$ and mean values of the atmospheric density. For the white zones,
the existence condition \eqref{existence_condition} is not satisfied,
which implies that the equilibrium points do not exist.} \label{fig:eq_a_axis}
\end{figure}

\begin{figure}[h]
\centering \vglue0.1cm \hglue0.2cm
\includegraphics[width=6truecm,height=5truecm]{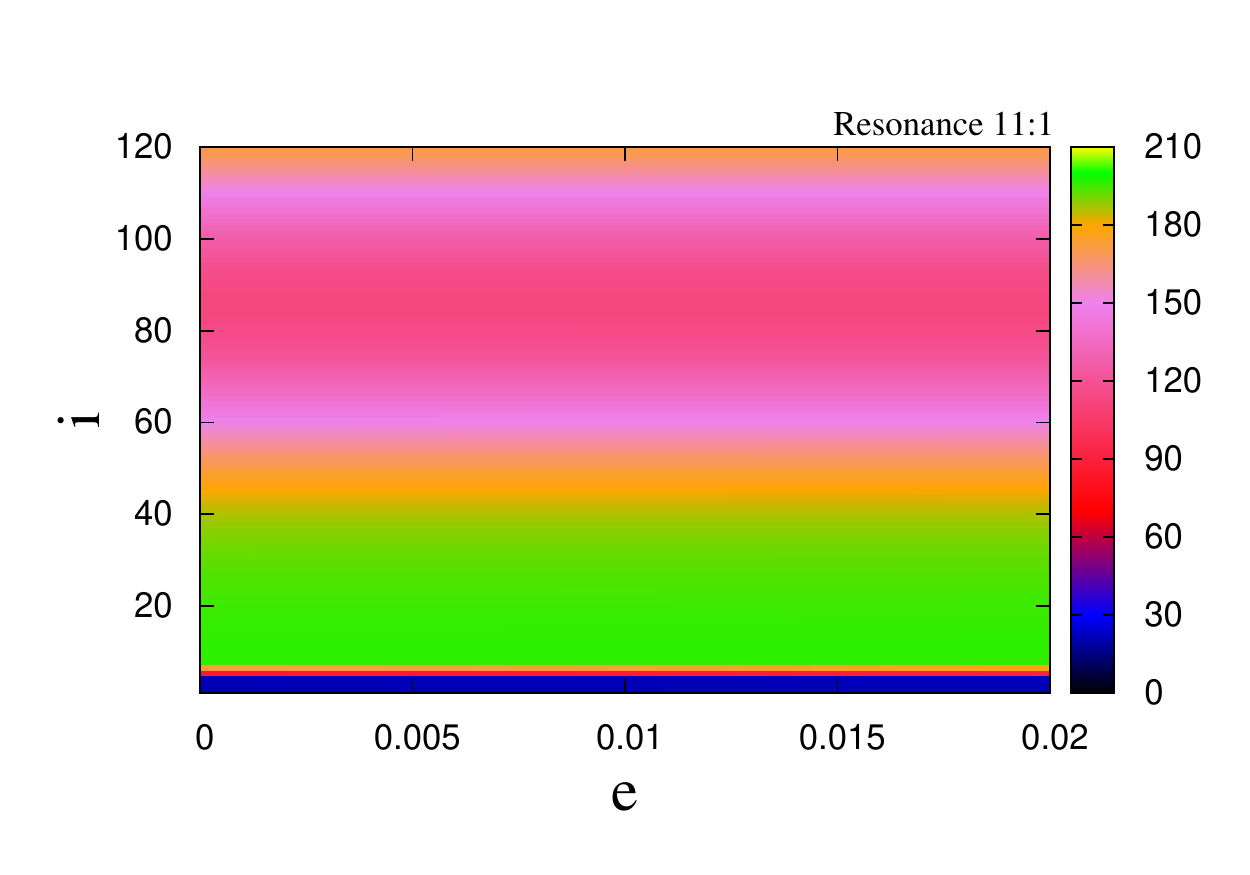}
\includegraphics[width=6truecm,height=5truecm]{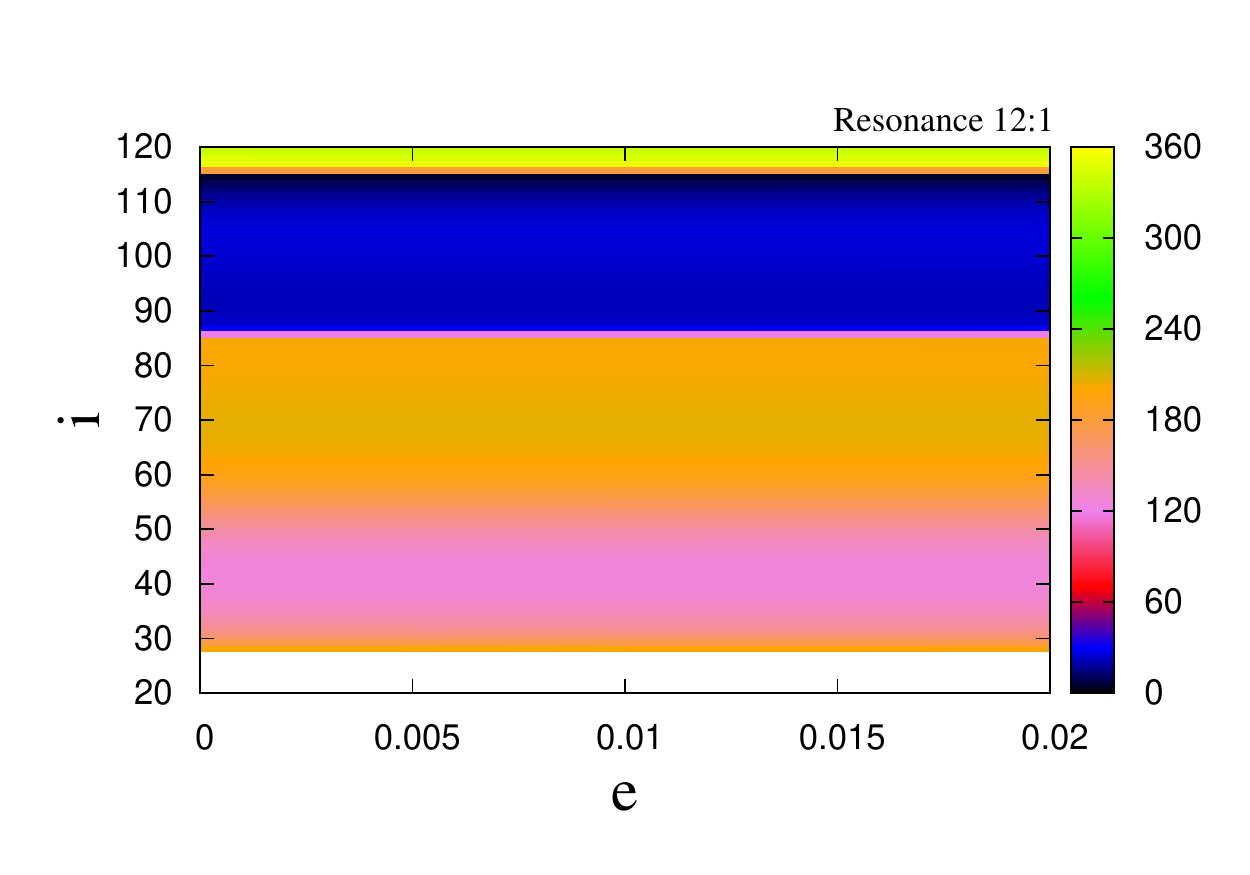}
\includegraphics[width=6truecm,height=5truecm]{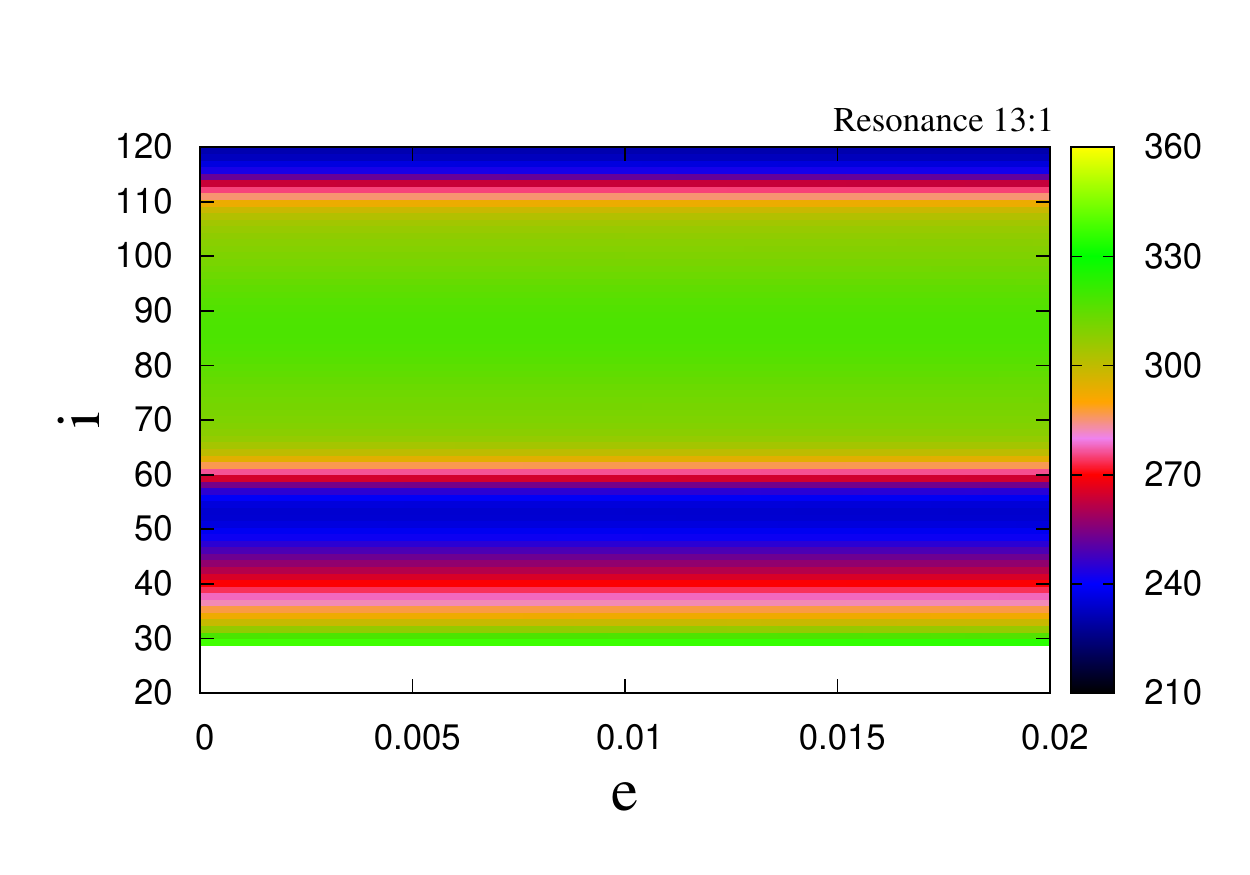}
\includegraphics[width=6truecm,height=5truecm]{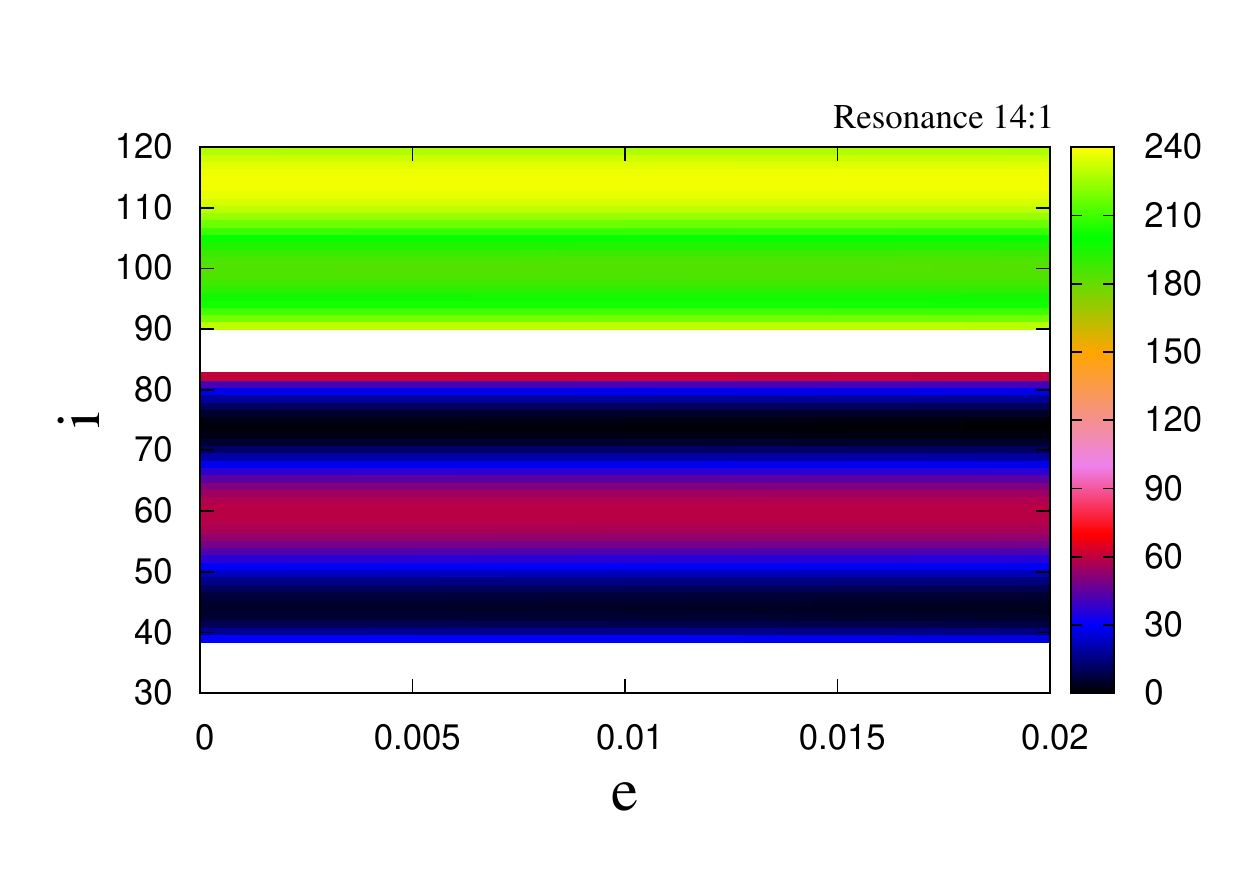}
\vglue0.5cm
\caption{Location the
equilibrium points (centers for the 11:1 resonance and spirals for the other resonances) on the $\sigma_{m1}$ axis
obtained using \equ{eq_L_sigma}. The color bar provides the position of equilibria in degrees. From top left to bottom right: 11:1, 12:1, 13:1, 14:1
resonances. Excluding the 11:1 resonance obtained within the conservative setting, all other plots are derived for  $B=220\, [cm^2/kg]$ and mean values of the atmospheric density. For the white zones, the existence condition \eqref{existence_condition}
is not satisfied, which implies that the equilibrium points do not exist.} \label{fig:eq_sigma_axis}
\end{figure}

\begin{figure}[h]
\centering \vglue0.1cm \hglue0.2cm
\includegraphics[width=6truecm,height=5truecm]{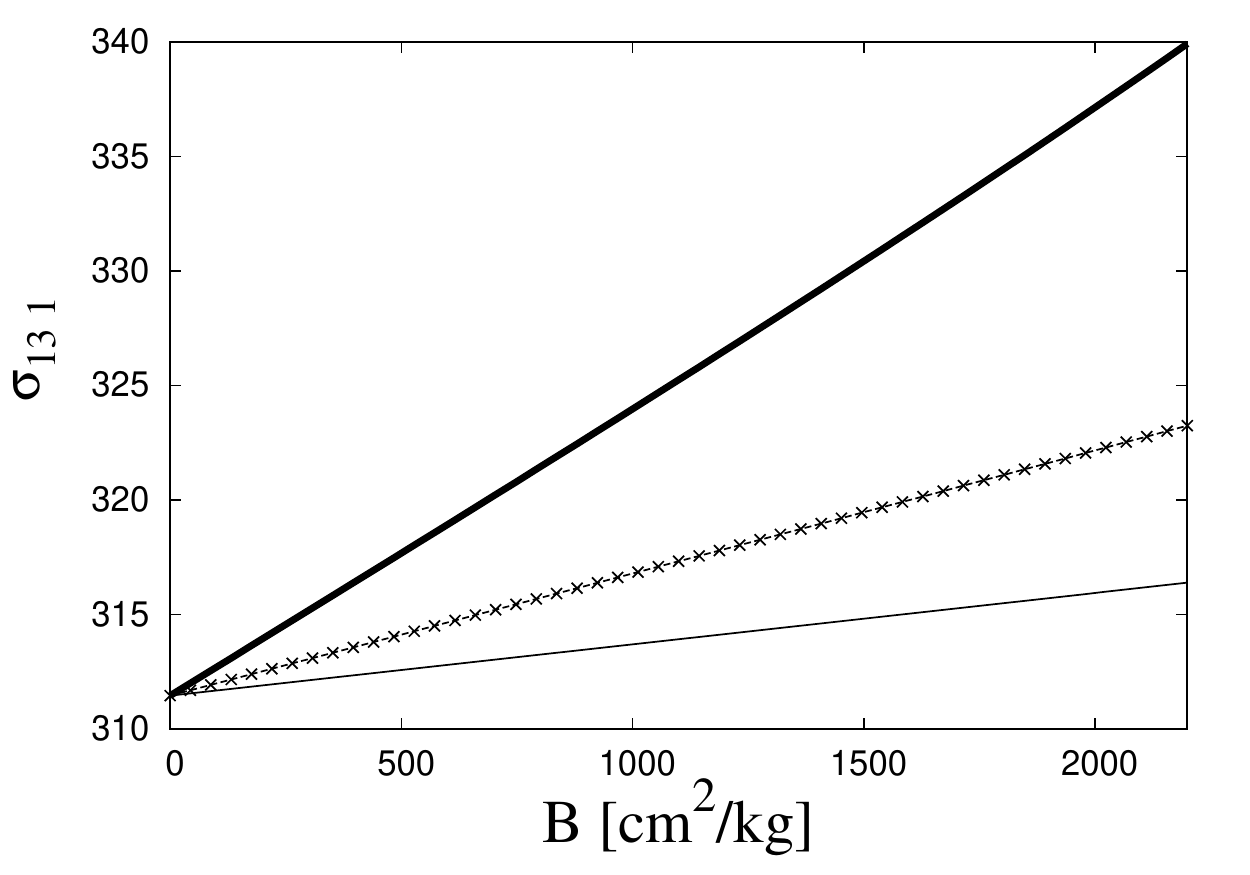}
\includegraphics[width=6truecm,height=5truecm]{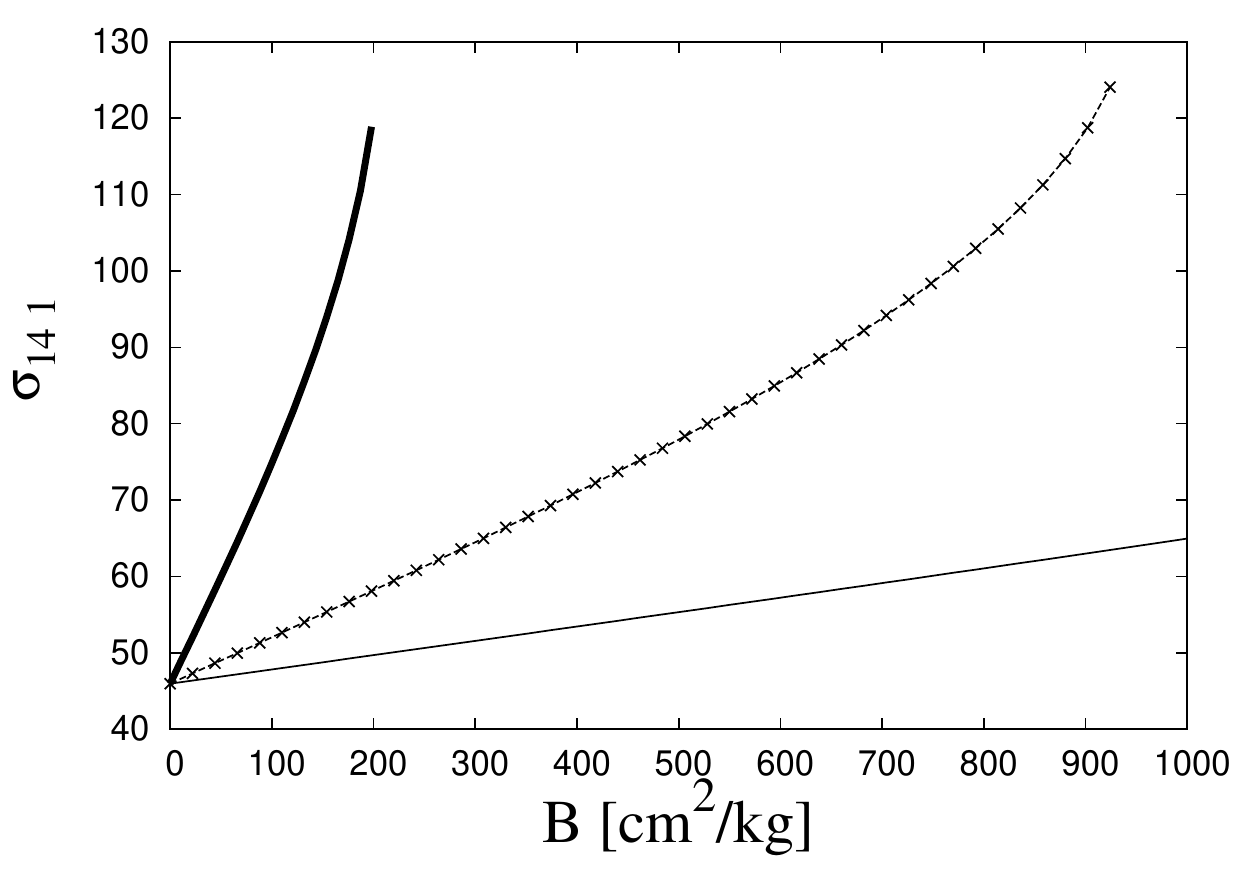}
\vglue0.8cm \caption{Location of the spiral equilibrium points on the
$\sigma_{m1}$ axis, expressed in degrees, as a function of the
ballistic coefficient $B$. The thinner lines are obtained for minimum
values of the atmospheric density, the dotted lines correspond to
mean atmospheric densities, while the thicker curves provide the
results for maximum values of the atmospheric density (see
Table~\ref{table:rho}). Left: the 13:1 resonance for $i=75^o$,
$e=0.005$, $\omega=0^o$, $\Omega=0^o$. Right: the 14:1 resonance
for $i=60^o$, $e=0.005$, $\omega=0^o$, $\Omega=0^o$.}
\label{fig:shift_sigma}
\end{figure}

Using the toy model introduced in Section~\ref{sec:toy},
we investigate the existence and location of the equilibrium points for each resonance and for
all values of eccentricity, inclination, ballistic coefficient and atmospheric density. We should stress that,
although all equilibria are unstable for the dissipative model,  the instability effects are small in the case of spiral points,
in the sense that a body placed close to this point will remain a long time in a neighborhood.
This will become evident in Section~\ref{sec:results} where various simulations are presented.

We report in Figures~\ref{fig:eq_a_axis}, \ref{fig:eq_sigma_axis} the locations of the semimajor axis and
the values of the angles $\sigma_{m1}$ for the equilibrium points (the centers for the 11:1 resonance and the spiral points
for the other resonances), as a function of eccentricity and inclination.
The plots corresponding to the 11:1 resonance are obtained within the conservative framework, while all
other plots are obtained by using the dissipative toy model with  $B=220\, [cm^2/kg]$ and for mean values
of the atmospheric density. The white color in Figures~\ref{fig:eq_a_axis}, \ref{fig:eq_sigma_axis} shows the
regions for which the existence condition \eqref{existence_condition} is not satisfied. In other words,
the dissipative effects are larger than the resonant ones, which implies that the equilibrium points do not exist.

Some transcritical bifurcation phenomena, as described in \cite{CGmajor,CGminor}, occur for the 12:1 and 14:1 resonances at
$i_0=85.99^o$ and at $i_0=86.18^o$, respectively (see the location of the unstable spiral equilibrium points on the $\sigma_{m1}$ axis close to these inclinations on the right plots of Figure~\ref{fig:eq_sigma_axis}). For example, in the case of the 14:1 resonance, the spiral point is located somewhere between $0^o$ and $30^o$ for $i \in [70^o , 80^o]$, while for $i>90^o$ the position of the spiral point is close to $200^o$. A similar remark can be made for the  resonance 12:1. The reason for the occurrence of this phenomenon is the change of sign of a specific resonant term. More precisely, from the set
$\mathcal{M}_0^{14}$, the resonant term with the largest magnitude  at high inclinations is $\mathcal{T}_{1\!5 \, 1\!4\,7\,0}$. This term changes its sign, precisely at $i_0=86.18^o$.
Therefore, in the neighborhood of $i_0$ it happens that for $i<i_0$ the spiral points are located
close to the solution of $14\, \lambda_{15,14} -90^o = 12^o$, while for $i>i_0$ the equilibrium points
are located at about $\lambda_{15,14}=192^o$. Of course, the equilibria are not located exactly at these positions,
since $\mathcal{M}_0^{14}$ contains five terms, but very close to them. In the case of the 12:1 resonance,
$\mathcal{T}_{1\!5 \, 1\!2\,7\,0}$ is the resonant term with the greatest magnitude for large inclinations and it changes its sign at $i_0=85.99^o$.

In view of Theorem~\ref{Theorem:existence}, it follows that for increasing values of $\eta$ (equivalently the ballistic coefficient and/or the atmospheric density) the white regions increase their area, while the {\it surviving} equilibria shift on the $\sigma_{m1}$ axis. For each inclination and eccentricity, one can compute the maximum value of $\eta$ up to which the inequality \eqref{existence_condition} is satisfied. In the case of
the resonances 12:1 and 13:1, the simulations show that the existence condition \eqref{existence_condition} is usually fulfilled for inclinations larger than about $40^o$, even if the ballistic coefficient is large.
On the other hand, since the atmospheric density is much larger at the altitude of 880 $km$, with notable variations during a solar cycle, the dissipative effect has an important contribution for the 14:1 resonance.
Figure~\ref{fig:shift_sigma} shows the location of spiral points on the $\sigma_{m1}$ axis as a function of the ballistic coefficient, for minimum (thin line), mean (dotted curve) and maximum (thick curve) atmospheric density in the case of
the 13:1 resonance, for $i=75^o$ and $e=0.005$, as well as for the 14:1 resonance when $i=60^o$ and $e=0.005$.
The equilibrium points of \eqref{toy_canonical_eq_final} have been numerically obtained via the
bisection method.
For the 13:1 resonance, even though $B$ varies on a large interval, all three curves are straight line segments.
On the contrary, for the 14:1 resonance a curvature of the mean (dotted curve) and maximum (thick curve)
atmospheric density is clearly visible for increasing values of $B$,  thus pointing out the
limits of the approximations \eqref{eq_L_sigma}, corresponding to the toy model.
Besides, as the right panel of Figure~\ref{fig:shift_sigma} shows, the equilibrium points do not exist
for $B>200$ $cm^2/kg$ and a maximum value for $\rho$, and respectively for $B>924$ $cm^2/kg$
and a medium value of the atmospheric density.

We remark that plots like those in Figure~\ref{fig:shift_sigma} can be used to analyze the shift of the equilibrium points
on the $\sigma_{m 1}$ axis during a solar cycle. For instance, supposing that an infinitesimal body has the
ballistic coefficient $B=150\, cm^2/kg$ then, within an interval of 11 years, the location of the spiral point
varies between $48^o$ and $92^o$ for the 14:1 resonance, when $e=0.005$ and $i=60^o$.
A satellite placed at, let say, $\sigma_{1\!4 \,  1}=70^o$ will stay very close to the spiral point,
otherwise one should slightly correct its position to remain at the equilibrium point.

\section{Solar cycle and third body effects}\label{sec:results}

In this Section we consider a more complete model, which also takes into account the variation of the local density of the atmosphere as effect of the solar cycle, as well as the perturbations induced by
Sun and Moon. We provide numerical evidence that the analytical results obtained in the previous Sections are valid when a more complete physical model is considered. In particular, we
show that an object (satellite)
placed at an equilibrium point remains there for a long time (of the order of dozens of years), even if solar cycle and third body effects are
taken into account. Thus, we show a strong evidence that these points can be exploited in practice by parking satellites in their close vicinity. We exemplify just the case of the 14:1 resonance. Since the dissipative effects gradually decrease in magnitude with the altitude, for the other resonances studied in this paper the results are definitely better.

\begin{figure}[h]
\centering \vglue0.1cm \hglue0.2cm
\includegraphics[width=6truecm,height=5truecm]{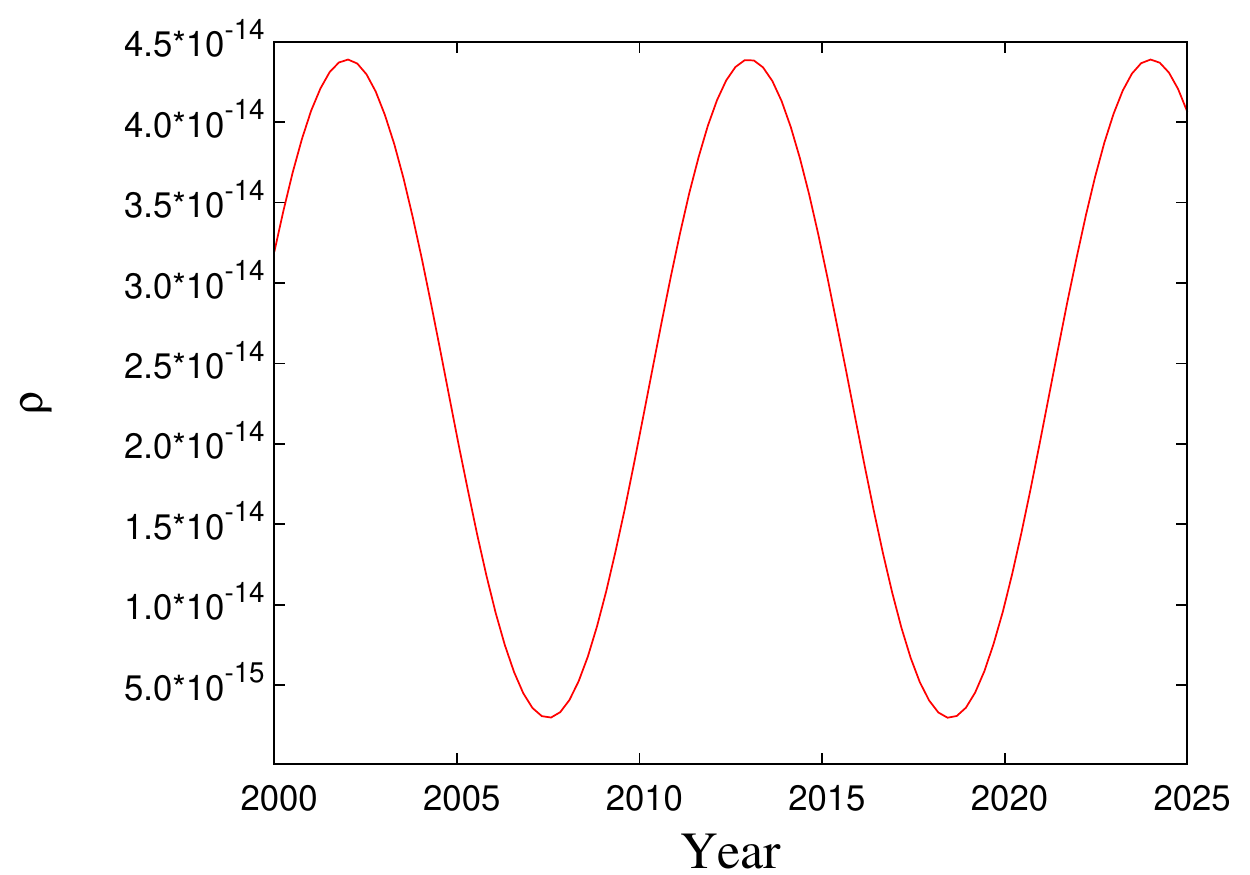}
\includegraphics[width=6truecm,height=5truecm]{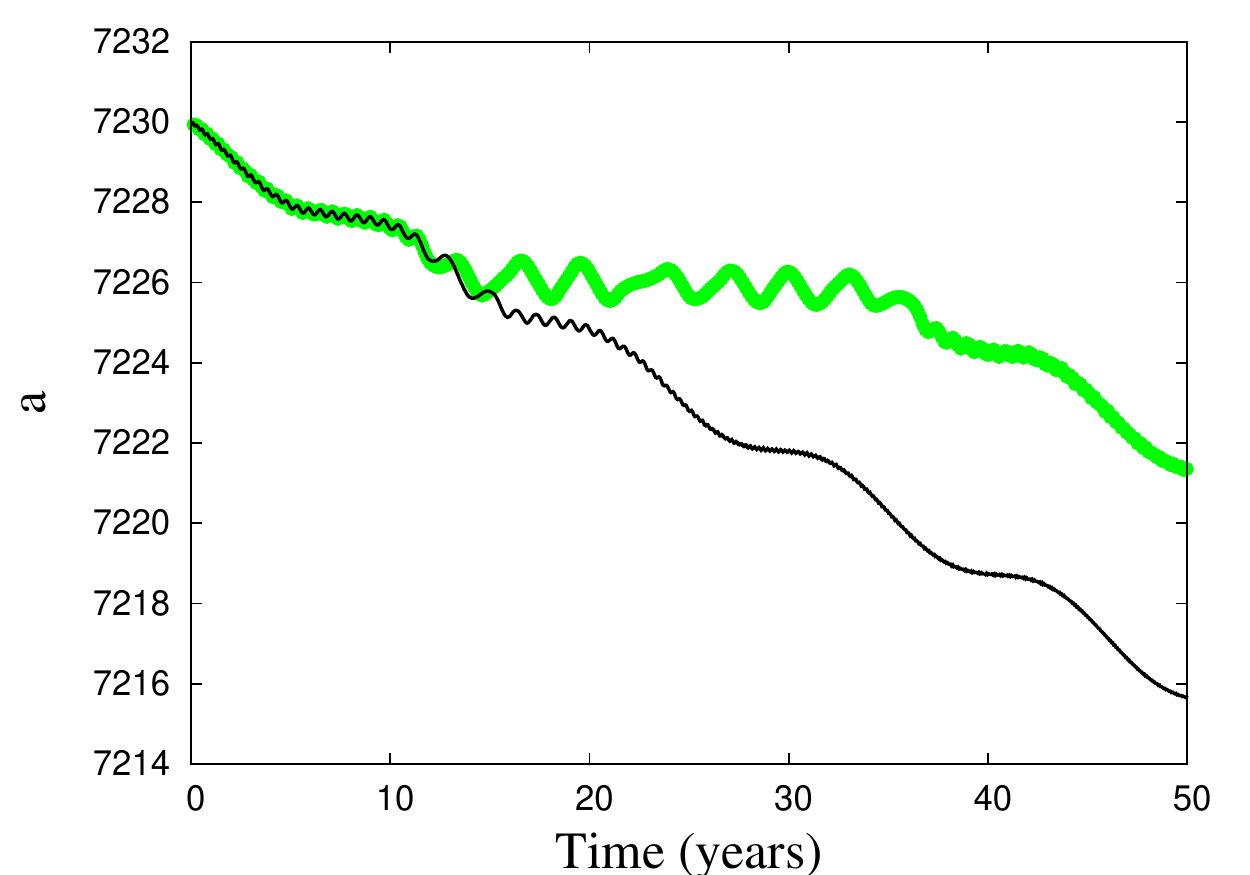}
\vglue0.8cm
\caption{Left: variation of density in $kg/m^3$ at the altitude of $800$ $km$, between the years 2000 and 2025, computed with the formula \eqref{rho_variation_cycle}.\\
Right: behavior of semi-major axis  for $B=100$ $[cm^2/kg]$ and the initial conditions $a=7230$ $km$, $e=0.005$, $i=70^o$, $\omega=0^o$, $\Omega=0^o$ and $\sigma_{1\!4\,1}=80^0$. The results obtained for the model that disregards the influence of
Sun and Moon are represented with the  green color, while the black color is used for the model that includes the attraction of
Sun and Moon. The initial epoch  is J2000 (January 1, 2000, 12:00 GMT).  } \label{fig:rho_solar_cycle}
\end{figure}

\begin{figure}[h]
\centering \vglue0.1cm \hglue0.2cm
\includegraphics[width=6truecm,height=5truecm]{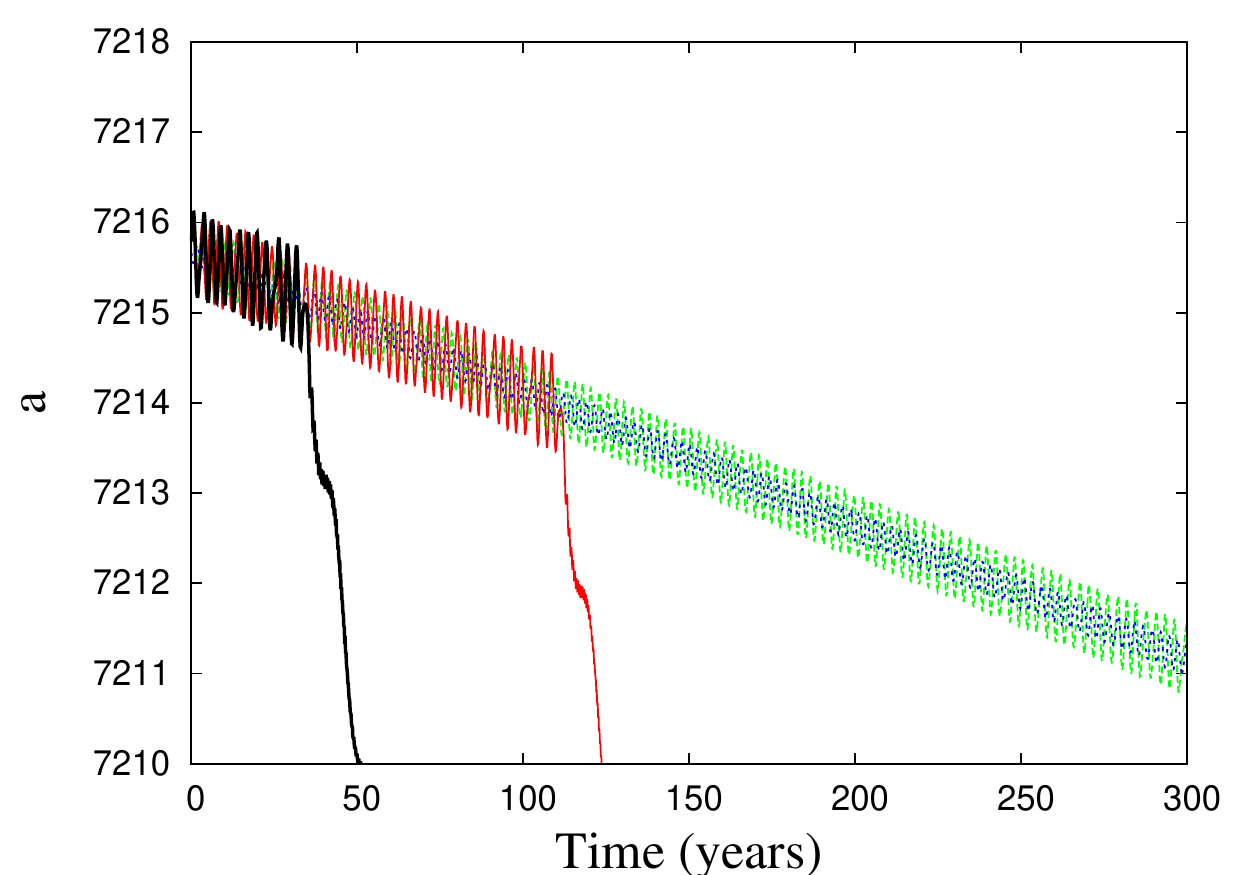}
\includegraphics[width=6truecm,height=5truecm]{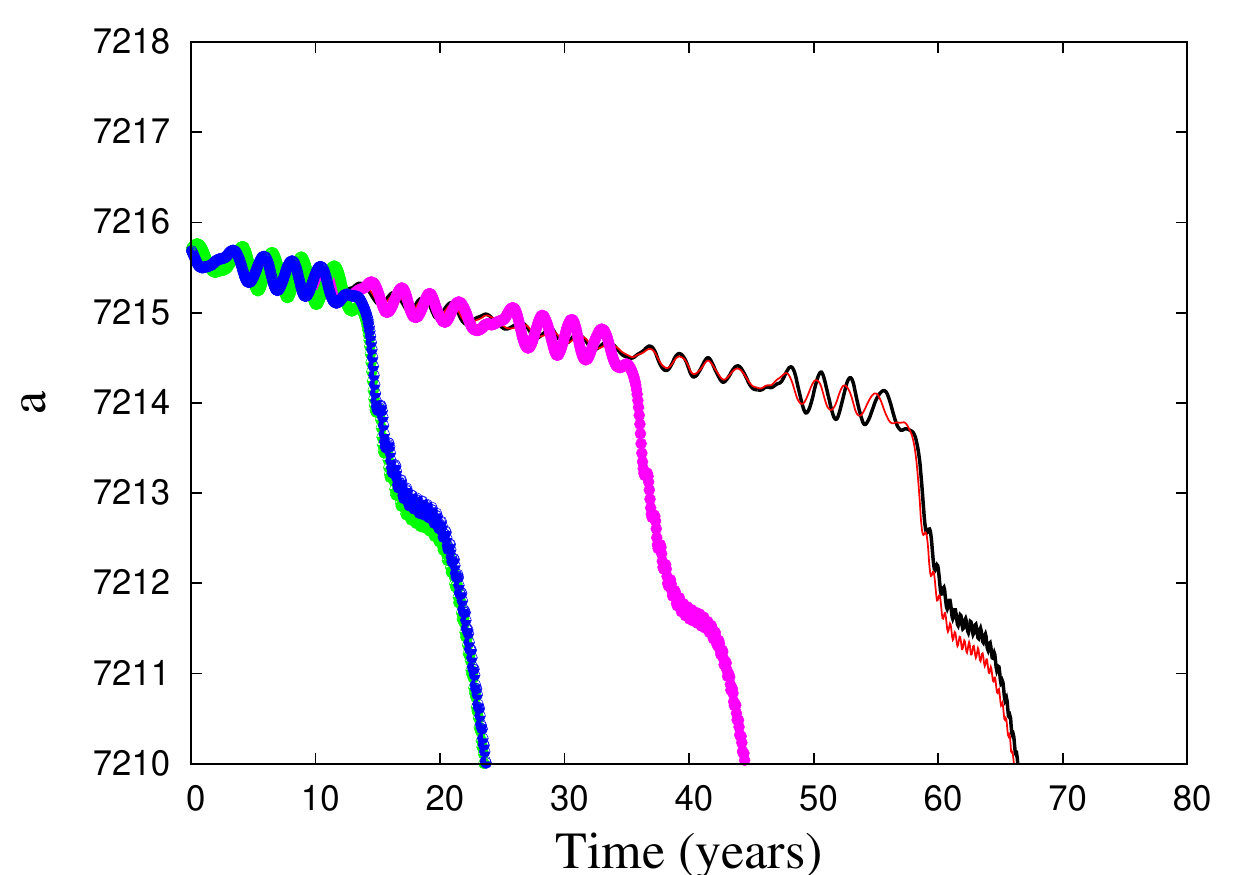}
\vglue0.8cm
\caption{Integration of several orbits showing the behavior of the semi-major axis inside the 14:1 resonance, for $B=100$ $[cm^2/kg]$ (left) and $B=200$ $[cm^2/kg]$ (right). The initial conditions, at the initial Epoch J2000 (January 1,
2000, 12:00 GMT), are $a=7215.7$ $km$, $e=0.005$, $i=60^o$, $\omega=0^o$ and $\Omega=0^o$, while for the resonant angle we used the following values. Left: $\sigma_{1\!4\,1}=50^0$ (blue), $\sigma_{1\!4\,1}=110^0$ (green), $\sigma_{1\!4\,1}=130^0$ (red), $\sigma_{1\!4\,1}=150^0$ (black). Right: $\sigma_{1\!4\,1}=70^0$ (blue), $\sigma_{1\!4\,1}=80^0$ (black), $\sigma_{1\!4\,1}=90^0$ (red), $\sigma_{1\!4\,1}=100^0$ (purple), $\sigma_{1\!4\,1}=110^0$ (green).} \label{fig:14_1_inside}
\end{figure}

\begin{figure}[h]
\centering \vglue0.1cm \hglue0.2cm
\includegraphics[width=6truecm,height=5truecm]{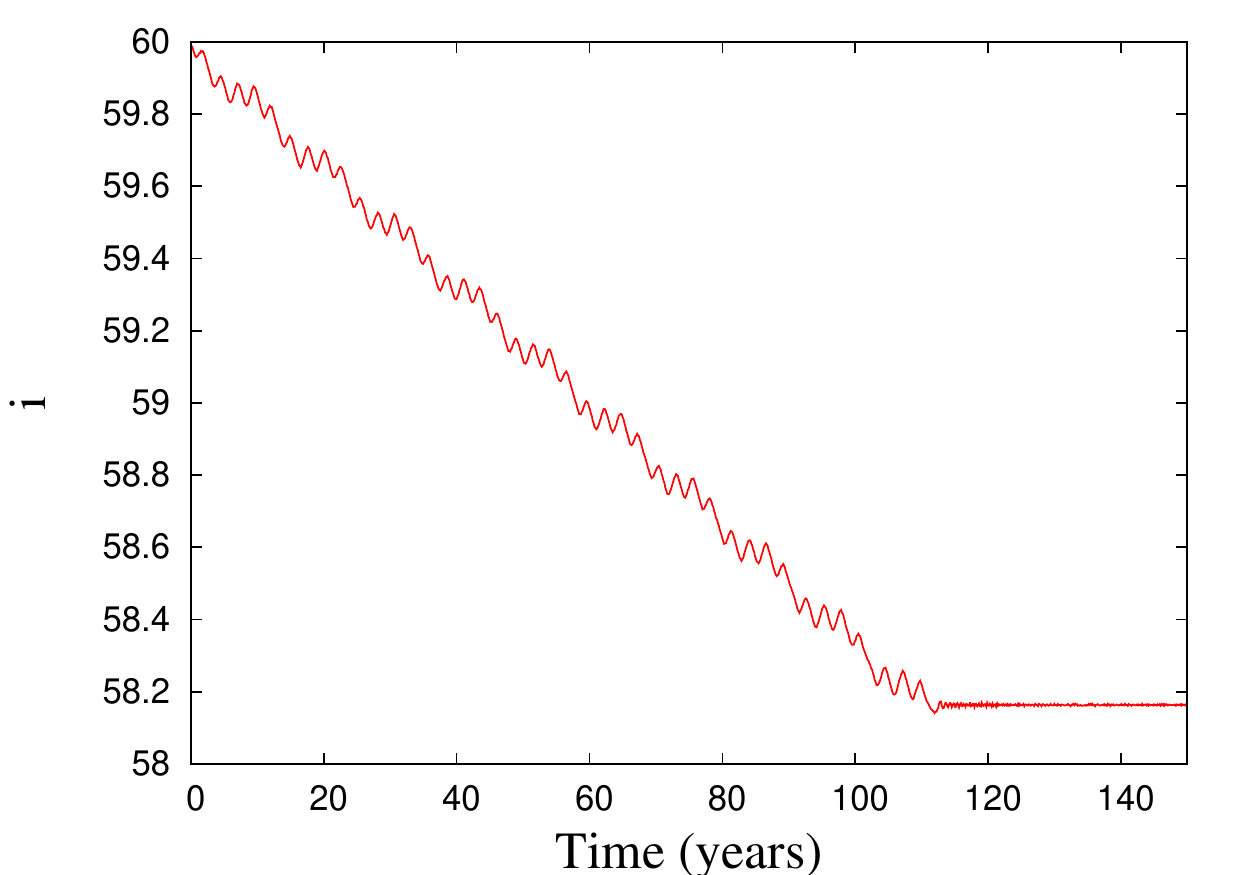}
\includegraphics[width=6truecm,height=5truecm]{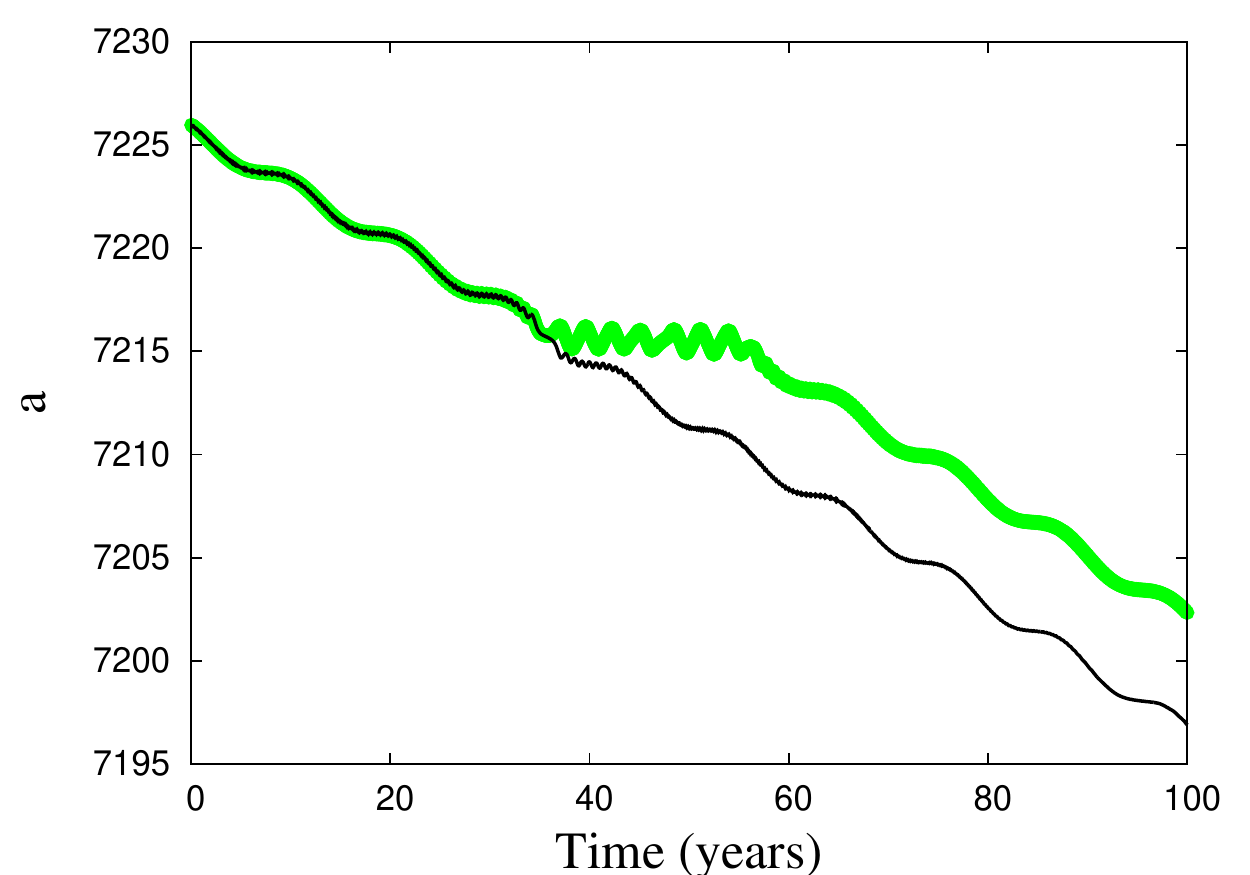}
\vglue0.8cm
\caption{Left: behavior of the inclination inside the 14:1 resonance, for $B=100$ $[cm^2/kg]$, $a=7215.7$ $km$, $e=0.005$, $i=60^o$, $\omega=0^o$, $\Omega=0^o$ and $\sigma_{1\!4\,1}=130^0$. For this orbit, the variation of
the semi-major axis is represented in red color in the left panel of Figure~\ref{fig:14_1_inside}. \\ Right: passage through the 14:1 resonance and
temporary capture into the 14:1 resonance. The plot is obtained for  $B=100$ $[cm^2/kg]$, $a=7226$ $km$, $e=0.005$, $i=60^o$, $\omega=0^o$, $\Omega=0^o$ and $\sigma_{1\!4\,1}=2^0$ for the black line (passage) and, respectively, $\sigma_{1\!4\,1}=1^0$ for the green line (capture). The initial epoch for all orbits is J2000 (January 1, 2000, 12:00 GMT).} \label{fig:14_1_outside}
\end{figure}

We suppose that the atmospheric density fluctuates with an 11-year cycle, as effect of the solar activity. To mimic the solar cycle, we shall use the following simple formula, which allows the density to vary periodically between its limits, minimum and maximum, at an altitude $h$:
\begin{equation}\label{rho_variation_cycle}
\rho(h)=\frac{\rho_{max}(h)+\rho_{min}(h)}{2}+\frac{\rho_{max}(h)-\rho_{min}(h)}{2} \cos \Bigl(\frac{2 \pi t}{T}-\phi_0\Bigr)\,,
\end{equation}
where $\rho_{max}(h)$ and $\rho_{min}(h)$ are computed by using the relation \eqref{rho_h}, $T$ is the period of the solar cycle equal to 11 years, $t$ is the time and $\phi_0$ is the phase angle. For instance, in the left panel of Figure~\ref{fig:rho_solar_cycle} we represent the variation of the density $\rho$  at the altitude of $h=800$ $km$, between the years 2000 and 2025. The solar activity depends on many factors and, of course,  one could refine or propose other equations to
model the variation of the density $\rho$. However, since our aim is to validate the analytical results presented in the previous Section, we shall keep the formulation as simple as possible.

Beside the influence of the solar cycle, we also take into account the lunisolar perturbations. In this case, the conservative part is described by the Hamiltonian
$$
\mathcal{K}=\mathcal{H}
-\mathcal{R}_{Sun}-\mathcal{R}_{Moon}\ ,
$$
where $\mathcal{H}$ is the geopotential Hamiltonian \eqref{H}, while $\mathcal{R}_{Sun}$ and $\mathcal{R}_{Moon}$ are the solar and lunar disturbing functions. We express these functions in terms of the orbital elements of both the perturbed and perturbing bodies by
considering the Kaula's expansion of the solar disturbing function (see \cite{Kaula1962}), and  the Lane's expansion of the lunar disturbing function
(see \cite{Lane1989, CGPR2016}). More precisely, the coefficients of $R_{Sun}$, $R_{Moon}$
expanded in Fourier series are functions of $(a/a_b)^n$, $e$, $e_b$, $i$, and $i_b$, while the trigonometric arguments are linear combinations of  $M$, $M_b$, $\omega$, $\omega_b$, $\Omega$, $\Omega_b$, where $n \in \mathbb{N}$, $n \geq 2$ and $a_b$, $e_b$, $i_b$, $M_b$, $\omega_b$ and $\Omega_b$ are the orbital elements of  the third body (Sun or Moon).  Since in computations we deal with finite expressions, we truncate the series expansions of the solar and lunar disturbing functions to a given order in the ratio of
the semi-major axes, and moreover we average over the fast angles.

As pointed out in various studies investigating the dynamics in the MEO region (see, e.g.,
\cite{CGfrontier, CGPbif, DRADVR15, CGPR2016, GDGR2016}), a reliable model is obtained by truncating the expansions to second order in the ratio of
the semi-major axes and averaging over both mean anomalies of the point mass and of the third body. Because LEO is closer to the Earth than MEO, then the ratio $a/a_b$ is smaller. Therefore, in LEO the lunisolar perturbations are smaller in magnitude than in MEO. In view of this argument, we shall truncate the series expansions to second order in the ratio of
the semi-major axes.

On the other hand, since in LEO the angles $\omega$ and $\Omega$ are much faster than in MEO, some resonances of the type (see
\cite{CEGGP2016} for further details)
$$
\alpha \dot\omega+ \beta \dot\Omega +\alpha_b \dot\omega_b+ \beta_b \dot\Omega_b-\gamma \dot{M}_b=0\ , \quad \alpha\,, \alpha_b \in \{\pm 2, 0\}\,,
 \hspace{0.3cm} \beta, \beta_b \in \{\pm 2, \pm 1, 0\}\ , \quad \gamma \in \mathbb{Z}\backslash\{0\}\ ,
$$
where the suffix is $b = S$  when the third-body perturber is the Sun and  it is $b = M$ for the Moon, called {\it (lunar or solar) semi-secular resonances}, might influence the long--term
evolution of the orbital elements. For small eccentricities and inclinations between 40 and 120 degrees, an analysis similar to that presented in
\cite{CEGGP2016} shows that lunar semi-secular resonances occur at an altitude below 600 $km$, while solar semi-secular resonances could occur at any altitude in LEO. For this reason, we average the Hamiltonian over the mean anomalies of the satellite and of the Moon, but  not over the mean anomaly of the Sun. In this way, we take into account the influence of some possible solar semi-secular resonances. We refer the reader to
\cite{CGPR2016} for the explicit expansions of the disturbing functions $\mathcal{R}_{Sun}$ and $\mathcal{R}_{Moon}$.

The numerical tests done so far show that lunisolar perturbations have a relatively small influence on the long-term evolution of
the semi-major axis. In the majority of the cases, we have  basically obtained the same behavior of the semi-major axis, either we have used the full model described above or we  have integrated a model that disregards the lunisolar perturbations.
However, there are some cases that show a remarkable difference. More precisely, as noticed in Section~\ref{sec:qualitative_resonance}, Figure~\ref{fig:pass_capture}, an orbit reaching the resonant region either it passes through resonance or it is
temporarily captured into resonance. This behavior has a strong stochastic feature.
Indeed, a small perturbation might lead to a different scenario than expected. For instance, in Figure~\ref{fig:rho_solar_cycle} (right panel), we describe the evolution of
the semi-major axis of an orbit, both under the model that disregards the lunisolar perturbations (green line) as well as under the full model that considers the attraction of Moon and Sun (black line). In the first case one gets the phenomenon of temporary capture into resonance, and in the second case the phenomenon of passage through the resonance. For other initial conditions the scenario could be opposite.
Therefore, even if the lunisolar perturbations are small in magnitude, they could be important in some cases, as the right panel of Figure~\ref{fig:rho_solar_cycle} shows. The study of lunisolar perturbations and of semi-secular resonances will be a subject of future work.

In Figure~\ref{fig:14_1_inside} we report some results obtained by propagating several initial conditions for a large time, starting from January 1.5, 2000 (J2000). All these initial data are located inside the libration regions of the 14:1
resonance or, to be precise, in the  basins of repulsion of the spiral equilibrium points. As the stability analysis presented in Section~\ref{sec:type} shows,  the equilibria of the dissipative toy model \eqref{toy_canonical_eq_final} are repellors.
From results of dynamical systems theory, the initial conditions located in the neighborhood
of these points do not evolve toward but rather away from them. Thus, within the framework of the dissipative system, the
libration regions of the conservative system should become sort of {\it basins of repulsion}. However, this effect is very small  even on long time scales. We will still use the terminology {\it libration regions}, even in the dissipative case, and not {\it basins of repulsion} as we should normally adopt in the framework of dissipative dynamical systems.

An important aspect, which enhances the complexity of the dynamics, is the variation of both the position of
the equilibrium points, as well as the position and width of the resonant regions, as effect of the solar cycle.
Indeed, we find that inside the libration region the initial conditions evolve (slowly) away from the spiral points,
as effect of the dissipation. Furthermore, the position of the equilibrium points, and as a consequence the position
and width of the libration regions, fluctuates with an 11-year cycle along the $\sigma_{1\!4 \,  1}$ axis. The amplitude of this variation depends on the value of the ballistic coefficient.

Figure~\ref{fig:14_1_inside} is described better if the results are corroborated with the analytical study presented in
Section~\ref{sec:qualitative_resonance}. In particular, the right panel of Figure~\ref{fig:shift_sigma} is relevant for our discussion, since it provides the shift of the equilibria on the $\sigma_{1\!4 \,  1}$ axis during a solar cycle. Thus, from Figure~\ref{fig:shift_sigma} it follows that for $B=100$ $cm^2/kg$, the position of the
spiral point oscillates between $47^o$ and $75^o$, while for $B=200$ $cm^2/kg$ between $50^o$ and $120^o$. In the left panel of Figure~\ref{fig:14_1_inside}, obtained for $B=100$ $cm^2/kg$, we integrate four orbits, characterized by the same initial conditions with the exception of the resonant angle $\sigma_{1\!4 \,  1}$ for which we took the following initial values: $50^o$ (blue), $110^o$ (green), $130^o$ (red) and $150^o$ (black). Being sufficiently close to the
spiral point, the first two initial conditions lead to {\it trapped motions} for more than $300$ years. Increasing the distance from the
spiral point, one obtains {\it escape motions} with increasingly smaller escape times.

For $i=60^o$, $e=0.005$ and ballistic coefficients larger than $200$ $cm^2/kg$, the right panel of Figure~\ref{fig:shift_sigma} shows that equilibrium points do not exist when the solar activity attains its maximum. Thus, for $B=200$ $cm^2/kg$, we do not expect to obtain trapped motions for hundreds of years. Indeed, the right panel of Figure~\ref{fig:14_1_inside} shows only {\it escape motions}, but even so, the escape time is very long in some cases.
It seems that the longest escape time is obtained for initial values of the resonant angle between $80^o$ (black line)  and $90^o$ (red line), namely at the middle of the interval $[50^o, 120^o]$, which represents the range of variation of the position of
the spiral point.

Another aspect to be noted is the fact that none of the curves drawn in Figure~\ref{fig:14_1_inside} is horizontal, but rather the semi-major axis slowly decreases in time for each orbit trapped into resonance.
For example, in the left plot of Figure~\ref{fig:14_1_inside}, the semi-major axis for the orbits represented by blue and green lines
decreases of about 4.5 $km$ within 300 years. This is due to the resonance, which slowly decreases the inclination. Indeed, the left plot of Figure~\ref{fig:14_1_outside} shows the
evolution of the inclination for the same orbit for which the variation of the semi-major axis is represented in red color in the left panel of Figure~\ref{fig:14_1_inside}.
For the trapped motion inside the resonance we notice a slow decrease of inclination from $60^o$ to $58.2^o$ within
about 100 years. Then, after the escape from the resonance, the inclination becomes nearly constant. Since the position of the equilibrium points on the semi-major axis  depends on the
inclination, see Figure~\ref{fig:eq_a_axis} and in particular the bottom right plot of Figure~\ref{fig:eq_a_axis} for the 14:1 resonance, a slow decrease of the inclination leads to a shift  of the position of equilibrium points along the semi-major axis.

Finally, the right panel of Figure~\ref{fig:14_1_outside} underlines again the stochastic behavior of the orbits reaching the resonant region. We propagate two orbits, whose initial angle $\sigma_{1\!4 \,  1}$ differs by only one degree. One orbit passes through the resonance and the other is
captured temporarily into the resonance.
At the light of the results presented in this work, we believe that it would be interesting to study passage or escape from
resonances in specific case studies as well as to move parameters or initial conditions to foster one
of the two situations, whose exploitation could be conveniently used to design disposal orbits.

\vskip.1in

\bf Acknowledgements. \rm
A.C. was partially supported by GNFM/INdAM. C.G. was supported by the Romanian Space Agency
(ROSA) within Space Technology and Advanced Research (STAR) Program
(Project no.: 114/7.11.2016).
\vglue1cm

\bibliographystyle{spmpsci}

\begin{thebibliography}{}

\bibitem[Alessi et al. (2016)]{ADRRVDQM}

E. M. Alessi, F. Deleflie, A.J. Rosengren, A. Rossi, G.B. Valsecchi, J. Daquin, K. Merz (2016),
\sl A numerical investigation on the eccentricity growth of GNSS disposal orbits, \rm
Celest. Mech. Dyn. Astr. {\bf 125}, n. 1, 71--90.


\bibitem[Bezdek \& Vokrouhlick{\'y} (2004)]{BV2004}
A. Bezdek, D. {Vokrouhlick{\'y}} (2004),
\sl Semianalytic theory of motion for close-Earth spherical satellites including drag and gravitational perturbation, \rm
Planetary and Space Science {\bf 52}, n. 14, 1233--1249.

\bibitem[Celletti (2010)]{Alebook}
A. Celletti (2010), {\sl Stability and Chaos in Celestial Mechanics}, Springer-Verlag,
Berlin; published in association with Praxis Publishing Ltd. (Chichester, ISBN:
978-3-540-85145-5).

\bibitem[Celletti \& Gale\c s (2014)]{CGmajor}
A. Celletti, C. Gale\c s (2014), \sl On the dynamics of space debris: 1:1 and 2:1 resonances, \rm
J. Nonlinear Science {\bf 24}, n. 6, 1231--1262.

\bibitem[Celletti \& Gale\c s (2015a)]{CGminor}
A. Celletti, C. Gale\c s (2015a), \sl Dynamical investigation of minor resonances for space debris, \rm Celest. Mech. Dyn. Astr.
{\bf 123},  203--222.

\bibitem[Celletti \& Gale\c s (2015b)]{CGexternal}
A. Celletti, C. Gale\c s (2015b), \sl A study of the main resonances outside the geostationary ring, \rm Advan. Space Res.
{\bf 56}, 388--405.

\bibitem[Celletti \& Gale\c s (2016)]{CGfrontier}
A. Celletti, C. Gale\c s (2016), \sl
A study of the lunisolar secular resonance $2\dot{\Omega}+\dot{\omega}=0$, \rm Frontiers in Astronomy and Space Sciences, {\bf 3}, 11 pages.

\bibitem[Celletti et al. (2016)]{CGPbif}
A. Celletti, C.  Gale\c s, G. Pucacco (2016), {\sl Bifurcation of lunisolar secular resonances for space debris orbits},
SIAM J. Appl. Dyn. Syst. {\bf 15}, 1352--1383.

\bibitem[Celletti et al. (2017a)]{CGPR2016}
A. Celletti, C. Gale\c s, G. Pucacco, A. Rosengren (2017a), \sl Analytical development of the lunisolar disturbing function and the critical
inclination secular resonance, \rm  Celest. Mech. Dyn. Astr. {\bf 127}, n.3, 259--283.

\bibitem[Celletti et al. (2017b)]{CEGGP2016}
A. Celletti, C. Efthymiopoulos, F. Gachet, C. Gale\c s, G. Pucacco (2017b), \sl Dynamical models and the
onset of chaos in space debris, \rm Int. J. Nonlinear Mechanics {\bf 90}, 147--163.

\bibitem[Chao (2005)]{Chao} C.C. Chao (2005), {\em Applied Orbit Perturbation and Maintenance},
Aerospace Press Series, AIAA (Reston, Virgina).

\bibitem[Daquin et al. (2016)]{DRADVR15}
J. Daquin, A.J. Rosengren, E.M. Alessi, F. Deleflie, G.B. Valsecchi, A. Rossi (2016), \sl The dynamical structure of the MEO region: long-term
stability, chaos, and transport, \rm Celest. Mech. Dyn. Astr. {\bf 124}, 335--366.

\bibitem[Deienno et al. (2016)]{DSBS}
R. Deienno, D. Merguizo Sanchez, A.F. Bertachini de Almeida Prado, G. Smirnov (2016),
\sl Satellite de-orbiting via controlled solar radiation pressure, \rm
Celest. Mech. Dyn. Astr. {\bf 126}, n. 4, 433--459.


\bibitem[Delhaise (1991)]{Del1991}
F. Delhaise (1991), \sl Analytical treatment of air drag and earth oblateness effects upon an artificial satellite, \rm
Celest. Mech. Dyn. Astr. {\bf 52}, n. 1, 85--103.

\bibitem[EGM (2008)]{EGM2008} Earth Gravitational Model 2008, $http://earth-info.nga.mil/GandG/wgs84/gravitymod/egm2008/$

\bibitem[Ely \& Howell (1997)]{EH}
T.A. Ely, K.C. Howell (1997), \sl Dynamics of artificial satellite orbits with
tesseral resonances including the effects of luni-solar perturbations, \rm
Dynamics and Stability of Systems {\bf 12}, n. 4, 243--269.

\bibitem[Formiga \& Vilhena del Moraes (2011)]{FV}
J.K.S. Formiga, R. Vilhena de Moraes (2011), \sl 15:1 Resonance effects on the orbital motion
of artificial satellites, \rm J. Aerospace Techn Man. {\bf 3}, n. 3, 251--258.

\bibitem[Froeschl\'e et al. (1997)]{froes} C. Froeschl\'e, E. Lega, R. Gonczi (1997),
\sl Fast Lyapunov indicators. Application to asteroidal
motion, \rm Celest. Mech. Dyn. Astr. {\bf 67}, 41--62.

\bibitem[Gaias et al. (2015)]{Gaias2015}
G. Gaias, J.-S. Ardaens, O. Montenbruck (2015), \sl Model of $J_2$
perturbed satellite relative motion with time-varying differential drag, \rm
Celest. Mech. Dyn. Astr. {\bf 123}, n. 4, 411--433.

\bibitem[Gedeon (1969)]{Gedeon}
G. Gedeon (1969), \sl Tesseral resonance effects on satellite orbits, \rm
Cel. Mech. {\bf 1}, n. 2, 167--189.

\bibitem[Gkolias (2016)]{GDGR2016}
I. Gkolias, J. Daquin, F. Gachet, A.J. Rosengren (2016), \sl From order to chaos in Earth satellite orbits, \rm
Astron. J. {\bf 152}, 5.

\bibitem[Guzzo et al. (2002)]{GLF2002}
M. Guzzo, E. Lega, Froeschl\'e (2002), \sl On the numerical detection of the effective stability of chaotic
motions in quasi-integrable systems, \rm Physica D. {\bf 163}, 1--25.

\bibitem[Guzzo \& Lega (2013)]{GL2013}
M. Guzzo, E. Lega (2013), \sl The numerical detection of the Arnold web and its use for long-term diffusion
studies in conservative and weakly dissipative systems, \rm Chaos {\bf 23}, 023124.

\bibitem[Hedin (1986)]{Hedin0}
A.E. Hedin (1986), \sl MSIS-86 thermospheric model, \rm J. Geophys. Res. {\bf 92}, 4649--4662.

\bibitem[Hedin (1991)]{Hedin}
A.E. Hedin (1991), \sl Extension of the MSIS thermosphere model into the middle and lower atmosphere, \rm
J. Geophys. Res. {\bf 96}, 1159--1172.

\bibitem[ISO 27852 (2010)]{ISO}
ISO 27852:1020(E) (2010), \sl Space systems -- Estimation of orbit lifetime. \rm

\bibitem[Jacchia (1971)]{Jacchia}
L.G. Jacchia (1971), \sl Revised static models of the thermosphere and exosphere with empirical
temperature profiles, \rm Smithsonian Astrophysical Observatory, Science Report No. 332, Cambridge, MA.

\bibitem[Kaula (1962)]{Kaula1962}
W. M. Kaula (1962), \sl Development of the lunar and solar disturbing functions for a close satellite, \rm
Astron. J. {\bf 67}, 300--303.

\bibitem[Kaula (1966)]{Kaula}
W.M. Kaula (1966), \em Theory of Satellite Geodesy, \rm Blaisdell Publ. Co.

\bibitem[Lane (1989)]{Lane1989}
M. T. Lane (1989),
\sl On analytic modeling of lunar perturbations of artificial satellites of the Earth, \rm
Celest. Mech. Dyn. Astr. {\bf 46}, 287--305.

\bibitem[Larson \& Wertz (1999)]{LW}
W. Larson, J. Wertz (1999), \sl Space mission analysis and design, \rm Kluwer publ.

\bibitem[Lema\^{\i}tre et al. (2009)]{LDV}
A. Lema\^{\i}tre, N. Delsate, S. Valk (2009),
\sl A web of secondary resonances for large $A/m$ geostationary debris, \rm
Celest. Mech. Dyn. Astr., 104, 383--402.

\bibitem[Liu \& Alford (1980)]{Liu}
J.J.F. Liu and R.L. Alford (1980), \sl Semianalytic theory for a close--Earth artificial satellite, \rm
J. Guidance and Control {\bf 3}, n. 4, 304--311.

\bibitem[Montenbruck \& Gill (2000)]{MG}
O. Montenbruck, E. Gill (2000), \em Satellite orbits, \rm Springer.

\bibitem[Rosengren \& Scheeres (2013)]{Rosengren2013}
A.J. Rosengren, D.J. Scheeres (2013), {\sl Long-term dynamics of high
area-to-mass ratio objects in high-Earth orbit,} Adv. Space Res.
\textbf{52}, 1545--1560.

\bibitem[Rosengren et al. (2014)]{RS2}
A.J. Rosengren, D.J. Scheeres, J.W. McMahon (2014), {\sl The classical
Laplace plane as a stable disposal orbit for geostationary
satellites,} Adv. Space Res. {\bf 53}, Issue 8, 1219--1228.

\bibitem[Roy (2004)]{Roy}
A. Roy (2004), \em Orbital motion, \rm (Fourth Edition) Taylor \& Francis.

\bibitem[Valk et al. (2009)]{VDLC}
S. Valk, N. Delsate, A. Lema\^{\i}tre, T. Carletti (2009), \sl Global dynamics of high area-to-mass ratios
geosynchronous space debris by means of the MEGNO indicator, \rm Advances in Space Research {\bf 43}, 1509--1526.


\end{thebibliography}

\end{document}